\documentclass{aa}
\usepackage{graphicx}
\usepackage{txfonts}
\usepackage{caption}
\usepackage{placeins}
\usepackage{amsmath}    
\usepackage{amssymb}
\usepackage{threeparttablex}
\usepackage{hyperref}
\usepackage{booktabs}
\hypersetup{
  colorlinks   = true, %Colours links instead of ugly boxes
  urlcolor     = blue, %Colour for external hyperlinks
  linkcolor    = blue, %Colour of internal links
  citecolor   = blue %Colour of citations
}
\usepackage{natbib}
\usepackage{orcidlink}
\bibpunct{(}{)}{;}{a}{}{,}
\begin{document}

\title{Companion architectures of sub-Saturns: Distinct migration pathways across the Neptunian landscape}
   
   \author{
   Luis Thomas\orcidlink{0009-0006-1571-0306}\inst{1,2}\thanks{\email{lthomas@mpe.mpg.de}}
    \and
    Louise D. Nielsen \orcidlink{0000-0002-5254-2499} \inst{1}
    \and
    Alex J. Cridland \orcidlink{0000-0003-1777-9930}  \inst{1}
    \and
    Martin Schlecker \orcidlink{0000-0001-8355-2107}  \inst{3}
    \and
    Sydney Vach \orcidlink{0000-0001-9158-9276} \inst{3}
}

   \institute{
   Universit\"ats-Sternwarte M\"unchen, Fakult\"at f\"ur Physik, Ludwig-Maximilians-Universit\"at M\"unchen, Scheinerstr.~1, 81679 M\"unchen, Germany
    \and
    Max-Planck-Institut f\"ur extraterrestrische Physik, Gie\ss enbachstra\ss e 1, 85748 Garching, Germany
    \and
    European Southern Observatory (ESO), Karl-Schwarzschild-Str. 2, 85748 Garching bei München, Germany
    }
   \date{Received 10 July 2026 / Accepted dd October 2026}

% \abstract{}{}{}{}{} 
% 5 {} token are mandatory
\abstract
  % context
  {Close-in sub-Saturns ($4.0\,\mathrm{R}_\oplus \leq R_\mathrm{p} \leq 8.5\,\mathrm{R}_\oplus$) are distributed unevenly across
the period--radius plane: they are depleted in the Neptunian desert at periods $P \lesssim 3.2$\,d, accumulate 
in a narrow overdensity at $P \simeq 3.2$--$5.7$\,d (the Neptunian ridge), and thin out in the savanna at $P \gtrsim 5.7$\,d.}
  % aims
  {We test whether sub-Saturns in different landscape regions have
  systematically different planetary companion architectures, as expected if desert
  and ridge planets arrived through high-eccentricity migration and
  savanna planets migrated quiescently.}
  % methods
  {We compiled 90 sub-Saturns in 86 systems with both transit and radial
  velocity (RV) data, constructed per-system RV and transit injection-recovery
  completeness maps, and combined them into detection probability surfaces for
  companions. From these maps we derived completeness-corrected occurrence rates for different companion types with a joint Poisson occurrence model.}
  % results
  {Companion architectures differ significantly across the landscape. Nearby companions ($P_\mathrm{comp} < 200$\,d) accompany $91.1_{-4.3}^{+3.2}\%$ of savanna sub-Saturns but only $36_{-14}^{+16}\%$ of desert and ridge sub-Saturns. This contrast is driven almost entirely by small ($M_\mathrm{comp} < 20\,\mathrm{M}_\oplus$) companions, accompanying $88.2_{-5.5}^{+4.2}\%$ of savanna but only $29_{-14}^{+17}\%$ of desert and ridge sub-Saturns, while medium-mass and giant companions within 200\,d are rare ($\lesssim 13\%$) everywhere, so the sub-Saturn typically dominates its inner system. Savanna sub-Saturns moreover reside in compact multi-planet systems. These contrasts are robust to crosscuts in sub-Saturn radius, bulk density, eccentricity, and host-star properties.}
  % conclusions
  {Desert and ridge sub-Saturns reside in dynamically emptied systems whose nearby companion rates resemble those of hot Jupiters, while savanna sub-Saturns inhabit compact multi-planet systems resembling those of warm Jupiters. This parallel supports two migration channels within a single population: high-eccentricity migration delivering planets to the desert and ridge, and quiescent disk migration or in-situ formation populating the savanna.}


   \maketitle
% 
%-------------------------------------------------------------------

\section{Introduction}
The sub-Saturn exoplanet population ($4.0\,\mathrm{R}_\oplus \leq R_\mathrm{p} \leq 8.5\,\mathrm{R}_\oplus$) exhibits distinct features that constrain planet formation and evolution, most notably the Neptunian desert, a prominent paucity of close-in planets \citep{lecavelier2007,szabo2011,nesvorn2013,Sanchis2014,lundkvist2016,Mazeh2016}. Possible explanations for this scarcity of planets include atmospheric escape \citep{kurokawa2014,jackson2017,owen2019,koskinen2022}, migration mechanisms \citep{koenig2016,castrogonzales2026,Zanazzi2026}, or a combination of both \citep{ionov2018,owenlai2018,Vissapragada2022,Lazovik2023}.
The discovery of an overdensity of sub-Saturns at orbital periods between 3.2 and 5.7 days \citep[coined the Neptunian ridge,][]{castro2024a} and a slight underdensity at orbital periods $P\gtrsim5.7\,$days \citep[coined the Neptunian savanna,][]{Bourrier2023} has complicated the picture of the sub-Saturn population. Recent works have shown that close-in (hot) sub-Saturns in the desert and ridge might have different properties from the farther-out (warm) sub-Saturns in the savanna. 

\cite{castro2024b} analyzed the bulk density of sub-Saturns and found that savanna sub-Saturns have densities that rarely surpass 1.0\,g\,cm$^{-3}$ while a significant number of sub-Saturns in the desert and ridge have densities between 1.5 and 2.0\,g\,cm$^{-3}$. Another difference between hot and warm sub-Saturns was found in the metallicity distribution of their host stars. \cite{Vissapragada2025} found that the metallicity of hot sub-Saturn hosts is significantly higher than for warm sub-Saturns \citep[also ][]{dong2018}. Furthermore, the host star metallicity of hot sub-Saturns in the desert and ridge is statistically indistinguishable from the host star metallicity of hot Jupiters. 

Given the similarity in host star metallicity as well as the coincidence in orbital-period space between the Neptunian ridge and the hot Jupiter pile-up \citep{udry2003}, \cite{Vissapragada2025} suggest that sub-Saturns in the desert and ridge might form ``top-down''. These sub-Saturns would then be the exposed interiors of hot Jupiters that have undergone atmospheric escape. That would naturally explain the higher bulk densities of ridge and desert sub-Saturns. This scenario requires an escape mechanism more efficient than photoevaporation alone, which is not efficient enough in stripping planets at the upper edge of the Neptunian desert of their atmosphere to produce smaller planets \citep{Vissapragada2022}. Hot Jupiters that migrated via high-eccentricity migration (HEM) could be subjected to partial envelope disruption during close pericenter passage \citep{Faber2005,liu2013}.  An alternative way for hot Jupiters that undergo a quiescent migration to lose a significant portion of their atmosphere is through Roche Lobe overflow \citep{Valsecchi2014,Valsecchi2015,Jackson2016}. 

However, it is not out of the question that hot sub-Saturns form similarly to the warm sub-Saturns, i.e.\ not undergoing or prematurely stopping runaway accretion, explaining their smaller size compared to the Jovian planets \citep[e.g.\ ][]{Ida2004b,Suzuki2018,Moldenhauer2021a,Schlecker2021b,Schlecker2022}. Observational support for this interior-structure perspective comes from
the envelope mass fractions of sub-Saturns. \cite{thomas20252} found
their distribution to be bimodal, plausibly marking the onset of runaway
gas accretion and dividing the population into Neptune-like planets with
modest envelopes and Jovian-like planets dominated by their gaseous
envelope. The ridge being at the same periods as the hot Jupiter pile-up could just be a consequence of the specific migration mechanism, mirroring the one of hot Jupiters, without the sub-Saturns evolving directly from evaporating hot Jupiters. A higher core mass and atmospheric metallicity would help these sub-Saturns survive photoevaporation from their host stars through deeper gravitational wells and more efficient atmospheric cooling \citep{owenmurray2019}, which might produce the observed density and host star metallicity distributions. Indeed recent works have shown that HEM followed by tidal circularization naturally reproduces the location of the ridge itself \citep{castrogonzales2026,Zanazzi2026}. Dynamical evidence points in the same direction. Several of the best-characterized planets in the ridge and desert occupy eccentric, near-polar orbits \citep[e.g.\ HAT-P-11\,b, GJ~436\,b, GJ~3470\,b, TOI-2374\,b;][]{sanchis2011,bourrier2018,stefansson2022,Yee2025}, and close-in Neptunes retain significant eccentricities more often than tidal circularization alone would suggest \citep{correia2020}.

In this work, we use the planetary companion architectures of sub-Saturn systems as
an additional, dynamical tracer of their formation and migration pathways.
The different migration scenarios make distinct predictions for the
systems' present-day architectures. HEM is
dynamically violent: the migrating planet crosses the orbits of any nearby
siblings and typically ejects or accretes them, so HEM predicts a deficit
of nearby companions and low system multiplicity
\citep{rasio1996dynamical,wu2011secular,Mustill2015,petrovich2015hot,bitsch2020eccentricity}. The paucity of nearby companions to hot Jupiters is widely read as evidence for this history \citep{Latham2011,Steffen2012,Knutson2014,Dawson2018}.
Quiescent disk migration or in-situ formation, in contrast, preserves
compact, dynamically cold multi-planet configurations like those of the
\textit{Kepler} multis \citep{goldreich1980disk,ida2008toward,Dittkrist2014,Fabrycky2014,Xie2016,Weiss2018}.

This paper is organized as follows. Sect.~2 describes the sub-Saturn sample. Sect.~3 presents the radial velocity (RV), transit, and combined detection completeness analysis. Sect.~4 presents the completeness-corrected occurrence rates across the landscape and their crosscuts in planetary and stellar properties. We discuss the implications for sub-Saturn migration and the hot Jupiter connection in Sect.~5 and summarize our conclusions in Sect.~6.

%--------------------------------------------------------------------
\section{Sub-Saturn sample} 
In accordance with \cite{castro2024a}, who defined the boundaries of the Neptunian desert, ridge, and savanna, we considered all planets with radii between $4\,\mathrm{R}_\oplus$ and $8.5\,\mathrm{R}_\oplus$ as sub-Saturns. For our sample, we queried the NASA Exoplanet Archive composite table \citep{Christiansen2025}\footnote{\url{https://exoplanetarchive.ipac.caltech.edu/cgi-bin/TblView/nph-tblView?app=ExoTbls&config=PSCompPars} (last accessed 31 March 2026)} for sub-Saturns with radii and masses measured to a precision of 33\% or better. The criterion was applied to both the upper and the lower $1\sigma$ uncertainty of each parameter, and the sample includes planets with masses measured through RVs as well as through transit timing variations although we dropped systems without any RV observations. The precision requirement was set to make certain that only well-characterized sub-Saturns are in the sample, limiting the probability of false positives while also ensuring that there is sufficient photometric and spectroscopic data to detect companion planets. In total, our sample consists of 90 sub-Saturns orbiting 86 host systems, with four systems hosting two sub-Saturns each. We classified each sub-Saturn according to its position in the exo-Neptunian landscape defined by \cite{castro2024a}: the Neptunian desert ($P < 3.2$\,d), the Neptunian ridge ($3.2 \leq P \leq 5.7$\,d), and the Neptunian savanna ($P > 5.7$\,d). Our sample contains 9 desert, 29 ridge, and 52 savanna sub-Saturns, with four two-sub-Saturn systems contributing counts to the respective bins of each planet individually. Three of the two-sub-Saturn systems host both sub-Saturns within the same landscape region, while one system hosts sub-Saturns on either side of the ridge--savanna boundary.

\begin{figure*}
    \centering
    \includegraphics[width=1.90\columnwidth]{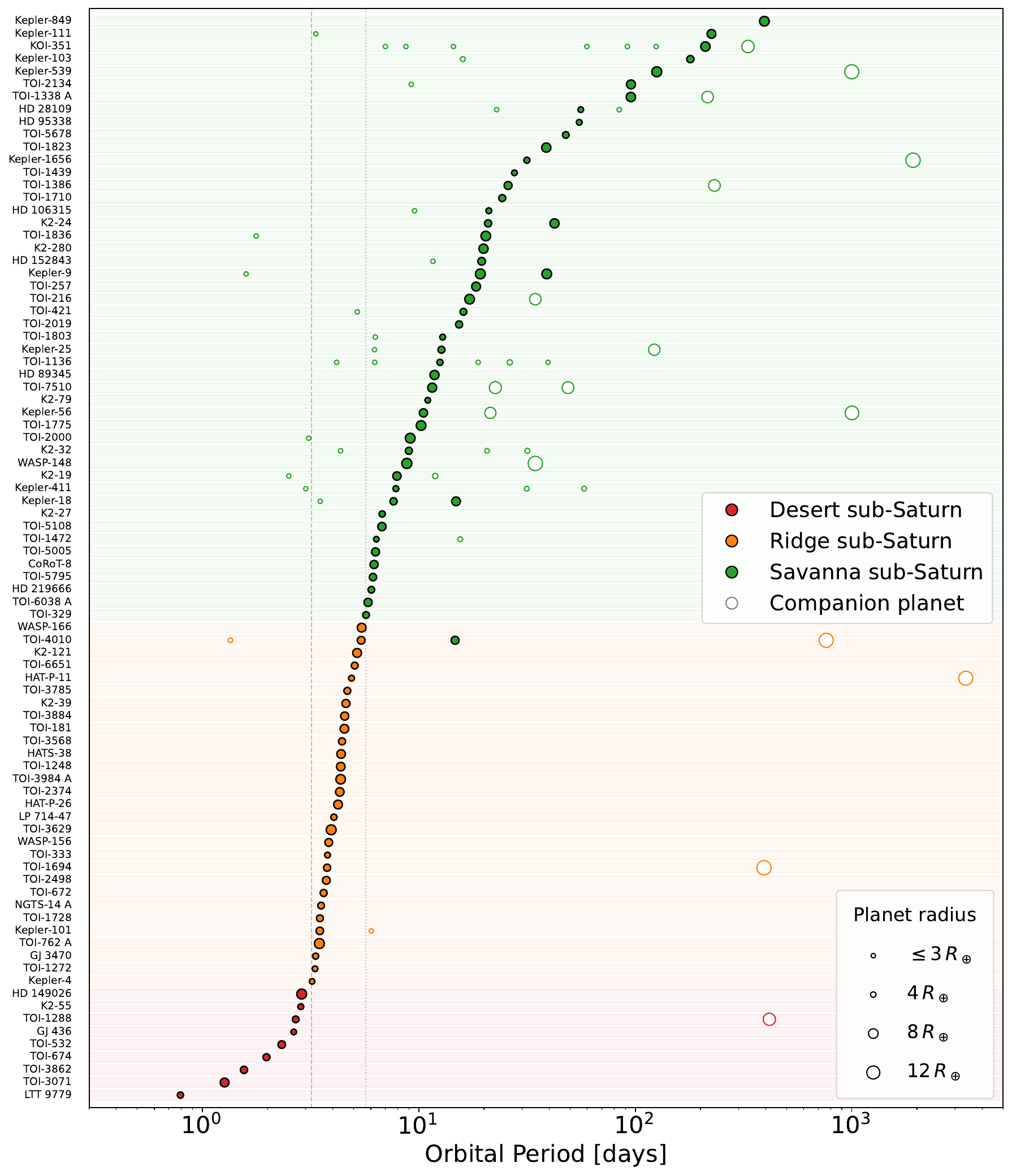} 
    \caption{Orbital architecture of the 86 systems in our sample, ordered by orbital period of the shortest-period sub-Saturn. Filled circles are sub-Saturns colored by their position within the Neptunian landscape. Unfilled circles indicate companion planets with the size of the circle indicating the radius of the companion.}
    \label{fig:sample}
\end{figure*}

Figure \ref{fig:sample} shows the system architectures of our sample.  Only 1 of the 9 desert sub-Saturns has a known companion ($11\%$). Out of the 29 ridge sub-Saturns, 4 have companions ($14\%$), and from the 52 savanna sub-Saturns, 30 are in multi-planet systems ($58\%$). The companion rate thus increases sharply from the desert and ridge to the savanna \citep{Almenara2026}. However, this trend might be an artifact of observational biases. Planets that are close to their host stars are easier to detect and therefore require a shorter baseline and lower precision in their observations. Therefore, it is possible that the lack of companions for the close-in desert and ridge sub-Saturns is due to the limited sensitivity of the available data. In order to correct for this bias, we analyzed the completeness of the photometric and spectroscopic datasets for each sub-Saturn in our sample.

\section{Methodology}
\subsection{RV completeness}
We collected all available published RV measurements for each of the 86 sub-Saturn systems. Table~\ref{tab:rv_observations} summarizes the number of RV points for each system. To assess the completeness of the available RV observations, we performed injection-recovery tests using the Python code \texttt{RVSearch} \citep{Rosenthal2021}. \texttt{RVSearch} takes into account offsets between data from different instruments during the analysis, allowing us to combine all available datasets for the planet search.

First, we used the iterative periodogram search implemented in \texttt{RVSearch} to recover the known planets. The false-alarm probability threshold for the discoveries was set to 1\%. \texttt{RVSearch} computes goodness-of-fit periodograms by comparing the Bayesian information criterion (BIC) between the best-fit $(n+1)$-planet model and the $n$-planet fit to the data. In the first iteration, the zero-planet model is just a flat line. We ran this iterative planet search until no more significant signals are detected by the periodogram search. We then performed an injection-recovery analysis, injecting 10,000 synthetic planets with orbital periods between 1 and 10,000 days, eccentricities between 0 and 0.9, and RV semi-amplitudes between 0.1 and 1000\,m\,s$^{-1}$. The search algorithm was then run one more time, comparing the $n$-planet model containing all the known planets of the system to an $(n+1)$-planet model to check if we can detect the injected planet. This procedure allowed us to construct a completeness map for each star based on their full RV dataset.

For some of the known planets, the planetary signal was not detected with the required sensitivity in the periodogram analysis. These are planets detected by the transit method, where subsequent RV measurements are used to confirm the planets and measure their masses (or in the case of some small companions do not have a precise mass measurement at all), but lack the precision to detect the planets on their own. In these cases, we supplied the literature orbital parameters of the known planets (orbital period, time of conjunction, eccentricity, argument of periastron, and RV semi-amplitude) to \texttt{RVSearch} as a fixed Keplerian component before running the search algorithm, so that the $n$-planet model in the injection-recovery analysis contains all known planets.

Because the synthetic planets are injected into the observed RVs, stellar activity and instrumental red noise enter the completeness estimates through the data. Unmodelled variability raises the residual noise floor and thus the detection threshold of the periodogram search, lowering the recovered completeness. \texttt{RVSearch} additionally fits a white-noise jitter term for each instrument. A more sophisticated treatment of the stellar activity in the detection stage could allow for a slightly better detection efficiency. As such our RV completeness maps are on the conservative side.

\subsection{Transit completeness}

All 86 systems in our sample lie within the \textit{TESS} \citep{Ricker2015} footprint and have been observed in at least one \textit{TESS} sector, providing a minimum baseline of $\sim$27\,d per system. In addition, 13 systems were observed by the \textit{Kepler} \citep{Borucki2010kepler} primary mission with a nearly continuous four-year baseline, and 11 systems were observed by the K2 mission with $\sim$80-day campaign baselines. We constructed separate photometric completeness maps for each mission, yielding a total of 110 completeness surfaces. Light curves were retrieved from the Mikulski Archive for Space Telescopes (MAST) using the \texttt{lightkurve} package \citep{2018ascl.soft12013L}. For each system and mission, we downloaded the pre-search data conditioning simple aperture photometry (PDCSAP) light curves, i.e.\ the time stamps, fluxes, and flux uncertainties, which have been corrected for instrumental systematics by the respective mission pipelines \citep{smith2012kepler,stumpe2012kepler, stumpe2014multiscale}. For the \textit{TESS} photometry, we used the 120\,s cadence data from the Science Processing Operations Center \citep[SPOC;][]{jenkins2016tess} for all sectors where it was available; otherwise, we used the full-frame image extractions \citep{caldwell2020tess,huang2020photometry,huang2020b}. For \textit{Kepler} targets, we downloaded the long-cadence ($\sim$30\,min) PDCSAP light curves for all available quarters (Q1--Q17). For K2 targets, we downloaded the EVEREST K2 light curves for the relevant campaigns. Before injecting synthetic planet signals, we masked the signals of all known transiting planets in each system. We opted to mask the signals instead of subtracting them in order to avoid biases from imperfect subtraction.

We injected $N_\mathrm{inj} = 3000$ synthetic transit signals into each masked, undetrended light curve. The parameters of each injected planet were drawn independently from a log-uniform distribution in the range $P \in [1, 1000]$\,d and $R_\mathrm{p} \in [0.5, 20]\,\mathrm{R}_\oplus$, converted to $R_\mathrm{p}/R_\star$. Impact parameters were drawn from a uniform distribution $b \in [0, 1]$, and the transit epochs were distributed uniformly within one orbital period from the start of observations. Each synthetic transit was computed using \texttt{batman} \citep{kreidberg2015batman} on a circular orbit, with a quadratic limb-darkening law with fixed coefficients $(u_1, u_2) = (0.3, 0.2)$ and the stellar mass and radius of the host from the archive table, and multiplied into the normalized, undetrended flux. The injected light curve was detrended with the biweight time-windowed filter implemented in the \texttt{wotan} package \citep{Hippke2019}, with a filter window of $0.5$\,d. We then attempted to recover these signals through a Box Least Squares \citep[BLS;][]{Kovacs2002} transit search using the \texttt{astropy} implementation \citep{astropy2013}. For each injection, the BLS periodogram was computed over a period window of $[P_\mathrm{inj}\times 0.37, P_\mathrm{inj} \times 2.25]$ with 5000 linearly spaced trial periods, and a grid of ten trial transit durations spanning $0.5$--$2\times$ the injected duration. The edges of the period window were chosen to avoid placing a harmonic of the injected period at the edge of the searched range. Centering the search on the injected ephemeris keeps the $3000$ injections per surface computationally tractable; we verified on a small subset that a fully blind search with a fixed period and duration grid recovers the same fraction of injections. An injection was considered recovered if the peak BLS signal satisfied a signal-to-pink noise ratio $\mathrm{S/N}_\mathrm{pink} \geq 8$ \citep{Hartman2016,Vach2024} and the recovered period matched the injected period to within $5\%$, allowing for the $1/2$ and $2/1$ harmonics within the searched window. We imposed no explicit requirement on the number of observed transits; injections with fewer than two transits inside the observing windows are almost never recovered, since a single transit does not produce a matching BLS period, and a stricter criterion that discards them explicitly changes the resulting occurrence rates by at most $0.4$ percentage points.

\subsection{Combined completeness} \label{sec:combined_completeness}
The occurrence-rate framework of Sect.~\ref{sec:occurrence_rate} interprets each system's detections and non-detections against the sensitivity of their available datasets. Since the known companions were discovered through either channel (e.g.\ non-transiting outer companions through their RV signals alone, and small close-in companions through their transits alone) the relevant quantity is the probability that a hypothetical companion would have been recovered by at least one of the two datasets. The two methods moreover probe complementary regions of the companion parameter space: the photometry provides sensitivity to small planets at short orbital periods, where RV detection is limited by the measurement precision, while the RV data retain sensitivity at long periods, where the geometric transit probability vanishes. For these reasons, we combined the RV and transit completeness surfaces into a single completeness map per system, quantifying the detectability of companions for each sub-Saturn.

The RV completeness maps are defined in $(P, K)$ space, while the transit completeness maps are defined in $(P, R_\mathrm{p})$ space. For each system $j$ in our sample and each point on a grid in companion parameter space $(P, M_\mathrm{comp})$, we drew $N_\mathrm{mc} = 1000$ synthetic companions. To evaluate both completeness maps for the same physical companion, we used the mass--radius relation of \cite{muller2024}:
\begin{equation}\label{eq:mr_muller}
    R_\mathrm{p}(M_\mathrm{comp}) = \begin{cases}
        (1.02\pm 0.03)\,M_\mathrm{comp}^{0.27 \pm0.04} & M_\mathrm{comp} < 4.37\,\mathrm{M}_\oplus \\
        (0.56 \pm 0.03)\,M_\mathrm{comp}^{0.67 \pm 0.05} & 4.37 \leq M_\mathrm{comp} < 127\,\mathrm{M}_\oplus \\
        (18.6\pm6.7)\,M_\mathrm{comp}^{-0.06 \pm0.07} & M_\mathrm{comp} \geq 127\,\mathrm{M}_\oplus
    \end{cases}
\end{equation}
where $R_\mathrm{p}$ is in Earth radii and $M_\mathrm{comp}$ is in Earth masses. The coefficients of Eq.~\ref{eq:mr_muller} were drawn for every Monte Carlo companion from Gaussian distributions with the quoted uncertainties, propagating the scatter of the relation into the completeness maps. As the large uncertainty for the relation of the giant planets can produce unphysically small radii, we chose to impose a lower limit of $8\,\mathrm{R}_\oplus$ for the simulated giant planets ($M_\mathrm{comp} \geq 127\,\mathrm{M}_\oplus$). The companion radius is used to evaluate the transit completeness map and to compute the geometric transit probability. The RV completeness map was evaluated at the semi-amplitude $K$ computed from $M_\mathrm{comp} \sin i_\mathrm{comp}$, with the inclination of the individual draw, the stellar mass from the archive table, and a circular orbit. Companions detected only in RV enter the analysis with their minimum masses, consistent with this convention. For each draw, we also assigned a mutual inclination between the companion's orbit and the known, transiting sub-Saturn's orbit and computed the companion's true inclination with respect to the line of sight. The two orbits are related by a mutual inclination $\Delta i$ and a nodal azimuth $\varphi$:
\begin{equation}\label{eq:cos_icomp}
    \cos i_\mathrm{comp} = \cos i_\mathrm{ss}\,\cos \Delta i +  \sin i_\mathrm{ss}\,\sin \Delta i\,\cos \varphi,
\end{equation}
where the subscripts $_{\mathrm{ss}}$ and $_\mathrm{comp}$ denote properties of the sub-Saturn and companion planets, respectively, and $\varphi$ is the longitude of the relative node measured in the orbital plane of the sub-Saturn, i.e.\ the azimuthal angle of the companion's orbital angular momentum around that of the sub-Saturn. $\varphi$ is not the difference of the sky-projected longitudes of ascending node, which is related to $\Delta i$ through both inclinations. $\varphi$ was drawn uniformly from $[0, 2\pi)$, corresponding to an isotropic orientation of the relative node. We verified that this parametrization recovers an isotropic distribution of $i_\mathrm{comp}$ when $\Delta i$ itself is drawn isotropically, for any $i_\mathrm{ss}$. We adopted $i_\mathrm{ss} = 90$\,deg for all systems, for which Eq.~\ref{eq:cos_icomp} simplifies to $\cos i_\mathrm{comp} = \sin \Delta i\,\cos \varphi$; a variant of the analysis using the measured sub-Saturn inclinations instead changes the resulting occurrence rates by at most 1 percentage point. 

For our default model, we adopted the empirical mutual inclination distribution calibrated by \cite{he2020} from the \textit{Kepler} multi-transit population. They found that the median mutual inclination scales with the intrinsic planet multiplicity $n$ as
\begin{equation}\label{eq:sigma_i}
    \tilde{\mu}_i(n) = \tilde{\mu}_{i,5}\left(\frac{n}{5}\right)^{-1.73},
\end{equation}
with $\tilde{\mu}_{i,5} = 1.10^{+0.15}_{-0.11}$\,deg \citep{he2020}. For each system, we evaluated the relation at the intrinsic multiplicity implied by the simulation, $n = n_\mathrm{known} + 1$, i.e.\ the number of known planets plus the hypothetical companion, and converted the median to a Rayleigh scale parameter via $\sigma_i = \tilde{\mu}_i / \sqrt{\ln 4}$, giving $\sigma_i \approx 0.93$\,deg at $n = 5$. The mutual inclination $\Delta i$ was then drawn from $\mathrm{Rayleigh}(\sigma_i)$, and $\varphi$ was drawn uniformly from $[0, 2\pi)$. The relation of \cite{he2020} is calibrated on the compact \textit{Kepler} multi-planet systems, and its applicability to massive companions at long orbital periods is not established; there, however, the choice of mutual-inclination model has little influence on the combined completeness, which is dominated by the RV data owing to the vanishing transit probability, and we verify in Appendix~\ref{app:val_maps} that alternative prescriptions, including the extreme case of isotropic companion orbits, leave all population comparisons unchanged.

A companion transits if its orbital inclination satisfies $|\cos i_\mathrm{comp}| < (R_\star + R_\mathrm{p})/a$, where $a$ is the semi-major axis computed from $P$ and $M_\star$ via Kepler's third law. For each Monte Carlo draw, we only checked for the detection probability in our transit completeness maps if this criterion was met. Otherwise, the probability of detecting the companion in the photometry was set to zero. The criterion assumes circular companion orbits, consistent with the circular orbits of the transit injections. The dependence of the detectability on the impact parameter is contained in the per-system transit completeness surfaces, which are constructed from injections with impact parameters drawn uniformly from $[0, 1]$ and are evaluated here as a function of $(P, R_\mathrm{p})$ only, i.e.\ marginalized over the impact parameter.

The combined completeness at each grid point is the fraction of Monte Carlo draws for which the companion would be detected by at least one method:
\begin{equation}\label{eq:combined_mc}
    \langle C_j^\mathrm{combined} \rangle = \frac{1}{N_\mathrm{mc}} \sum_{k=1}^{N_\mathrm{mc}} \left[ 1 - \big(1 - p_{\mathrm{RV},k}\big)\big(1 - p_{\mathrm{tr},k}\big) \right],
\end{equation}
where $p_{\mathrm{RV},k}$ and $p_{\mathrm{tr},k}$ are the RV and transit detection probabilities of the $k$th draw, evaluated from smooth detection-probability models fitted to the injection-recovery results of each system in the coordinates in which the respective surveys operate: logistic regressions on fourth-degree polynomial features of $(\log P, \log K)$ for the RV and of $(\log P, \log R_\mathrm{p})$ for the transit surveys, fitted to the binary recovery outcomes. Where the combined grid extends beyond the injection ranges, the models were evaluated at the nearest range boundary rather than extrapolated (at $P = 1$\,d for RV evaluations at shorter periods, and at $P = 1000$\,d for transit evaluations at longer periods). The Monte Carlo average of Eq.~\ref{eq:combined_mc} was evaluated on a $100 \times 100$ grid in $(\log P, \log M_\mathrm{comp})$ spanning $P \in [0.5, 10^{4}]$\,d and $M_\mathrm{comp} \in [0.5, 2000]\,\mathrm{M}_\oplus$ to give the final completeness map.

\begin{figure*}
    \centering
    \includegraphics[width=1.90\columnwidth]{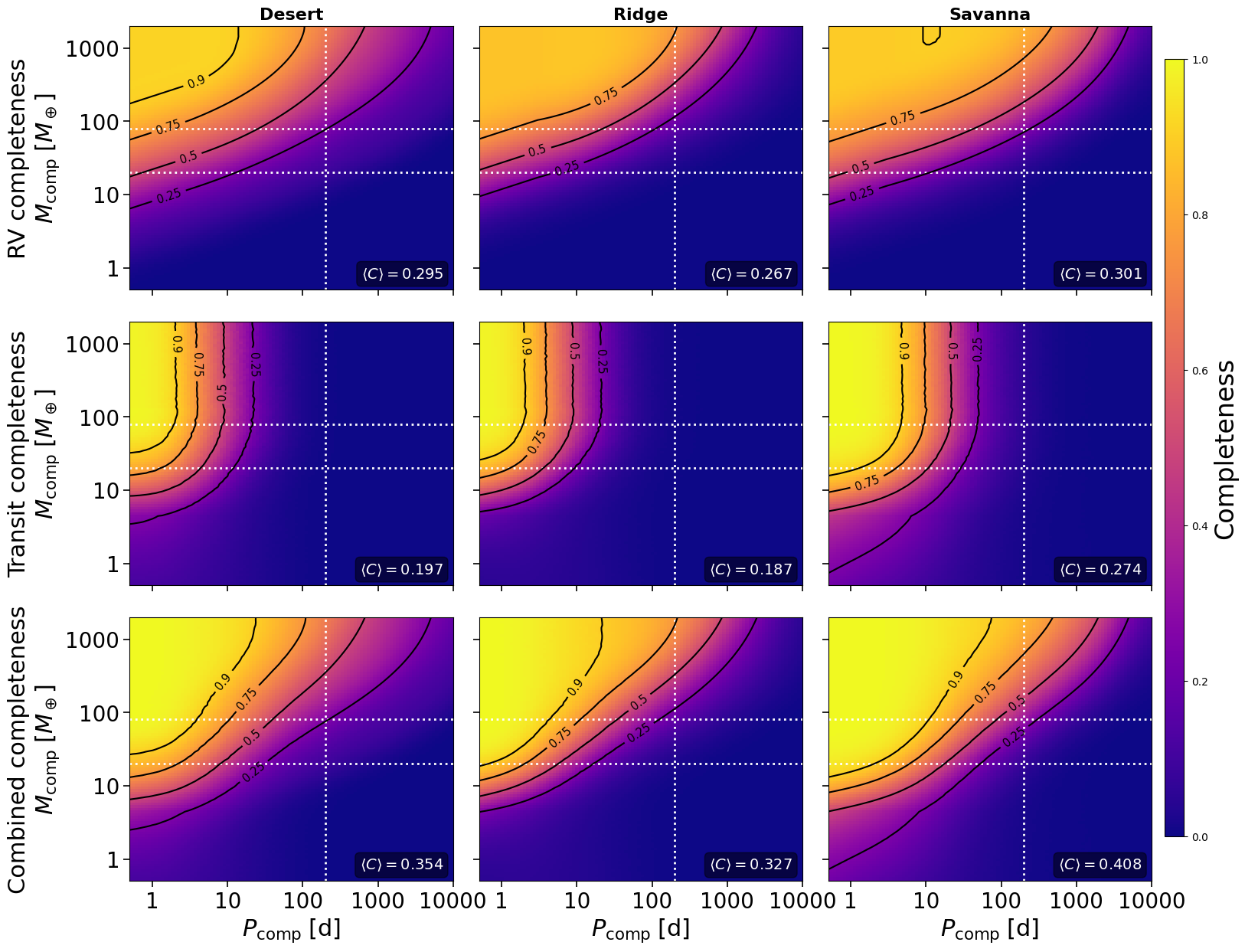} 
    \caption{Completeness maps of our injection-recovery analysis for the RV data (top row), the photometric data (middle row) and the combination of both datasets (bottom row). Plotted are the average completeness values as a function of orbital period and planetary mass for the sub-Saturn systems in the sample divided into the desert (left column), ridge (middle column) and savanna (right column) subsamples. The white dotted lines indicated the cells we divide the parameter space into for the occurrence rate calculations.}
    \label{fig:comp}
\end{figure*}

Figure \ref{fig:comp} shows the average completeness of the sub-Saturn systems across the desert, ridge, and savanna. For each Neptunian landscape bin, we plot the averaged completeness maps for the RV-only, transit-only, and combined cases. At large orbital periods, the completeness is dominated by the RV datasets due to the shrinking transit probability, while at short periods, the addition of the transit data increases the detection sensitivity towards smaller planets. As expected, the average completeness is higher for the systems with a savanna sub-Saturn since the detection of the sub-Saturn in the savanna requires a larger data baseline than for sub-Saturns that are closer-in. We find average completeness values across the whole log-log parameter space in the combined maps of $35\,\%$, $33\,\%$, and $41\,\%$ for the desert, ridge, and savanna subsamples, respectively.

\subsection{Occurrence rate calculation}\label{sec:occurrence_rate}

We computed completeness-corrected companion occurrence rates with a joint Poisson occurrence model defined on disjoint cells of the companion parameter space. In contrast to fitting each companion category with an independent rate parameter, a single generative model for the whole parameter space guarantees that the inferred rates are mutually consistent: the rate of a composite category (e.g.\ all nearby companions) follows directly from the rates of its constituent cells, and the fractions of systems with at least one and with more than one companion derive from the same underlying multiplicity distribution.

We partitioned the companion parameter space, $P \in [0.5, 10^{4}]$\,d and $M_\mathrm{comp} \in [0.5, 2000]\,\mathrm{M}_\oplus$, into six disjoint cells illustrated in Fig.~\ref{fig:comp}. Three mass classes: small ($M_\mathrm{comp} < 20\,\mathrm{M}_\oplus$), medium ($20\,\mathrm{M}_\oplus \leq M_\mathrm{comp} < 80\,\mathrm{M}_\oplus$), and giant ($M_\mathrm{comp} \geq 80\,\mathrm{M}_\oplus$), each divided at $P = 200$\,d into a nearby and a far period range. Within each cell $i$, the number of companions of system $j$ is modeled as $k_{j,i} \sim \mathrm{Poisson}(\mu_i)$, where $\mu_i$ is the expected number of companions per system in that cell, assumed to be the same for all systems within a given population (e.g.\ the desert+ridge or the savanna sub-Saturns). Each companion, if present, is detected independently with probability $C_{j,i}$, the mean completeness of system $j$ over cell $i$,
\begin{equation}\label{eq:cell_completeness}
    C_{j,i} = \frac{1}{A_i} \int\!\!\int_{\mathcal{R}_i}
    C_j(\log P, \log M)\,\mathrm{d}\log P\,\mathrm{d}\log M\,,
\end{equation}
where $C_j(\log P, \log M) \equiv \langle C_j^\mathrm{combined} \rangle$ is the combined completeness map of Sect.~\ref{sec:combined_completeness} and $A_i$ is the cell area in dex$^2$. The unweighted mean corresponds to our baseline assumption of a log-uniform intrinsic distribution of companions within each cell. By Poisson thinning, the number of detected companions in cell $i$ then follows $n_{j,i} \sim \mathrm{Poisson}(\mu_i\,C_{j,i})$.

Rather than conditioning on the raw detected counts, we used the truncated per-system outcome $o_{j,i} \in \{0,\, 1,\, {\geq}2\}$ (no, exactly one, or at least two detected companions in the cell), with probabilities
\begin{equation}\label{eq:cell_outcome}
\begin{aligned}
    p(n_{j,i}=0) &= e^{-\mu_i C_{j,i}}\,,\\
    p(n_{j,i}=1) &= \mu_i\,C_{j,i}\; e^{-\mu_i C_{j,i}}\,,\\
    p(n_{j,i}\geq 2) &= 1 - e^{-\mu_i C_{j,i}}\left(1 + \mu_i\,C_{j,i}\right)\,.
\end{aligned}
\end{equation}
The truncation at two detections makes the inference robust to the few systems with many detected companions and to possible correlations between the detections of companions in the same system, while retaining the information that connects the single- and multiple-companion rates. This observation model interpolates between two established frameworks through the amount of count information retained. With the untruncated counts, the likelihood factorizes into per-system Poisson terms whose
product is exactly the point-process likelihood of \citet{Rosenthal2022} with a piecewise-constant intensity. Truncating instead to the binary outcome $\{0, {\geq}1\}$ recovers a binary detection model of the same type as the Beta-Binomial framework of \citet{Bryan2024}, with which it coincides in the rare-companion limit, where $1 - e^{-\mu_i C_{j,i}} \approx \mu_i C_{j,i}$; at higher rates the two differ, as the Poisson form accounts for the enhanced detectability of systems hosting several companions. We verify in Appendix~\ref{app:validation} that our results are insensitive to this choice.

The log-likelihood of the data in one cell is
\begin{equation}\label{eq:cell_loglik}
    \ln\mathcal{L}(\mu_i) = \sum_{j} w_j \ln p\!\left(o_{j,i} \mid \mu_i\,C_{j,i}\right)\,,
\end{equation}
where the weight $w_j = 1/n_{\mathrm{ss},j}$, with $n_{\mathrm{ss},j}$ the number of sub-Saturns in the system to which sub-Saturn $j$ belongs, accounts for the four systems hosting two sub-Saturns. For systems hosting a single sub-Saturn, $w_j = 1$ and the standard unweighted likelihood is recovered; for the four two-sub-Saturn systems each sub-Saturn receives $w_j = 1/2$, preserving the effective number of independent systems at 86. Because the two rows of a two-sub-Saturn system share the same companions and datasets, Eq.~\ref{eq:cell_loglik} is formally a pseudo-likelihood. Since the cells are disjoint and the companion processes independent, the joint likelihood factorizes, and the posterior of each $\mu_i$ was computed separately on a dense grid of $60{,}000$ equally spaced values of $\mu_i$ from $10^{-4}$ to $20$. We adopted a prior that is uniform in the per-cell occurrence fraction $f_i = 1 - e^{-\mu_i}$, i.e.\ $p(\mu_i) = e^{-\mu_i}$, because a prior flat in $\mu_i$ places substantial mass at fractions near unity and systematically inflates the rates of poorly constrained cells. We drew $2 \times 10^5$ posterior samples of each $\mu_i$ via inverse transform sampling and quantify the prior dependence in Appendix~\ref{app:val_prior}. 

All reported occurrence fractions derive from the same posterior draws of $(\mu_1, \ldots, \mu_6)$, using the additivity of independent Poisson processes. The fraction of systems hosting at least one companion in a union of cells $\mathcal{S}$ is
\begin{equation}\label{eq:cell_derived}
    f_{\mathcal{S}} = 1 - \exp\Big(-\textstyle\sum_{i \in \mathcal{S}} \mu_i\Big)\,.
\end{equation}
For $\mathcal{S}$ comprising the three nearby cells, Eq.~\ref{eq:cell_derived} yields the nearby-companion rate; for $\mathcal{S}$ comprising all six cells, it yields the any-companion rate. The fraction of systems with more than one companion follows from the total rate $\mu_\mathrm{tot} = \sum_i \mu_i$ as $f_{\geq 2} = 1 - e^{-\mu_\mathrm{tot}}(1 + \mu_\mathrm{tot})$. Nested categories are therefore consistent by construction, $f_\mathrm{small,nearby} \leq f_\mathrm{nearby} \leq f_\mathrm{any}$, and the single- and multiple-companion fractions follow from one multiplicity distribution. We report the median and $68\%$ credible interval of each fraction.

The integrated completeness $C_j$ is the mean of the detection-probability surface under an assumed intrinsic distribution of companions that is uniform in $(\log P, \log M)$ within the region $\mathcal{R}$, and the corrected occurrence rates inherit this assumption. As we show in Appendix~\ref{app:val_intrinsic}, the absolute rates of broad, low-completeness categories depend strongly on it whereas the desert+ridge versus savanna comparisons are robust, because both populations share the same maps and assumption. All rates quoted in this paper adopt the log-uniform intrinsic distribution and should be read with this caveat in mind.

Companion categories defined relative to the sub-Saturn's orbit were fitted with the same truncated-count likelihood on the corresponding system-specific period ranges spanning the full mass range. Because these two regions mix mass ranges of very different completeness, we used them only for the relative comparison of the inner versus the outer rate, not for absolute statements.

Population comparisons are quantified by the posterior probability that one occurrence fraction exceeds another, $p(f_A > f_B)$, computed by drawing independently from the two posteriors. A value of $p = 0.5$ corresponds to no evidence for a difference, and we regard $p > 0.95$ ($p > 0.99$) as moderate (strong) evidence. We validate the occurrence rate framework in Appendix~\ref{app:validation} by verifying the
calibration of the inference on simulated surveys drawn through the real
per-system completeness maps, re-deriving the rates in the beta-binomial and
untruncated-count limiting cases and with Approximate Bayesian Computation
\citep[ABC-PMC;][]{Kunimoto2020}, and quantifying the sensitivity to the prior and to the assumed intrinsic companion distribution.

\section{Results} 
 
\subsection{Companion rates across the Neptunian landscape}\label{sec:rates_landscape}
With the combined RV and photometric detection sensitivity maps of the available data, we quantify the occurrence of companions across the 86 sub-Saturn systems. We report the completeness-corrected occurrence fractions of the following companion categories (Sect.~\ref{sec:occurrence_rate}), motivated by different formation channel diagnostics:
\begin{enumerate}
    \item Any and multiple companions: derived from the total rate $\mu_\mathrm{tot}$, the fractions of systems hosting at least one companion anywhere in the parameter space and more than one companion (total number of planets $\rm{N}\geq3$), respectively.
    \item Nearby planets: the union of the three nearby cells, i.e.\ all planetary companions with orbital periods $P_\mathrm{comp} < 200$\,d. These would have likely been cleared in the case of dynamically violent (HEM) pathways.
    \item Companion type: the fraction of systems hosting at least one small ($M_\mathrm{comp} < 20\,\mathrm{M}_\oplus$), medium ($20\,\mathrm{M}_\oplus \leq M_\mathrm{comp} < 80\,\mathrm{M}_\oplus$), or giant ($M_\mathrm{comp} \geq 80\,\mathrm{M}_\oplus$) companion in the nearby ($P_\mathrm{comp} < 200$\,d) period range. 
    \item Massive long-period companions: all companions with a
    planetary mass $M_\mathrm{comp} \geq 1\,\mathrm{M}_\mathrm{Jup}$
    ($318\,\mathrm{M}_\oplus$) and orbital period $P_\mathrm{comp} > 365$\,d
    (i.e.\ $a \gtrsim 1\,$au for a solar-mass host). These are the distant
    massive bodies that may have driven HEM. This category is a sub-region of the giant far cell and is fitted with the same likelihood on its own region.
    \item Inner and outer nearby planets: for the planetary companions with orbital periods $P_\mathrm{comp} < 200$\,d we further examine whether they orbit inside of the orbit of the sub-Saturn ($P_\mathrm{comp} < P_\mathrm{ss}$) or in a wider orbit than the sub-Saturn ($P_\mathrm{ss}<P_\mathrm{comp}  <200\,$d).

\end{enumerate}

Throughout this work, a companion is any confirmed planet in the system
other than the sub-Saturn under consideration. In the four two-sub-Saturn
systems, the sibling sub-Saturn counts as a companion of the other. To avoid double-counting systems, each sub-Saturn in a two-sub-Saturn system enters all analyses with weight $1/2$ (Sect.~\ref{sec:occurrence_rate}), so the weighted effective sample size equals the 86 host systems; one of the four systems hosts sub-Saturns on either side of the ridge--savanna boundary, yielding non-integer weighted bin sizes (desert 9, ridge 28.5, savanna 48.5). 

Table~\ref{tab:rates_corrected} presents the completeness-corrected occurrence rates for each companion category, divided into the Neptunian landscape classes (desert, ridge, and savanna). For each category, the table lists the weighted number of systems with at least one detection (at least two for the multiple category) over the weighted bin size, the mean completeness of the category region, and the posterior median with its $68\%$ credible interval. The nearby, any, and multiple rows derive from the same posterior draws as the cell rates (Sect.~\ref{sec:occurrence_rate}), whereas the massive long-period category is a sub-region of the giant far cell fitted with the same likelihood, and the inner and outer nearby categories are evaluated on the subsample with $P_\mathrm{ss} < 200$\,d (savanna $N=45.5$). As expected, the completeness correction substantially increases the inferred occurrence rates relative to the raw detection fractions, especially in cells with limited sensitivity. The small and medium long-period cells are a limiting case: their mean completeness is at most a few percent, so their posteriors largely return the prior, and the any- and multiple-companion fractions derived from the total rate inherit this prior dependence (Appendix~\ref{app:val_prior}). As such we focus on the more constrained categories such as the nearby companion categories for the interpretation of the results. Additionally, we find that the posterior distributions of the companion fraction after completeness correction for the desert and ridge sample agree well across the companion classes. Given the evidence that the ridge and desert populations follow the same underlying trends \citep[e.g.\ in bulk density and host star metallicity;][]{castro2024b,Vissapragada2025}, we combine the two populations to compare to the savanna sub-Saturns for the following analysis.

\begin{table*}
\centering
\caption{Completeness-corrected companion fractions by Neptunian landscape class from the joint Poisson occurrence model with a prior uniform in the per-cell fraction $f_i$.}
\label{tab:rates_corrected}
\resizebox{1.99\columnwidth}{!}{
\begin{tabular}{lcccccccc}
\hline\hline
\noalign{\smallskip}
Category & \multicolumn{2}{c}{Desert ($N=9$)} & \multicolumn{2}{c}{Ridge ($N=28.5$)} & \multicolumn{2}{c}{Desert+ridge ($N=37.5$)} & \multicolumn{2}{c}{Savanna ($N=48.5$)}\\
 & Raw (compl.) & Corrected & Raw (compl.) & Corrected & Raw (compl.) & Corrected & Raw (compl.) & Corrected\\
\noalign{\smallskip}\hline\noalign{\smallskip}
Any companion & $1/9$ ($35\%$) & $92.3_{-12.4}^{+5.8}\%$ & $3.5/28.5$ ($33\%$) & $91.6_{-11.8}^{+6.2}\%$ & $4.5/37.5$ ($33\%$) & $87.7_{-13.6}^{+8.5}\%$ & $28.5/48.5$ ($41\%$) & $98.72_{-1.67}^{+0.86}\%$ \\[3pt]
Multiple companions & $0/9$ ($35\%$) & $73_{-25}^{+18}\%$ & $0.5/28.5$ ($33\%$) & $71_{-23}^{+19}\%$ & $0.5/37.5$ ($33\%$) & $62_{-23}^{+22}\%$ & $11.5/48.5$ ($41\%$) & $93.1_{-6.5}^{+4.2}\%$ \\[3pt]
Massive long-period ($M_\mathrm{comp} \geq 1\,\mathrm{M}_\mathrm{Jup}$, $P_\mathrm{comp}>365$\,d) & $0/9$ ($31\%$) & $16_{-12}^{+22}\%$ & $1.5/28.5$ ($24\%$) & $24_{-12}^{+15}\%$ & $1.5/37.5$ ($26\%$) & $18.3_{-9.1}^{+12.6}\%$ & $2.5/48.5$ ($37\%$) & $15.4_{-6.7}^{+8.9}\%$ \\[3pt]
Nearby ($P_\mathrm{comp}<200$\,d) & $0/9$ ($51\%$) & $42_{-19}^{+22}\%$ & $1.5/28.5$ ($48\%$) & $45_{-16}^{+18}\%$ & $1.5/37.5$ ($49\%$) & $36_{-14}^{+16}\%$ & $24.5/48.5$ ($59\%$) & $91.1_{-4.3}^{+3.2}\%$ \\[3pt]
Inner nearby ($P_\mathrm{comp}<P_\mathrm{ss}$) & $0/9$ ($69\%$) & $9.1_{-6.7}^{+13.1}\%$ & $0.5/28.5$ ($65\%$) & $5.9_{-3.8}^{+6.6}\%$ & $0.5/37.5$ ($66\%$) & $4.5_{-2.9}^{+5.1}\%$ & $17/45.5$ ($69\%$) & $47.2_{-7.4}^{+7.4}\%$ \\[3pt]
Outer nearby ($P_\mathrm{ss}<P_\mathrm{comp}<200$\,d) & $0/9$ ($45\%$) & $12.7_{-9.4}^{+17.6}\%$ & $1.5/28.5$ ($39\%$) & $16.4_{-8.2}^{+11.4}\%$ & $1.5/37.5$ ($41\%$) & $12.6_{-6.4}^{+9.2}\%$ & $12.5/45.5$ ($42\%$) & $58.3_{-9.0}^{+8.5}\%$ \\[3pt]
Small nearby ($M_\mathrm{comp}<20\,\mathrm{M}_\oplus$) & $0/9$ ($18\%$) & $23_{-17}^{+27}\%$ & $1.5/28.5$ ($13\%$) & $37_{-18}^{+20}\%$ & $1.5/37.5$ ($14\%$) & $29_{-14}^{+17}\%$ & $19.5/48.5$ ($26\%$) & $88.2_{-5.5}^{+4.2}\%$ \\[3pt]
Medium nearby ($20\,\mathrm{M}_\oplus \leq M_\mathrm{comp}<80\,\mathrm{M}_\oplus$) & $0/9$ ($61\%$) & $10.2_{-7.5}^{+14.6}\%$ & $0.5/28.5$ ($57\%$) & $6.6_{-4.2}^{+7.3}\%$ & $0.5/37.5$ ($58\%$) & $5.1_{-3.2}^{+5.7}\%$ & $4.5/48.5$ ($72\%$) & $13.4_{-4.8}^{+6.0}\%$ \\[3pt]
Giant nearby ($M_\mathrm{comp}\geq 80\,\mathrm{M}_\oplus$) & $0/9$ ($84\%$) & $7.7_{-5.7}^{+11.4}\%$ & $0/28.5$ ($85\%$) & $2.7_{-2.0}^{+4.3}\%$ & $0/37.5$ ($85\%$) & $2.1_{-1.6}^{+3.3}\%$ & $4/48.5$ ($91\%$) & $11.8_{-4.1}^{+5.2}\%$ \\[3pt]
\noalign{\smallskip}\hline
\end{tabular}
}
\tablefoot{Raw: weighted number of systems with at least one detection (at least two for the multiple category) over the weighted bin size, with the mean completeness of the category region in brackets. Corrected: posterior median and $68\%$ credible interval. Sub-Saturns in two-sub-Saturn systems enter with weight $1/2$, yielding non-integer counts and bin sizes.}
\end{table*}

\begin{figure*}
    \centering
    \includegraphics[width=1.99\columnwidth]{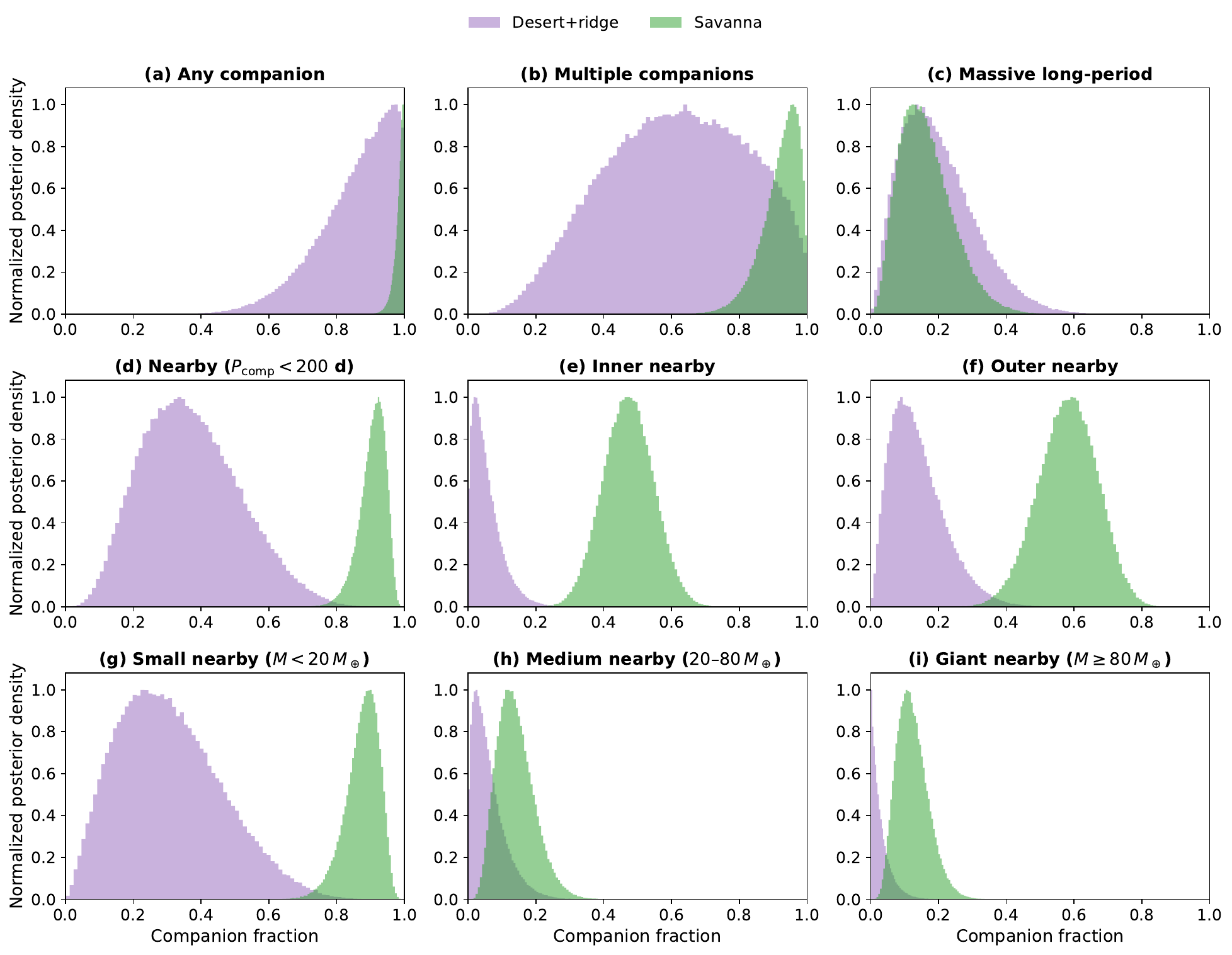} 
    \caption{Posterior distributions of the completeness-corrected occurrence rates of different types of companions for systems with a known sub-Saturn in the desert or ridge (purple) or savanna (green). Each posterior distribution is normalized to unit peak height for visibility.}
    \label{fig:post_land}
\end{figure*}

The posterior distributions for the different companion categories of desert+ridge and savanna sub-Saturns are shown in Fig.~\ref{fig:post_land}. When comparing the existence of companions in close orbits ($P_\mathrm{comp} < 200$\,d), we find a striking difference between the two populations: for sub-Saturns in the savanna, companion planets within orbital periods of 200\,days are ubiquitous, with an occurrence rate of $91.1_{-4.3}^{+3.2}\%$, compared to only $36_{-14}^{+16}\%$ for sub-Saturns residing in the desert or ridge ($p(f_\mathrm{sav} > f_\mathrm{d+r}) = 0.9998$). The any- and multiple-companion fractions derived from the total rate are high for both populations ($98.72_{-1.67}^{+0.86}\%$ versus $87.7_{-13.6}^{+8.5}\%$, and $93.1_{-6.5}^{+4.2}\%$ versus $62_{-23}^{+22}\%$), with the savanna nominally higher in both cases ($p = 0.93$); as these two quantities integrate over the poorly constrained far cells, their absolute values carry the prior dependence discussed above, and we base no population comparison on them alone. In contrast, massive long-period companions are inferred at consistent rates in the two populations: $18.3_{-9.1}^{+12.6}\%$ for the desert+ridge and $15.4_{-6.7}^{+8.9}\%$ for the savanna.
 
To explore the system architectures of sub-Saturns further, we look at the companion types and locations in more detail. The relative position of the nearby companions ($P<200$\,days) with respect to the sub-Saturn is probed by the inner and outer nearby categories, and the companion types by the mass-resolved cell rates, comparing super-Earth and sub-Neptune companions ($M_\mathrm{comp}<20\,\mathrm{M}_\oplus$) to other sub-Saturns ($20\,\mathrm{M}_\oplus\leq M_\mathrm{comp}<80\,\mathrm{M}_\oplus$) and giant planets ($M_\mathrm{comp} \geq 80\,\mathrm{M}_\oplus$). For the companion types, we focus on the nearby cells, as the completeness for the smaller planets at large orbital periods is very limited, and we already presented the occurrence rates for giant long-period companions. Figure \ref{fig:post_land} shows the resulting posterior distributions for the ridge+desert and savanna sub-Saturns. For the desert+ridge sub-Saturns, we find a similarly low occurrence rate for outer companions ($12.6_{-6.4}^{+9.2}\%$) as for inner companions ($4.5_{-2.9}^{+5.1}\%$). For the savanna sub-Saturns, the occurrence rates of outer and inner companions are also similar ($58.3_{-9.0}^{+8.5}\%$ and $47.2_{-7.4}^{+7.4}\%$, indicating no preferred companion position relative to the sub-Saturn. Looking at the types of companions, we find that other sub-Saturns or giant planets within 200\,days are rare for both the desert+ridge and savanna sub-Saturns, with rates between $2.1\%$ and $13.4\%$ for sub-Saturn and giant companions. The big difference in nearby companions between the desert+ridge and savanna is therefore driven by the lack of small planet companions to desert+ridge sub-Saturns (occurrence rate of $29_{-14}^{+17}\%$) compared to the prevalence of these types of companions for savanna sub-Saturns with an occurrence rate of $88.2_{-5.5}^{+4.2}\%$ ($p(f_\mathrm{sav} > f_\mathrm{d+r}) = 0.9996$).
 
\subsection{Companion rates by sub-Saturn radius}\label{sec:rates_radius}
 
Recently, the analysis of envelope mass fractions of sub-Saturns by \cite{thomas20252} revealed a bimodal distribution which might originate from the onset of runaway gas accretion, dividing the sub-Saturn population into Neptune-like and Jovian planets. To test whether the planetary multiplicity depends on the sub-Saturn size, we split the sample at $R_p = 6\,\mathrm{R}_\oplus$ into a Neptune-like bin ($4\,\mathrm{R}_\oplus \leq R_p < 6\,\mathrm{R}_\oplus$, $N=44.5$, 46 planets) and a giant-like bin ($6\,\mathrm{R}_\oplus \leq R_p \leq 8.5\,\mathrm{R}_\oplus$, $N=41.5$, 44 planets) according to the radius boundaries that corresponded to the envelope mass fraction gap of \cite{thomas20252}. 
 
\begin{figure*}
    \centering
    \includegraphics[width=1.99\columnwidth]{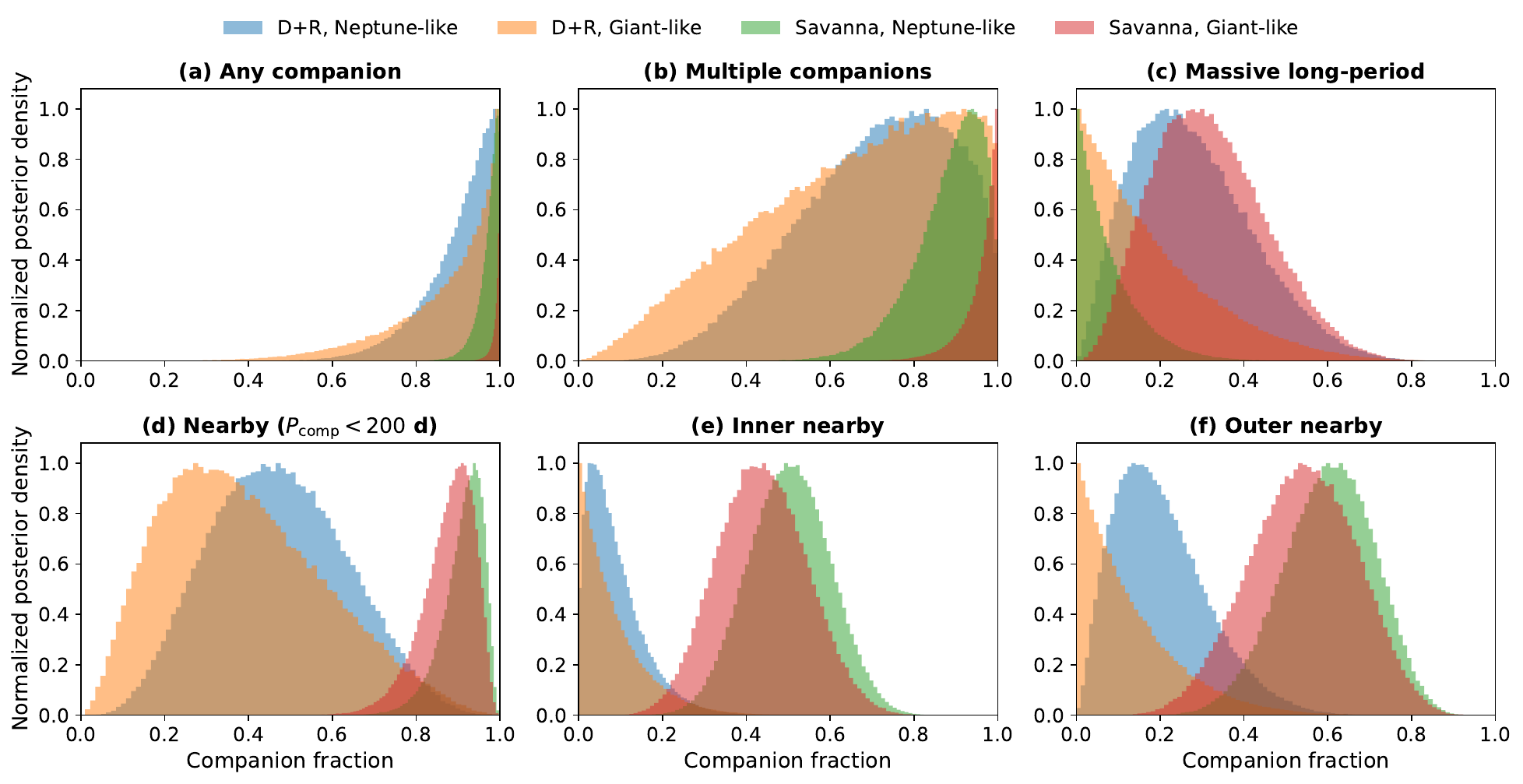} 
    \caption{Posterior distributions of the completeness-corrected occurrence rates of different types of companions for systems with a Neptune-like ($4\,\mathrm{R}_\oplus \leq R_p < 6\,\mathrm{R}_\oplus$) or giant-like ($6\,\mathrm{R}_\oplus \leq R_p \leq 8.5\,\mathrm{R}_\oplus$) sub-Saturn. The different sub-Saturn classes are further divided by their location within the Neptunian landscape. Each posterior distribution is normalized to unit peak height for visibility.}
    \label{fig:post_rad_land}
\end{figure*}

Comparing the two radius classes (Appendix~\ref{app:rad}), we find that they exhibit very similar companion rates across most categories. The nearby companion rate is $83.5_{-7.3}^{+5.7}\%$ for the Neptune-like and $81.2_{-8.8}^{+6.9}\%$ for the giant-like sub-Saturns, with similarly consistent rates for the small and medium nearby cells and the position-resolved categories. The only exception is the giant nearby cell, in which companions are more common around the giant-like sub-Saturns ($14.4_{-4.9}^{+6.1}\%$ versus $1.7_{-1.2}^{+2.6}\%$, $p = 0.99$), a difference carried by the savanna members of the two classes ($21.6_{-7.2}^{+8.6}\%$ versus $3.0_{-2.2}^{+4.7}\%$). When we further subdivide by landscape (Fig.~\ref{fig:post_rad_land}), the landscape gradient remains the dominant signal in both radius bins, with a nearby companion rate contrast of $p(f_\mathrm{sav} > f_\mathrm{d+r}) = 0.997$ ($0.989$) for the Neptune-like (giant-like) class and insignificant differences between the radius classes within a given landscape. This indicates that the landscape-dependent architecture trend is not an artifact of one particular sub-Saturn size dominating a given landscape and it persists independently of the planet radius.
 
\subsection{Companion rates by sub-Saturn density}\label{sec:rates_density}
 
\cite{castro2024b} and \cite{castrogonzales2026} identified a density dichotomy between ridge and savanna sub-Saturns with the ridge planets showing a bimodal density distribution with peaks at $\sim 0.7$\,g\,cm$^{-3}$ and $\sim 1.7$\,g\,cm$^{-3}$ while savanna planets are typically less dense ($\rho \lesssim 1$\,g\,cm$^{-3}$) and do not show a secondary peak. One potential explanation for the dense ridge and desert planets is that they are the stripped remnants of initially larger bodies that underwent high-eccentricity migration \citep{castro2024a,Vissapragada2025}. We split our sample at $\rho = 1.4$\,g\,cm$^{-3}$ which is the minimum of the bimodality identified by \cite{castrogonzales2026}, yielding a low-density bin ($N = 60$, 64 planets) and a high-density bin ($N = 26$, 26 planets).

The occurrence rates of the density split are summarized in Appendix~\ref{app:den}. The nearby companion rate is $89.3_{-5.1}^{+3.9}\%$ for the low-density compared to $62_{-13}^{+12}\%$ for the high-density sub-Saturns ($p(f_\mathrm{low} > f_\mathrm{high}) = 0.98$), a difference that is again carried by the small nearby companions ($86.8_{-6.2}^{+4.7}\%$ versus $54_{-15}^{+14}\%$) and also visible in the inner nearby category ($37.6_{-6.6}^{+6.9}\%$ versus $14.1_{-6.6}^{+9.1}\%$, $p = 0.98$). 

\begin{figure*}
    \centering
    \includegraphics[width=1.99\columnwidth]{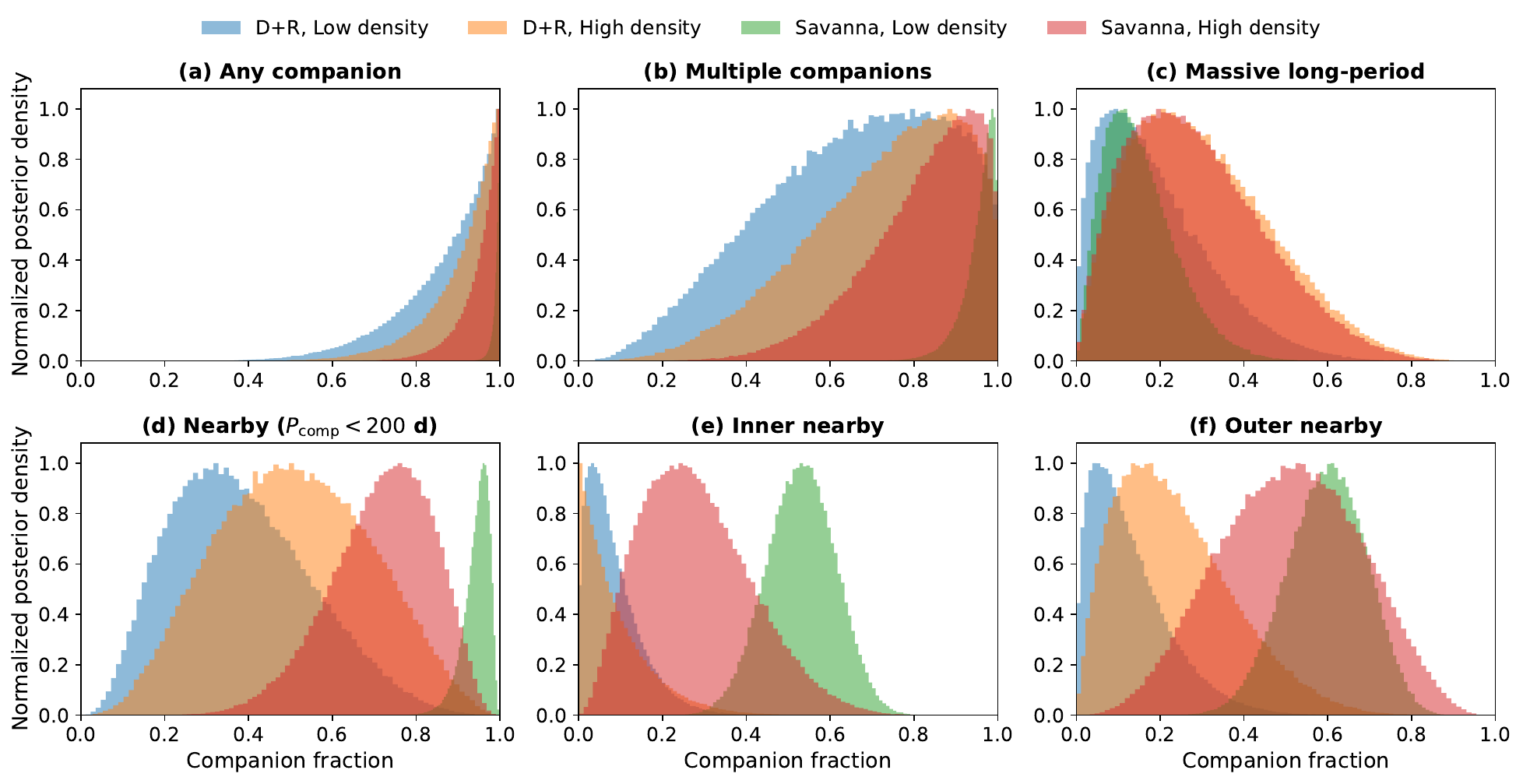} 
    \caption{Same as Fig.~\ref{fig:post_rad_land}, but for the two density classes: low-density ($\rho< 1.4$\,g\,cm$^{-3}$) and high-density ($\rho \geq 1.4$\,g\,cm$^{-3}$).}
    \label{fig:post_den_land}
\end{figure*}

When the two density classes are further divided by landscape (Fig.~\ref{fig:post_den_land}), we find that this density dependence resides mainly in the savanna. Among savanna sub-Saturns, the nearby companion rate of the low-density class exceeds that of the high-density class ($95.0_{-3.5}^{+2.3}\%$ versus $73_{-14}^{+11}\%$, $p = 0.99$; small nearby companions $93.1_{-4.7}^{+3.1}\%$ versus $60_{-17}^{+15}\%$, $p = 0.99$), which constitutes moderate to
strong evidence under the convention of Sect.~\ref{sec:occurrence_rate}. Within the desert+ridge, in contrast, the two density classes are statistically indistinguishable
($p = 0.26$--$0.53$ across the companion categories), with the high-density class nominally the higher of the two in the nearby category ($51_{-19}^{+20}\%$ versus $36_{-15}^{+19}\%$).

The landscape gradient itself remains highly significant for the low-density sub-Saturns, whose nearby companion rate is $95.0_{-3.5}^{+2.3}\%$ in the savanna versus $36_{-15}^{+19}\%$ in the desert+ridge ($p(f_\mathrm{sav} > f_\mathrm{d+r}) = 0.9997$). For the high-density sub-Saturns the gradient points in the same direction
but is much weaker ($73_{-14}^{+11}\%$ versus $51_{-19}^{+20}\%$, $p = 0.83$), reflecting both the smaller subsamples ($N = 13$ each) and the reduced savanna rate of this class.
 
\subsection{Companion rates by sub-Saturn eccentricity}\label{sec:rates_eccentricity}
 
The orbital eccentricity of the sub-Saturn itself might encode additional information on its dynamical evolution. High-eccentricity migration could produce residual eccentricities that may not be fully circularised, particularly for planets in the savanna, where tidal damping timescales are longer. For warm Jupiters, a large fraction ($\sim 70\%$) has been found to reside in dynamically hot ($e>0.1$) orbits \citep{Schlecker2020,Dong20212,Eberhardt2022,Morgan2026}.  Of our 90 sub-Saturns, 86 have a reported eccentricity constraint; we classified 58 as circular ($e<0.1$) and 28 as eccentric ($e \geq 0.1$) (weighted: $N = 54$ and $28$), and excluded the four sub-Saturns without an eccentricity measurement from this crosscut.
 
As shown in Appendix~\ref{app:ecc}, companion rates are mostly higher for circular than for eccentric sub-Saturns. Nearby companions are more common around circular than around eccentric sub-Saturns ($90.0_{-5.2}^{+3.9}\%$ versus $68_{-12}^{+10}\%$, $p(f_\mathrm{circ} > f_\mathrm{ecc}) = 0.98$), again driven by the small nearby companions ($87.5_{-6.4}^{+4.8}\%$ versus $61_{-13}^{+12}\%$). The comparatively companion-poor systems of the eccentric sub-Saturns are in line with the well-established anticorrelation between orbital eccentricity and the richness of planetary systems \citep{Limbach2015,vaneylen2019}.

\begin{figure*}
    \centering
    \includegraphics[width=1.99\columnwidth]{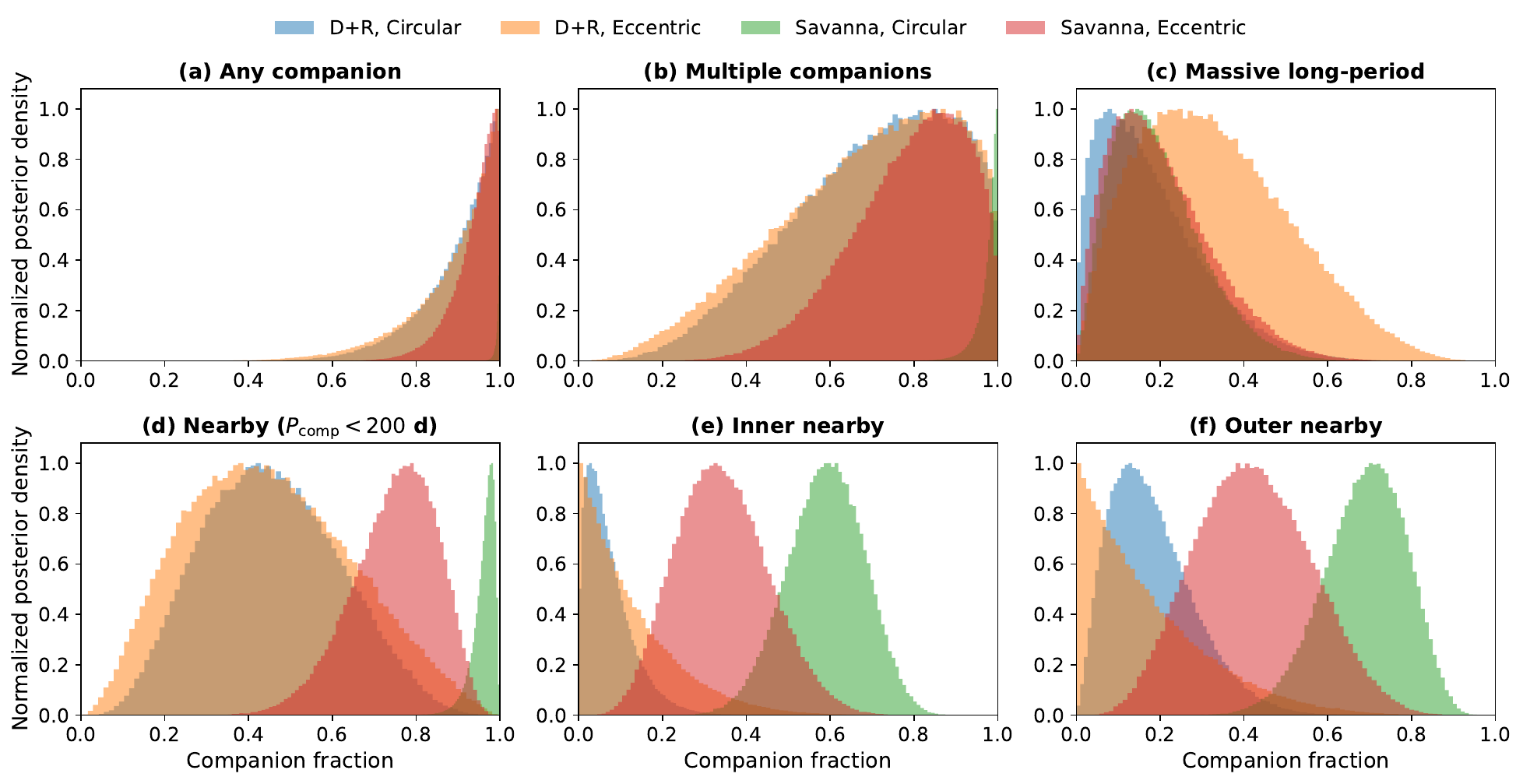} 
    \caption{Same as Fig.~\ref{fig:post_rad_land}, but for the two eccentricity classes: circular ($e< 0.1$) and eccentric ($e \geq 0.1$).}
    \label{fig:post_ecc_land}
\end{figure*}

When further divided by landscape (Fig.~\ref{fig:post_ecc_land}), the eccentricity dependence, like the density dependence, is a feature of the savanna: circular savanna sub-Saturns have higher nearby ($96.8_{-2.9}^{+1.7}\%$ versus $75.9_{-11.2}^{+9.1}\%$, $p = 0.99$), small nearby ($95.2_{-4.2}^{+2.5}\%$ versus $68_{-14}^{+12}\%$), and inner nearby companion rates ($59.2_{-9.5}^{+9.0}\%$ versus $34_{-11}^{+12}\%$, $p = 0.95$) than their eccentric counterparts, whereas the two classes are indistinguishable within the desert+ridge ($p = 0.24$--$0.64$; $45_{-16}^{+18}\%$ versus $44_{-19}^{+22}\%$ for the nearby category). For circular sub-Saturns the landscape gradient is very strong ($96.8_{-2.9}^{+1.7}\%$ versus $45_{-16}^{+18}\%$, $p(f_\mathrm{sav} > f_\mathrm{d+r}) = 0.9998$); for eccentric sub-Saturns it points in the same direction without reaching significance ($p = 0.90$), owing to the small eccentric desert+ridge subsample ($N = 9$).

\subsection{Companion rates by host star metallicity and mass}\label{sec:rates_stellar}
 
\begin{figure}
    \centering
    \includegraphics[width=0.99\columnwidth]{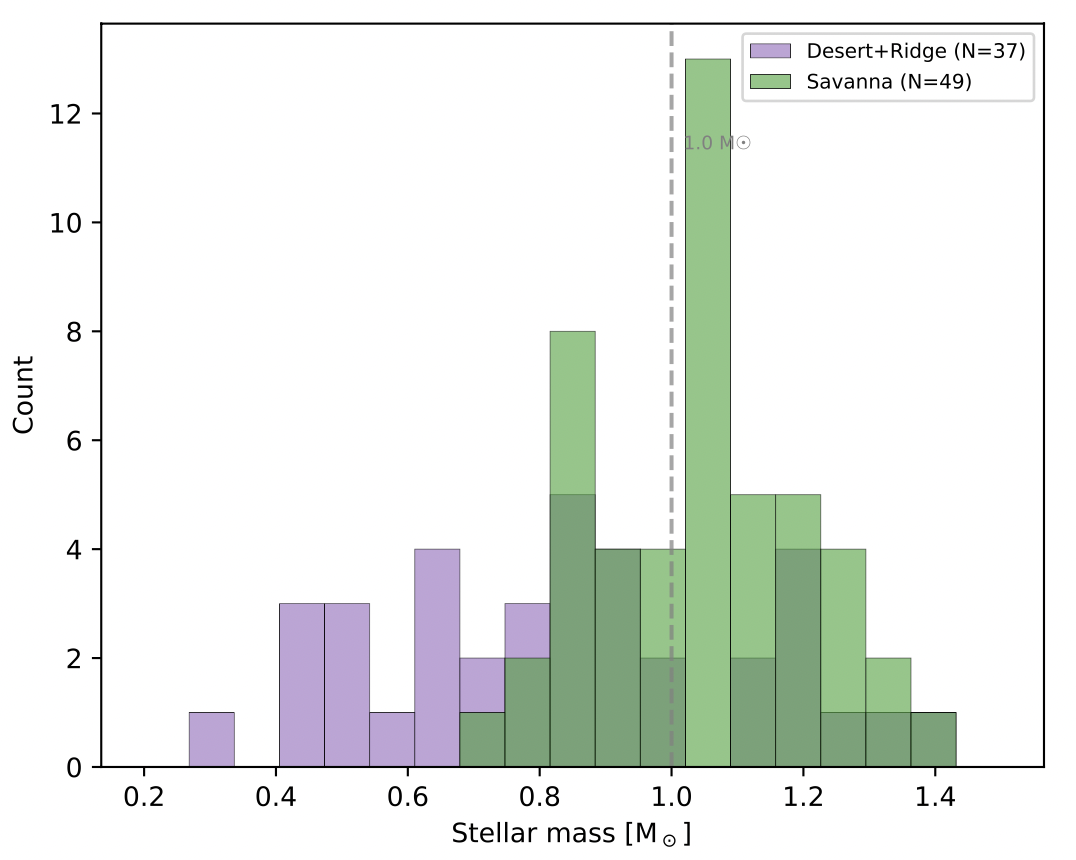} 
    \caption{Distribution of host star masses for sub-Saturns in the desert and ridge (purple) compared to the savanna (green). Savanna sub-Saturn hosts cluster around $\sim 1\,\text{M}_\odot$ while desert and ridge hosts extend to M-dwarfs.}
    \label{fig:hoststar_dist}
\end{figure}

In previous sections, we have investigated whether the properties of the sub-Saturns influence the occurrence rate of companions. Here, we look at potential influences of the properties of the host stars. The host star metallicity of sub-Saturn systems has been shown to differ across the Neptunian landscape, with desert and ridge hosts being significantly more metal-rich than savanna hosts \citep{Vissapragada2025}. Similarly, the stellar mass distribution differs between the landscape groups. Figure \ref{fig:hoststar_dist} shows that the savanna sub-Saturns cluster mostly around solar-mass stars, while the distribution of host stars for desert and ridge sub-Saturns is much broader, extending to low-mass M-dwarfs. To test whether these stellar properties independently influence the companion occurrence rate, or whether they merely trace the landscape classification, we performed binary split analyses analogous to Sects.~\ref{sec:rates_landscape}--\ref{sec:rates_eccentricity}.
 
We split the sample at the median host star metallicity of $\mathrm{[Fe/H]} = 0.17$\,dex into a metal-poor bin ($N = 43$, 44 planets) and a metal-rich bin ($N = 43$, 46 planets). For the stellar mass, we split at $M_\star = 1.0\,\mathrm{M}_\odot$, yielding a low-mass ($N = 46$, 48 planets) and a high-mass ($N = 40$, 42 planets) subsample. The posterior distributions for both splits are shown in Appendices~\ref{app:stellar_mass} and \ref{app:stellar_met}.
 
Neither stellar property shows a significant overall effect on companion occurrence. For the metallicity split, the nearby companion rate is $86.5_{-6.5}^{+4.9}\%$ for metal-poor hosts compared to $75.8_{-9.8}^{+8.2}\%$ for metal-rich hosts ($p(f_\mathrm{poor} > f_\mathrm{rich}) = 0.84$). The only metallicity contrast is for inner nearby companions, which are more common around metal-poor hosts ($41.2_{-7.8}^{+8.0}\%$ vs $19.4_{-6.3}^{+7.5}\%$, $p = 0.98$). Given the well-documented difference in host-star metallicity between the desert+ridge and savanna sub-Saturns \citep{Vissapragada2025}, this difference likely just traces the landscape split. The stellar mass split yields similarly inconclusive results. The nearby companion rate is $80.3_{-8.8}^{+7.0}\%$ for low-mass hosts compared to $84.4_{-7.2}^{+5.6}\%$ for high-mass hosts.

When further subdivided by landscape, the landscape gradient remains the dominant signal. The savanna nearby companion rate significantly exceeds the desert+ridge rate for metal-poor ($91.4_{-5.3}^{+3.6}\%$ vs $36_{-17}^{+22}\%$, $p = 0.996$), metal-rich ($88.4_{-8.3}^{+5.6}\%$ vs $47_{-17}^{+18}\%$, $p = 0.99$), and low-mass hosts ($94.9_{-5.0}^{+2.9}\%$ vs $31_{-13}^{+17}\%$, $p = 0.9999$) and marginally exceeds for the  high-mass hosts ($86.8_{-6.8}^{+5.1}\%$ vs $65_{-22}^{+19}\%, p = 0.87$), potentially reflecting the small high-mass host star desert+ridge subsample ($N = 10$). Within a given landscape, the companion rates of the different metallicity and stellar-mass bins are consistent, confirming that neither stellar property carries significant predictive power for companion occurrence beyond what the landscape classification already captures.

\subsection{Association between stellar and planetary properties and companion status}\label{sec:ad_tests}
 
To complement the occurrence rate analysis presented in Sects.~\ref{sec:rates_landscape}--\ref{sec:rates_stellar}, we performed two-sample Anderson-Darling (AD) and Kolmogorov-Smirnov (KS) tests to assess whether the distributions of stellar and planetary properties differ systematically between companion-hosting and companion-free systems. Unlike the binary splits used above, these tests operate on the full continuous distributions without imposing arbitrary boundaries, providing an independent assessment of which properties are most strongly associated with companion status. For each property and companion category, we compared the property distributions of systems with and without a detected companion using the two-sample AD test, with $p$-values computed by permutation ($10^{4}$ label permutations), and the KS statistic as a cross-check. We evaluated six properties against six companion categories. Since each test family evaluates 36 property--category combinations, we controlled the false discovery rate with the Benjamini--Hochberg procedure and interpret only associations with $q < 0.05$.
 
\begin{figure*}
    \centering
    \includegraphics[width=1.99\columnwidth]{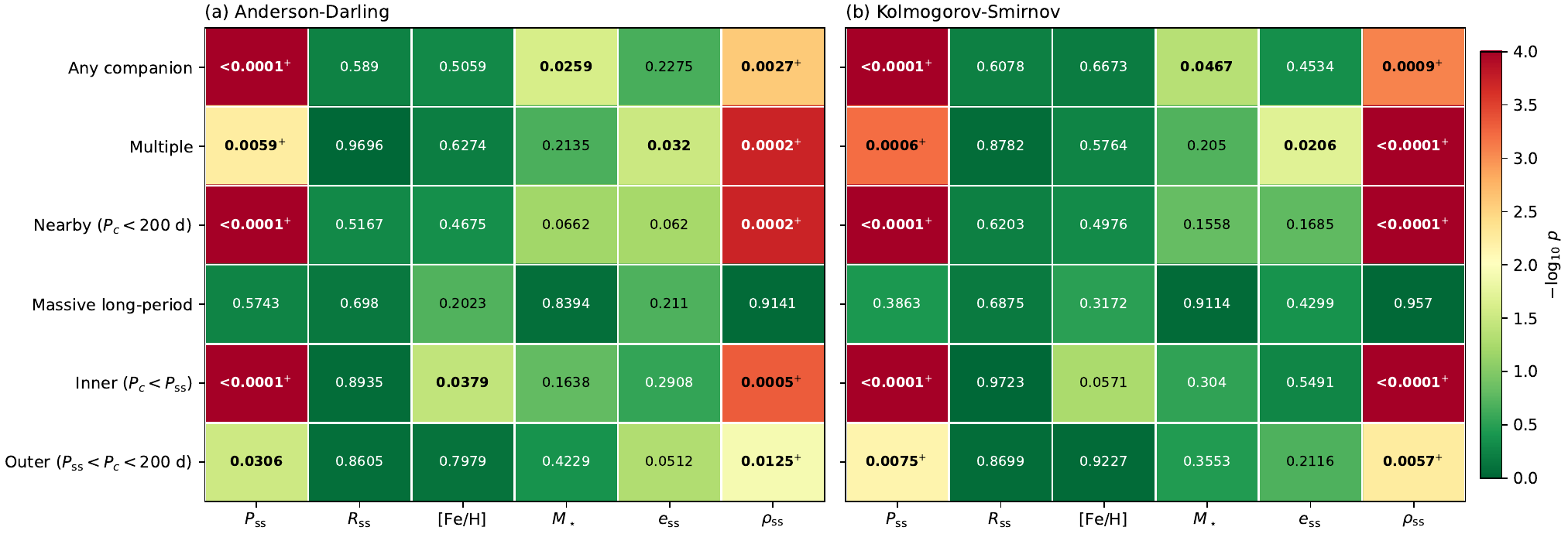}
    \caption{Association between system properties and the presence of
    detected companions from the model-free two-sample tests. Panel (a):
    Anderson-Darling; panel (b): Kolmogorov-Smirnov (permutation $p$-values
    on the raw detection labels). Bold entries mark $p < 0.05$; a `$+$'
    marks associations that remain significant after false-discovery-rate
    control at $q < 0.05$. }
    \label{fig:heatmap}
\end{figure*}

The results are summarized in Fig.~\ref{fig:heatmap}. Only two properties are associated with companion presence after false-discovery control, consistently in both test families: the sub-Saturn orbital period (the any, multiple, nearby, and inner companion categories; $q \leq 0.027$ AD) and its bulk density (the multiple, nearby, and inner categories at $q \leq 0.003$, the any category at $q = 0.014$, and the outer category marginally at $q = 0.050$; all six categories at $q \leq 0.023$ under the KS statistic). This suggests that density may carry genuine information about companion architecture that is at least partially independent of the landscape classification. We find that systems with nearby companions have systematically lower bulk densities (median $\rho = 0.6$\,g\,cm$^{-3}$) than systems without (median $\rho = 1.0$\,g\,cm$^{-3}$). This association is corroborated by the completeness-corrected rates of Sect.~\ref{sec:rates_density}, where the nearby and inner nearby companion rates of the low-density class exceed those of the high-density class at $p = 0.99$ and $p = 0.98$, respectively. This suggests that the binary split at $\rho = 1.4$\,g\,cm$^{-3}$ adopted there is not optimal for tracing this association, as the median densities of both the companion-hosting and the companion-free systems lie below the boundary. The AD and KS tests, which operate on the full continuous distributions without imposing a boundary, retain greater statistical power for such a gradual trend. This is also why these tests can discriminate between the density and eccentricity associations, which are of comparable strength in the binned rates
(Sect.~\ref{sec:rates_eccentricity}). At the same time, the tests presented here operate on the raw detection labels and therefore conflate the occurrence of companions with their detectability. For this reason, all quantitative occurrence statements in this work rest on the completeness-corrected analysis of Sects.~\ref{sec:rates_landscape}--\ref{sec:rates_stellar}, and we read the present tests as a complementary model-free screen.

For the same properties we also compared the desert+ridge and savanna populations directly, to identify potential confounders of the landscape classification. Besides the orbital period (which defines the classes), the host star mass differs significantly between the two populations (AD and KS $p = 10^{-4}$), and the bulk density marginally ($p \simeq 0.02$), whereas the sub-Saturn radius, host metallicity, and eccentricity are consistent ($p > 0.1$).
 
\section{Discussion}\label{sec:discussion}
 
\subsection{Implications for the migration of sub-Saturns}\label{sec:disc_unified}
 
The results of our occurrence rate analysis support different evolutionary origins for the desert+ridge and savanna sub-Saturns. Sub-Saturns in the savanna have companion architectures consistent with dynamically cold migration: nearby companions are nearly ubiquitous ($91.1_{-4.3}^{+3.2}\%$), and most savanna sub-Saturns reside in multi-planet systems ($93.1_{-6.5}^{+4.2}\%$). Companions orbiting well interior to the sub-Saturn might survive the sub-Saturn's HEM if its pericenter never approaches their orbits. However, the sub-Saturn should at least clear planets whose orbit it crosses during dynamical migration and thus they should lack close companions that orbit outside of the present-day sub-Saturn orbit. In our analysis, the nearby companions are found to be orbiting both inside and outside the sub-Saturn's orbit at similar occurrence rates. Thus, these systems are consistent with formation in a dynamically quiescent protoplanetary disk, either in situ or through gradual disk-driven migration. The population of sub-Saturns residing in the desert and ridge has a different companion architecture. Their nearby companion rate is a factor of $\sim 2.5$ lower ($36_{-14}^{+16}\%$, $p(f_\mathrm{sav} > f_\mathrm{d+r}) = 0.9998$). This pattern is the expected outcome of HEM, where gravitational interactions with a distant massive perturber pump the eccentricity of the migrating body, bringing it to a short-period orbit while dynamically clearing nearby small planets \citep{rasio1996dynamical,Mustill2015,Dawson2018}.

Because the landscape classes are defined solely by the orbital period of the sub-Saturn, the deficit of the desert and ridge planets might reflect a generic scarcity of companions to short-period planets instead of tracing their migration history. However, the broader close-in planet population argues against a generic scarcity. Small planets on comparably short or even shorter orbits are routinely accompanied: ultra-short-period planets ($P < 1$\,d) commonly have longer-period siblings, and once the declining transit probability is accounted for, a majority of them are inferred to host additional companions \citep{Sanchis2014,winn2018}. Likewise, super-Earths and sub-Neptunes with orbital periods inside the ridge boundary frequently reside in the compact, dynamically cold configurations characteristic of the \textit{Kepler} multi-planet systems \citep{Weiss2018,zhu2018}. A short present-day orbit alone therefore does not imply the absence of nearby companions. Instead, the companion deficit at $P \lesssim 5.7$\,d appears specific to intermediate-mass and giant planets (i.e.\ desert and ridge sub-Saturns and hot Jupiters) which points to a mass-dependent delivery mechanism, such as HEM, that operates on these bodies while the small-planet population reaches similar periods with its companions intact.

Notably, the massive long-period companions capable of having acted as perturbers are detected around desert+ridge and savanna sub-Saturns at consistently small rates: $18.3_{-9.1}^{+12.6}\%$ versus $15.4_{-6.7}^{+8.9}\%$. While planet--planet scattering migration scenarios might result in an ejected outer perturber \citep{Chatterjee2008, beauge2012} most other HEM mechanisms should leave the massive outer perturber intact \citep{Wu2003,Fabrycky2007,wu2011secular,petrovich2015hot}. While our analysis implies these planets to be rare for all sub-Saturns there are two caveats which need to be considered. First, our massive outer perturber category only extends to orbital periods of 10,000 days, which was chosen due to the reduced sensitivity at longer orbital periods. However, a massive perturber could reside in farther orbits which are not considered in our analysis.  Second, this category is by construction restricted to planetary companions: three of the 38 systems hosting a desert or ridge sub-Saturn (NGTS-14\,A, TOI-762\,A, TOI-3984\,A) and eight of the 49 hosting a savanna sub-Saturn are known to host at least one stellar companion, a fraction of which may themselves be capable of driving high-eccentricity migration through von Zeipel--Lidov--Kozai cycles \citep{Fabrycky2007,Naoz2016}, as invoked for hot Jupiters. As such the rate of perturbers capable of driving HEM is likely higher than our occurrence rate of massive companions and thus does not necessarily rule out that desert and ridge planets preferentially migrate via HEM.

Another key finding of this work is that the landscape gradient in nearby companion occurrence is driven almost entirely by small ($M<20\,\mathrm{M}_\oplus$) planets ($29_{-14}^{+17}\%$ versus $88.2_{-5.5}^{+4.2}\%$). Medium ($20$--$80\,\mathrm{M}_\oplus$) and giant ($\geq80\,\mathrm{M}_\oplus$) companions within 200\,days are rare in both populations, with occurrence rates of at most $\sim 13\%$ and at most marginal differences between the desert+ridge and savanna. This indicates that the sub-Saturn is typically the dominant body in its immediate orbital neighborhood, regardless of landscape. The distinguishing factor is whether small super-Earth and sub-Neptune companions coexist with the sub-Saturn. In the disk migration scenario, the sub-Saturn forms alongside smaller planets in a common protoplanetary disk, preserving a hierarchy of planet sizes. In the HEM scenario, the dynamical excitation disrupts or ejects these fragile small companions \citep{Mustill2015}, while the massive perturber responsible for triggering the migration resides at long periods and is largely unaffected.

To test whether the landscape gradient is driven by a confounding variable rather than reflecting a genuine architectural difference, we subdivided the sample by sub-Saturn radius, bulk density, eccentricity, host star metallicity, and host star mass, and repeated the occurrence rate analysis within each bin. In all five crosscuts, the landscape gradient remains the dominant signal. The savanna nearby companion rate exceeds that of the desert+ridge within every bin, with no crosscut variable eliminating or reversing the trend, although the contrast falls short of significance in the smallest desert+ridge subsamples (the high-density, eccentric, and high-mass bins, $p = 0.83$--$0.90$). The two crosscuts that show significant secondary structure are density and eccentricity, both expressed through the nearby, and in particular the small nearby, companion rates. Low-density sub-Saturns have a substantially higher nearby companion rate ($89.3_{-5.1}^{+3.9}\%$) than high-density sub-Saturns ($62_{-13}^{+12}\%$), and circular sub-Saturns a higher rate ($90.0_{-5.2}^{+3.9}\%$) than eccentric ones ($68_{-12}^{+10}\%$). In both cases this secondary structure resides in the savanna, while the density and eccentricity classes are indistinguishable within the desert+ridge (Sects.~\ref{sec:rates_density} and~\ref{sec:rates_eccentricity}).

Within the desert and ridge, our results provide no evidence for a
density-dependent division between migration channels. The recently
proposed unified view of the exo-Neptunian landscape by \cite{Bourrier2025} associates
low-density (``fluffy'') Neptunes in the ridge with early disk-driven
migration and denser desert and ridge planets with a late arrival through
HEM. In our analysis, however, the low- and high-density desert+ridge sub-Saturns show statistically indistinguishable companion architectures ($p = 0.31$--$0.51$ across the companion categories), with nearby companion rates of $36_{-15}^{+19}\%$ and $51_{-19}^{+20}\%$ that are both depressed relative to their savanna counterparts. The companion architectures therefore do not single out the
dense desert+ridge planets as the HEM arrivals. Instead, they are consistent with scenarios in which HEM operates across the desert+ridge population irrespective of bulk density, as in the population models of \cite{castrogonzales2026} and \cite{Zanazzi2026}, in which high-eccentricity tidal migration, accompanied by partial envelope loss for the planets reaching the shortest periods, naturally reproduces the ridge and the desert.

The savanna, by contrast, does show internal structure. If it hosts a
dominant population of dynamically cold migrators alongside a minority of
dynamically hot arrivals retaining residual eccentricity, then density and
eccentricity should correlate with companion architecture within the
savanna, even though neither property sets the landscape class \citep{Bourrier2025}. This is supported by the secondary structure identified in Sects.~\ref{sec:rates_density} and~\ref{sec:rates_eccentricity}: low-density savanna sub-Saturns have a higher nearby companion rate ($95.0_{-3.5}^{+2.3}\%$) than high-density ones ($73_{-14}^{+11}\%$), and circular savanna sub-Saturns ($96.8_{-2.9}^{+1.7}\%$) a higher rate than eccentric ones ($75.9_{-11.2}^{+9.1}\%$), with the multiple companion rates following the same pattern ($98.3_{-2.9}^{+1.2}\%$ versus $81_{-16}^{+12}\%$ for the eccentricity split). Additionally, the bulk density is the only property besides orbital period whose association with companion status survives false-discovery-rate control in our KS and AD tests (Sect.~\ref{sec:ad_tests}). These correlations are consistent with a minority population of dynamically hot systems within the savanna, although our data cannot distinguish whether such systems arrived through late HEM or were excited in situ. The eccentric savanna systems, which retain companions at a reduced but still substantial rate, are consistent with the variant proposed by \cite{Bourrier2025} for TOI-421, a member of our sample, in which disk migration of a compact pair is followed by an internal instability that excites eccentricities and mutual inclinations without expelling the companion.
 
\subsection{Implications for the hot Jupiter--sub-Saturn connection}\label{sec:disc_hj}
 
An evolutionary connection between the Neptunian ridge and the hot Jupiter population was proposed by \cite{castro2024a} on the basis of their alignment in orbital period and relative occurrence rates, and \cite{Vissapragada2025} provided independent support from the host-star metallicities. The resemblance in companion occurrence rates presented here is a new, dynamical piece of evidence for this connection. We find that the companion architectures of the sub-Saturn population mirrors that of the Jovian population. In both planet classes, the close-in population is companion-poor while the more distant population is companion-rich. \cite{Wu2023} found nearby companions rates of $\geq12\pm6\%$ for hot Jupiters and $\geq70\pm16\%$ for warm Jupiters based on the analysis of \textit{Kepler} TTVs. The search for nearby transiting companions to hot Jupiters in \textit{TESS} by \cite{Sha2026} found an occurrence rate of $7.6^{+5.5}_{-3.8}\%$. Our derived nearby companion occurrence rates for the sub-Saturns ($36_{-14}^{+16}\%$ for desert+ridge and $91.1_{-4.3}^{+3.2}\%$ for savanna) are slightly higher than those numbers albeit consistent with the results from \cite{Wu2023}. However, the different studies use different frameworks for the occurrence rate calculations and have different underlying assumptions. For example, \cite{Sha2026} use the Beta-Binomial approach for the calculation of the occurrence rates. As a comparison the inferred nearby companion occurrence rate for our desert+ridge sub-Saturn sample with a Beta-Binomial method is $11.1_{-5.7}^{+8.2}\%$ (Appendix~\ref{app:val_methods}) in agreement with their hot Jupiter nearby companion rate.

A notable difference in companion architecture between the desert+ridge sub-Saturns and hot Jupiters is seen in the occurrence rate of massive long-period companions. In our analysis we find them to be rare ($18.3_{-9.1}^{+12.6}\%$ for the desert+ridge) which stands in contrast to similar studies of hot Jupiter companions finding long-period giant companions to be common \citep{Knutson2014,Zink2023}. However, our analysis counts only companions with resolved orbits inside the completeness grid, i.e.\ $P < 10^{4}$\,d ($a \lesssim 9$\,au for a solar-mass host), whereas the hot Jupiter statistics rest substantially on long-term RV accelerations, extending their sensitivity to $\sim$20\,au and into the brown-dwarf regime \citep{Knutson2014,Zink2023}. In contrast to some of the hot Jupiter companion analyses, systems in our sample exhibiting an unresolved RV drift but no resolved outer orbit enter our analysis as non-detections. Our $\simeq 18\%$ is therefore a lower limit on outer-companion occurrence in the sense probed by those studies, and its apparent deficit relative to the $\sim$50--70\% reported for hot Jupiter hosts should not be over-interpreted.

While these occurrence rates are for the full desert+ridge sub-Saturn sample, they show consistent results when we further subdivide the desert+ridge sub-Saturns by their radius, density, or eccentricity. Therefore, we find no evidence for a different formation mechanism for a subset of this subpopulation, e.g.\ starting as hot Jupiters undergoing strong envelope loss. If a significant fraction of desert+ridge sub-Saturns were stripped hot Jupiters \citep{Vissapragada2025}, we might expect the denser, more stripped planets to show different companion properties from the less dense ones that would be true sub-Saturns. However, we find that the nearby companion deficit of desert+ridge sub-Saturns is present in both density classes: $51_{-19}^{+20}\%$ (high density) and $36_{-15}^{+19}\%$ (low density), compared to savanna rates of $73_{-14}^{+11}\%$ and $95.0_{-3.5}^{+2.3}\%$ respectively. The occurrence rate for nearby companions to high-density sub-Saturns in the desert+ridge is even marginally higher, i.e.\ less similar to the hot Jupiter companion rates, which is the opposite of what we would expect if they were stripped hot Jupiters.

A similar dichotomy might be expected when dividing the desert+ridge sub-Saturns into different radius classes. If we assume that there is a limit to how small a stripped hot Jupiter can get, we might expect the sub-Saturns with larger radii to exhibit a more hot Jupiter-like occurrence rate compared to the smaller sub-Saturns. We again find consistent nearby occurrence rates between the desert+ridge Neptune-like sub-Saturns ($47_{-16}^{+17}\%$) and the desert+ridge giant-like sub-Saturns ($37_{-17}^{+23}\%$), with a slightly lower rate for the giant-like sub-Saturns.  The absence of clear differences suggests that the desert+ridge sub-Saturns could all have formed as intermediate-mass planets and then migrated via HEM, with a majority of them naturally ending up in the ridge driven by the tidal disruption limit \citep{castrogonzales2026,Zanazzi2026}. If we assume that hot Jupiters also arrive at their present-day orbits via HEM, the observed similarities in nearby companion occurrence rates are then a natural consequence of the shared migration mechanism and not a direct evolutionary link in which sub-Saturns are the stripped remnants of former hot Jupiters.

\subsection{The radial velocity follow-up selection function}\label{sec:selection_bias}
 
In order to achieve maximum sensitivity to detect companion planets, we required the planets in our sample to have both transit and RV data instead of the usual approach of focusing on one homogeneous dataset. This introduces a potential selection effect: the sample is shaped not only by which planets exist, but by which transiting candidates were chosen for RV follow-up. If the decision to pursue RV follow-up depended on system multiplicity, for example, if multi-planet systems were preferentially targeted because they are scientifically richer, or conversely avoided because they require more observations to characterize, then the multiplicity distribution of our sample would not represent the underlying population, biasing our companion occurrence rates.
 
We compared our sample against the parent population of all transit-detected sub-Saturn hosts in our radius range ($4.0\,\mathrm{R}_\oplus \leq R_\mathrm{p} \leq 8.5\,\mathrm{R}_\oplus$), without the mass-precision requirement, yielding 294 host systems of which 86 (29\%) have sufficient RV data to enter our sample. The follow-up fraction shows no positive dependence on transiting multiplicity. In fact we find the opposite: 32\% of hosts with a single transiting planet versus 25\% of multi-transiting hosts are followed up (standardized mean difference $-0.16$), and a permutation Cochran--Armitage trend test gives $p = 0.45$ \citep{Cochran1954,Armitage1955}. The mild multiplicity imbalance traces to the discovery survey, \textit{Kepler} hosts being both multi-rich and faint. To control for such confounding we fit an unpenalized logistic regression of follow-up status on a binary multiplicity indicator (two or more transiting planets), the host-star \textit{Gaia} magnitude, the sub-Saturn orbital period and radius, and the discovery survey. The follow-up probability depends strongly on host-star brightness ($p < 10^{-4}$) and on the sub-Saturn properties ($p = 0.005$ for orbital period, $p = 0.011$ for radius), as expected: brighter stars, shorter periods, and larger planets yield larger and more readily confirmed RV signals. Crucially, the multiplicity coefficient is consistent with zero ($p = 0.86$), corresponding to a multi-versus-single follow-up odds ratio of $1.07$ with a 95\% confidence interval of $[0.50, 2.30]$. We further note that the residual raw imbalance points in the conservative direction: an under-selection of multi-planet systems would deflate the savanna companion rates and cause us to underestimate the landscape gradient.
 
To propagate the remaining uncertainty in the selection function into the occurrence rates, we constructed inverse-probability weights that remove the multiplicity-dependent part of the fitted selection model at fixed observability, $w_j^{IPW} = \hat\pi(x_j;\beta_\mathrm{multi}{\to}0)/\hat\pi(x_j)$, and repeated the occurrence rate inference with these weights combined with the $1/n_\mathrm{ss}$ weights of Sect.~\ref{sec:occurrence_rate}. Only the multiplicity channel is reweighted: the dependence of follow-up on brightness and planet size acts on observability, which is precisely what the per-system completeness maps already quantify and correct. The measured correction is negligible, with every reported rate shifting by less than 1 percentage point, and propagating the selection-model uncertainty by bootstrap resampling of the parent sample (200 refits) bounds the 68\% systematic within $\pm 3$ ($\pm 4$) percentage points for the savanna (desert+ridge) categories, well inside the statistical uncertainties. Even under stress scenarios in which multi-planet systems are assumed to be $1.5$ or $2$ times more likely to receive follow-up, the largest shift is 10 percentage points (the savanna small nearby rate under the $2\times$ scenario), leaving the landscape gradient intact. 
 
These results show that the RV follow-up selection function operates on observability (i.e.\ stellar brightness, orbital period, and planet size) rather than on system architecture. Since these observability factors are what our per-system completeness maps quantify and correct for, we conclude that the follow-up selection does not significantly bias our occurrence rate analysis. The strong dependence of follow-up on sub-Saturn radius is further reassuring in light of our finding (Sect.~\ref{sec:rates_radius}) that companion occurrence is itself independent of sub-Saturn radius: even though larger sub-Saturns are over-represented in our sample, this does not distort the inferred companion rates.

\section{Summary}

We have analyzed the planetary companion architectures of 90 sub-Saturns (4.0--8.5\,$\mathrm{R}_\oplus$) in 86 host systems across the Neptunian desert, ridge, and savanna, using combined radial velocity and transit completeness maps to derive completeness-corrected companion occurrence rates with a joint Poisson occurrence model. Our main results are as follows:

\begin{itemize}\setlength{\itemsep}{8pt}

\item The raw companion fraction rises sharply across the landscape, from $11\%$ in the desert and $14\%$ in the ridge to $58\%$ in the savanna. After correcting for the completeness of the datasets with the joint occurrence model, this gradient persists: the nearby companion rate ($P_\mathrm{comp} < 200$\,d) is $36_{-14}^{+16}\%$ for sub-Saturns in the desert and ridge compared to $91.1_{-4.3}^{+3.2}\%$ for savanna sub-Saturns ($p(f_\mathrm{sav} > f_\mathrm{d+r}) = 0.9998$).

\item In the savanna, nearby companions are found interior and exterior to
the sub-Saturn at similar rates ($47.2_{-7.4}^{+7.4}\%$ and $58.3_{-9.0}^{+8.5}\%$).

\item The difference in nearby companions is driven almost entirely by low-mass companions. Small ($M_\mathrm{comp} < 20\,\mathrm{M}_\oplus$) nearby companions are far more common around savanna sub-Saturns ($88.2_{-5.5}^{+4.2}\%$) than around desert+ridge sub-Saturns ($29_{-14}^{+17}\%$), whereas medium-mass and giant companions within 200\,d are rare ($\lesssim 13\%$) in both groups. The sub-Saturn is thus typically the dominant body in its immediate orbital neighborhood.

\item These signatures point to quiescent migration for the savanna and high-eccentricity migration for the desert and ridge. Savanna sub-Saturns reside in compact, dynamically cold multi-planet systems, while desert and ridge sub-Saturns show a marked deficit of nearby companions.

\item The landscape gradient is the dominant signal when the sample is subdivided by sub-Saturn properties. Splitting by radius, bulk density, or eccentricity leaves the savanna-versus-desert+ridge contrast intact within each bin. After orbital period, the sub-Saturn bulk density is the only property whose association with companion presence survives false-discovery control in the model-free two-sample tests: low-density sub-Saturns show a markedly higher nearby companion rate ($89.3_{-5.1}^{+3.9}\%$) than high-density ones ($62_{-13}^{+12}\%$), a contrast that resides in the savanna.

\item The host-star populations differ across the landscape: desert and ridge sub-Saturns orbit a broad range of hosts extending to low-mass M-dwarfs, while savanna sub-Saturns orbit predominantly solar-type stars, adding to the observed difference in host star metallicity between the desert+ridge and savanna sub-Saturns. However, neither host-star property carries predictive power for companion occurrence, with statistically indistinguishable companion rates when dividing by host star properties.

\end{itemize}

These findings demonstrate that sub-Saturns are not a dynamically unique population. Instead, they undergo similar evolutionary pathways as their gas-giant counterparts, with their final orbital configurations defined by distinct migration pathways and the broader dynamical environment of the system.

\begin{acknowledgements}
We thank the anonymous referee for the corrections and suggestions that helped us improve the presentation of the paper. Funding for the \textit{TESS} mission is provided by NASA's Science Mission Directorate. We acknowledge the use of public \textit{TESS} data from pipelines at the TESS Science Office and at the TESS Science Processing Operations Center. Resources supporting this work were provided by the NASA High-End Computing (HEC) Program through the NASA Advanced Supercomputing (NAS) Division at Ames Research Center for the production of the SPOC data products. This paper includes data collected by the \textit{TESS} mission, which are publicly available from the Mikulski Archive for Space Telescopes (MAST) operated by the Space Telescope Science Institute (STScI). This research has made use of the NASA Exoplanet Archive, which is operated by the California Institute of Technology, under contract with the National Aeronautics and Space Administration under the Exoplanet Exploration Program. This paper includes data collected by the Kepler mission and obtained from the MAST data archive at the Space Telescope Science Institute (STScI). Funding for the Kepler mission is provided by the NASA Science Mission Directorate. STScI is operated by the Association of Universities for Research in Astronomy, Inc., under NASA contract NAS 5–26555.
\end{acknowledgements}

\bibliographystyle{aa}
\bibliography{bib,bib_coauth} % if your bibtex file is called example.bib

\begin{appendix}
 
\section{Validation and robustness of the occurrence rate methodology}\label{app:validation}

The completeness-corrected occurrence rates of this work rest on three methodological choices: the truncated-count observation model of the joint Poisson framework (Sect.~\ref{sec:occurrence_rate}), the prior on the cell rates $\mu_i$, and the assumed intrinsic distribution of undetected companions within each cell. We verify the calibration of the inference on simulated surveys (Appendix~\ref{app:val_calibration}), re-derive the rates with independent statistical methods (Appendix~\ref{app:val_methods}), test the prior assumption (Appendix~\ref{app:val_prior}), and test the intrinsic companion distribution assumption (Appendix~\ref{app:val_intrinsic}). Lastly, we also verify the assumed mutual inclination distribution (Appendix~\ref{app:val_maps}). The population comparisons that carry the conclusions of this paper are robust in every case.

\subsection{Calibration on simulated surveys}\label{app:val_calibration}

We first verified that the inference machinery is statistically calibrated. For each of 250 synthetic surveys, we drew the true cell rates from the prior, populated every system of the desert+ridge and savanna samples with Poisson companion counts, placed each companion at a random position within its cell, and detected it with the probability given by the system's real combined completeness map at that position. The synthetic detections were then fitted exactly as the real data, using only the cell-mean completeness. If the machinery is correct, the credible intervals must attain their nominal coverage. We find $68\%$ ($95\%$) intervals that cover the truth in $64$--$75\%$ ($94$--$97\%$) of the replicas for every cell and every derived quantity, with median biases below $2.4$ percentage points. The simulation places companions by their individual positions while the fit uses only cell means, so this test also validates the reduction of the two-dimensional completeness maps to per-cell scalars under the baseline intrinsic distribution.

We further tested the Poisson multiplicity assumption with a posterior predictive check: simulating detected companion counts from the fitted rates through each system's completeness reproduces the observed multiplicity histogram of both populations (e.g.\ $20.5/16/7.5/4$ predicted savanna systems with $0/1/2/{\geq}3$ detected companions against $20/17/7/4.5$ observed), and a zero-inflated extension of the model, which allows for a separate fraction of barren systems, yields rates consistent with the baseline model. 

\subsection{Independent statistical methods}\label{app:val_methods}

We compared the joint-model occurrence fractions with three independent methods that make different statistical assumptions: the Beta-Binomial framework of \cite{Bryan2024}, in which each system contributes a binary detection outcome with an effective number of trials given by its regional mean completeness; the Poisson point-process likelihood of \cite{Rosenthal2022}, applied per category with a constant intensity per unit $(\log P, \log M)$ area; and Approximate Bayesian Computation \citep[ABC-PMC;][]{Kunimoto2020}, whose forward model simulates companions at individual positions in each system's completeness map and matches the weighted number of systems with detected companions. Table~\ref{tab:validation} summarizes the comparison for the combined desert+ridge and savanna populations.

The ordering and significance of the landscape contrast are preserved under every method: $p(f_\mathrm{sav} > f_\mathrm{d+r}) \geq 0.999$ for the nearby category throughout, and the massive long-period category is consistent between the populations under every method. The absolute fractions of the three independent methods agree well with one another but are systematically lower than the joint-model values for the
composite categories, with an understood origin: all three treat each category as a single region and correct the detections with its regional mean completeness, whereas the joint model evaluates the correction cell by cell, attributing it to the poorly-surveyed small companions that dominate the population. The agreement of ABC, whose forward model uses the full two-dimensional completeness information at the simulated companion positions, with the analytic region-based methods further verifies that the reduction of the completeness maps to regional means introduces no bias at the level of these treatments; the corresponding verification for the cell-level means of the joint model is provided by the simulated surveys of Appendix~\ref{app:val_calibration}.

\begin{table*}
\centering
\caption{Comparison of the joint occurrence model with three independent statistical methods for the combined desert+ridge (D+R, $N=37.5$) versus savanna (Sav, $N=48.5$) populations.}
\label{tab:validation}
\resizebox{1.99\columnwidth}{!}{
\begin{tabular}{lcccccccccccc}
\hline\hline
\noalign{\smallskip}
 & \multicolumn{3}{c}{Joint model} & \multicolumn{3}{c}{Beta-Binomial} & \multicolumn{3}{c}{Poisson process} & \multicolumn{3}{c}{ABC-PMC}\\
Category & D+R & Sav & $p$ & D+R & Sav & $p$ & D+R & Sav & $p$ & D+R & Sav & $p$\\
\noalign{\smallskip}\hline\noalign{\smallskip}
Any companion & $87.7_{-13.6}^{+8.5}\%$ & $98.72_{-1.67}^{+0.86}\%$ & $0.929$ & $37_{-12}^{+13}\%$ & $97.7_{-3.7}^{+1.7}\%$ & $1.000$ & $31_{-11}^{+12}\%$ & $76.5_{-6.7}^{+5.9}\%$ & $0.999$ & $36_{-14}^{+14}\%$ & $90.0_{-4.9}^{+3.7}\%$ & $1.000$ \\[2pt]
Multiple companions & $62_{-23}^{+22}\%$ & $93.1_{-6.5}^{+4.2}\%$ & $0.930$ & $8.6_{-5.4}^{+9.3}\%$ & $58_{-11}^{+10}\%$ & $0.999$ & $5.4_{-4.0}^{+8.3}\%$ & $44.5_{-9.3}^{+9.5}\%$ & $0.997$ & $9.6_{-8.4}^{+16.8}\%$ & $66.7_{-10.7}^{+8.9}\%$ & $0.997$ \\[2pt]
Massive long-period ($M_\mathrm{comp} \geq 1\,\mathrm{M}_\mathrm{Jup}$, $P_\mathrm{comp}>365$\,d) & $18.3_{-9.1}^{+12.6}\%$ & $15.4_{-6.7}^{+8.9}\%$ & $0.414$ & $19.6_{-9.8}^{+13.2}\%$ & $16.5_{-7.2}^{+9.4}\%$ & $0.410$ & $15.8_{-8.8}^{+12.8}\%$ & $13.9_{-6.5}^{+8.8}\%$ & $0.447$ & $20_{-13}^{+17}\%$ & $17.1_{-9.2}^{+10.9}\%$ & $0.433$ \\[2pt]
Nearby ($P_\mathrm{comp}<200$\,d) & $36_{-14}^{+16}\%$ & $91.1_{-4.3}^{+3.2}\%$ & $1.000$ & $11.1_{-5.7}^{+8.2}\%$ & $84.0_{-7.3}^{+5.8}\%$ & $1.000$ & $8.8_{-4.9}^{+7.7}\%$ & $57.8_{-7.4}^{+7.1}\%$ & $1.000$ & $11.0_{-7.5}^{+10.7}\%$ & $71.4_{-7.3}^{+7.0}\%$ & $1.000$ \\[2pt]
Inner nearby ($P_\mathrm{comp}<P_\mathrm{ss}$) & $4.5_{-2.9}^{+5.1}\%$ & $47.2_{-7.4}^{+7.4}\%$ & $1.000$ & $4.5_{-2.9}^{+5.2}\%$ & $54.3_{-8.7}^{+8.5}\%$ & $1.000$ & $2.8_{-2.1}^{+4.4}\%$ & $41.7_{-7.3}^{+7.5}\%$ & $1.000$ & $10.3_{-7.2}^{+9.0}\%$ & $50.6_{-9.3}^{+9.2}\%$ & $0.998$ \\[2pt]
Outer nearby ($P_\mathrm{ss}<P_\mathrm{comp}<200$\,d) & $12.6_{-6.4}^{+9.2}\%$ & $58.3_{-9.0}^{+8.5}\%$ & $0.999$ & $13.1_{-6.7}^{+9.5}\%$ & $63.8_{-10.8}^{+9.9}\%$ & $1.000$ & $10.4_{-5.9}^{+8.9}\%$ & $46.3_{-8.8}^{+8.8}\%$ & $0.996$ & $13.0_{-8.7}^{+12.9}\%$ & $55.6_{-9.9}^{+9.6}\%$ & $0.995$ \\[2pt]
Small nearby ($M_\mathrm{comp}<20\,\mathrm{M}_\oplus$) & $29_{-14}^{+17}\%$ & $88.2_{-5.5}^{+4.2}\%$ & $1.000$ & $33_{-16}^{+19}\%$ & $96.7_{-5.2}^{+2.5}\%$ & $1.000$ & $27_{-15}^{+19}\%$ & $78.6_{-8.0}^{+6.6}\%$ & $0.994$ & $34_{-17}^{+20}\%$ & $89.9_{-6.0}^{+4.4}\%$ & $0.999$ \\[2pt]
Medium nearby ($20\,\mathrm{M}_\oplus \leq M_\mathrm{comp}<80\,\mathrm{M}_\oplus$) & $5.1_{-3.2}^{+5.7}\%$ & $13.4_{-4.8}^{+6.0}\%$ & $0.867$ & $5.1_{-3.3}^{+5.7}\%$ & $14.3_{-5.1}^{+6.4}\%$ & $0.882$ & $3.1_{-2.3}^{+4.9}\%$ & $12.5_{-4.7}^{+6.0}\%$ & $0.913$ & $12.7_{-8.2}^{+11.2}\%$ & $14.9_{-8.9}^{+9.7}\%$ & $0.562$ \\[2pt]
Giant nearby ($M_\mathrm{comp}\geq 80\,\mathrm{M}_\oplus$) & $2.1_{-1.6}^{+3.3}\%$ & $11.8_{-4.1}^{+5.2}\%$ & $0.963$ & $2.1_{-1.6}^{+3.3}\%$ & $10.2_{-3.9}^{+5.0}\%$ & $0.942$ & $0.71_{-0.65}^{+2.33}\%$ & $9.0_{-3.6}^{+4.7}\%$ & $0.969$ & $12.7_{-8.3}^{+10.0}\%$ & $10.1_{-5.2}^{+7.1}\%$ & $0.416$ \\[2pt]
\noalign{\smallskip}\hline
\end{tabular}
}
\end{table*}

\subsection{Prior sensitivity}\label{app:val_prior}

The posterior of a cell rate depends on the prior wherever the likelihood is weak, and derived quantities accumulate the priors of their constituent cells. Table~\ref{tab:prior_sens} lists the posterior medians of all reported quantities under the adopted prior uniform in the per-cell fraction, a prior flat in $\mu_i$, and the Jeffreys prior $p(\mu_i) \propto \mu_i^{-1/2}$. The well-constrained quantities are robust. The nearby companion rate shifts by at most $8.4$ ($1.5$) percentage points for the desert+ridge (savanna), and the cell rates of the individual nearby cells by similar amounts. The any- and multiple-companion fractions, in contrast, inherit the prior dependence of the essentially unconstrained small and medium far cells, spanning up to $37$ percentage points for the desert+ridge multiple fraction. We therefore quote these two quantities as conditional on the adopted prior throughout, and base no population comparison on them alone; the comparisons themselves, $p(f_\mathrm{sav} > f_\mathrm{d+r})$, are unchanged under all three priors, as both populations share the same cells and priors.

The three priors of Table~\ref{tab:prior_sens} vary the shape of the per-cell prior but share the same construction. An independent prior is
placed on the rate of every cell, and each derived fraction inherits whatever
prior this construction induces on it.  A prior that is uninformative for each individual cell is not uninformative for a quantity derived from several cells. Under the baseline construction the total nearby rate $\mu_\mathrm{nb}$ is the sum of three cell rates with $p(\mu_i) = e^{-\mu_i}$ each, so it follows a Gamma$(3,1)$ prior, and the induced prior on the derived fraction $f_\mathrm{nearby} = 1 - e^{-\mu_\mathrm{nb}}$ has a median of $93\%$ and places only $2.5\%$ of its probability below $46\%$. The same mechanism acts more strongly still on the any- and multiple-companion fractions, which accumulate the priors of all six cells. This induced prior is harmless wherever the likelihood is informative, but it sets the posterior wherever the data are weak.

To isolate the effect of the construction, we refitted the model under the converse choice: a prior uniform in the derived fraction of a region, implemented as $p(\mu_\mathrm{T}) = e^{-\mu_\mathrm{T}}$ for the total rate $\mu_\mathrm{T}$ of the region with the cell shares distributed as Dirichlet$(1,\ldots,1)$, evaluating the posterior by importance sampling over draws from this prior. Under this construction the derived fraction has a uniform prior by definition, while the induced priors on the individual cells are no longer uniform; the two constructions are therefore complementary, and no prior can be simultaneously agnostic about the cells and their composites. Table~\ref{tab:construction_sens} compares the posterior medians. The
data-dominated savanna fractions are insensitive ($\leq 2$ percentage points,
except the multiple-companion fraction at $7$), whereas the weakly constrained
desert+ridge composites move substantially (nearby $35.8\% \to 19.3\%$, any
$87.7\% \to 53.9\%$, multiple $62.0\% \to 18.2\%$), which reconciles them with
the region-based estimators of Table~\ref{tab:validation}. The effect is most
severe for the desert-only bin: its nearby fraction of $42^{+22}_{-19}\%$, obtained from zero detections in nine systems, becomes $11.8^{+17.3}_{-8.8}\%$ with a $2.5$th percentile of $0.4\%$. The nonzero lower bound of the baseline value is thus set by the induced prior rather than by the data, as the small nearby cell retains only $1.6$ effective systems at its $18\%$ mean completeness.

Recovery tests on simulated surveys, with true rates swept across their full range and detection through the real completeness maps, show that neither construction dominates. The per-cell construction is nearly unbiased for the individual cell rates (within $\sim$$3$ percentage points for the small nearby cell) and for the data-dominated savanna composites, but overestimates weakly constrained composites; the composite-uniform construction behaves oppositely, underestimating the dominant small nearby cell by $4$--$13$ percentage points. At the present sample size the influence of the prior on sparsely populated bins is therefore irreducible and can only be relocated between the cells and their composites. We retain the per-cell construction as the baseline, since it is unbiased for the individual cell rates and the savanna fractions which are the quantities that are of most interest for our science question and we quote the composite-uniform values of Table~\ref{tab:construction_sens} as the
prior-construction bracket for the weakly constrained composites, in particular the desert+ridge any- and multiple-companion fractions and the desert-only bin. Every population comparison is preserved under both constructions, and the savanna versus desert+ridge nearby contrast strengthens under the alternative ($19.3\%$ versus $89.8\%$).

\begin{table*}
\centering
\caption{Prior sensitivity of the joint-model occurrence fractions (posterior medians, baseline intrinsic distribution).}
\label{tab:prior_sens}
\begin{tabular}{llcccc}
\hline\hline
\noalign{\smallskip}
Category & Population & Uniform $f$ & Flat $\mu$ & Jeffreys & $\Delta$ [pp] \\
\noalign{\smallskip}\hline\noalign{\smallskip}
Any companion & Desert+ridge & $87.7\%$ & $99.9\%$ & $96.6\%$ & 12.1 \\
Any companion & Savanna & $98.7\%$ & $99.9\%$ & $99.3\%$ & 1.2 \\
Multiple companions & Desert+ridge & $62.0\%$ & $99.0\%$ & $84.9\%$ & 37.0 \\
Multiple companions & Savanna & $93.1\%$ & $99.1\%$ & $96.1\%$ & 6.0 \\
Massive long-period ($M_\mathrm{comp} \geq 1\,\mathrm{M}_\mathrm{Jup}$, $P_\mathrm{comp}>365$\,d) & Desert+ridge & $18.3\%$ & $20.0\%$ & $15.8\%$ & 4.2 \\
Massive long-period ($M_\mathrm{comp} \geq 1\,\mathrm{M}_\mathrm{Jup}$, $P_\mathrm{comp}>365$\,d) & Savanna & $15.4\%$ & $16.2\%$ & $13.8\%$ & 2.4 \\
Nearby ($P_\mathrm{comp}<200$\,d) & Desert+ridge & $35.8\%$ & $40.2\%$ & $31.8\%$ & 8.4 \\
Nearby ($P_\mathrm{comp}<200$\,d) & Savanna & $91.1\%$ & $92.7\%$ & $92.2\%$ & 1.5 \\
Inner nearby ($P_\mathrm{comp}<P_\mathrm{ss}$) & Desert+ridge & $4.5\%$ & $4.7\%$ & $2.8\%$ & 1.9 \\
Inner nearby ($P_\mathrm{comp}<P_\mathrm{ss}$) & Savanna & $47.2\%$ & $48.3\%$ & $47.5\%$ & 1.1 \\
Outer nearby ($P_\mathrm{ss}<P_\mathrm{comp}<200$\,d) & Desert+ridge & $12.6\%$ & $13.4\%$ & $10.4\%$ & 2.9 \\
Outer nearby ($P_\mathrm{ss}<P_\mathrm{comp}<200$\,d) & Savanna & $58.3\%$ & $60.3\%$ & $59.2\%$ & 2.0 \\
Small nearby ($M_\mathrm{comp}<20\,\mathrm{M}_\oplus$) & Desert+ridge & $29.2\%$ & $33.7\%$ & $27.2\%$ & 6.5 \\
Small nearby ($M_\mathrm{comp}<20\,\mathrm{M}_\oplus$) & Savanna & $88.2\%$ & $90.2\%$ & $89.7\%$ & 2.0 \\
Medium nearby ($20\,\mathrm{M}_\oplus \leq M_\mathrm{comp}<80\,\mathrm{M}_\oplus$) & Desert+ridge & $5.1\%$ & $5.3\%$ & $3.1\%$ & 2.2 \\
Medium nearby ($20\,\mathrm{M}_\oplus \leq M_\mathrm{comp}<80\,\mathrm{M}_\oplus$) & Savanna & $13.4\%$ & $13.8\%$ & $12.5\%$ & 1.2 \\
Giant nearby ($M_\mathrm{comp}\geq 80\,\mathrm{M}_\oplus$) & Desert+ridge & $2.1\%$ & $2.1\%$ & $0.7\%$ & 1.4 \\
Giant nearby ($M_\mathrm{comp}\geq 80\,\mathrm{M}_\oplus$) & Savanna & $11.8\%$ & $12.1\%$ & $11.1\%$ & 1.0 \\
\noalign{\smallskip}\hline
\end{tabular}
\tablefoot{$\Delta$ is the maximum spread across the three priors in percentage points (pp).}
\end{table*}

\begin{table*}
\centering
\caption{Sensitivity of the joint-model occurrence fractions to the
construction of the prior (posterior medians).}
\label{tab:construction_sens}
\begin{tabular}{llccc}
\hline\hline
\noalign{\smallskip}
Category & Population & Per-cell uniform $f$ & Composite uniform & $\Delta$ [pp] \\
\noalign{\smallskip}\hline\noalign{\smallskip}
Any companion & Desert+ridge & $87.7\%$ & $53.9\%$ & 33.8 \\
Any companion & Savanna & $98.7\%$ & $96.8\%$ & 1.9 \\
Multiple companions & Desert+ridge & $62.0\%$ & $18.2\%$ & 43.8 \\
Multiple companions & Savanna & $93.1\%$ & $85.8\%$ & 7.3 \\
Nearby ($P_\mathrm{comp}<200$\,d) & Desert+ridge & $35.8\%$ & $19.3\%$ & 16.5 \\
Nearby ($P_\mathrm{comp}<200$\,d) & Savanna & $91.1\%$ & $89.8\%$ & 1.4 \\
Small nearby ($M_\mathrm{comp}<20\,\mathrm{M}_\oplus$) & Desert+ridge & $29.2\%$ & $13.2\%$ & 16.0 \\
Small nearby ($M_\mathrm{comp}<20\,\mathrm{M}_\oplus$) & Savanna & $88.2\%$ & $86.5\%$ & 1.7 \\
Medium nearby ($20\,\mathrm{M}_\oplus \leq M_\mathrm{comp}<80\,\mathrm{M}_\oplus$) & Desert+ridge & $5.1\%$ & $3.5\%$ & 1.6 \\
Medium nearby ($20\,\mathrm{M}_\oplus \leq M_\mathrm{comp}<80\,\mathrm{M}_\oplus$) & Savanna & $13.4\%$ & $13.1\%$ & 0.3 \\
Giant nearby ($M_\mathrm{comp}\geq 80\,\mathrm{M}_\oplus$) & Desert+ridge & $2.1\%$ & $1.6\%$ & 0.5 \\
Giant nearby ($M_\mathrm{comp}\geq 80\,\mathrm{M}_\oplus$) & Savanna & $11.8\%$ & $11.6\%$ & 0.2 \\

\noalign{\smallskip}\hline
\end{tabular}
\tablefoot{The baseline column repeats the values of
Table~\ref{tab:rates_corrected} (independent priors, uniform in each cell
fraction); the composite-uniform column places the uniform prior on the
derived fraction of the region instead (total rate with
$p(\mu_\mathrm{T}) = e^{-\mu_\mathrm{T}}$, Dirichlet$(1,\ldots,1)$ cell
shares). The nearby-region rows use the three nearby cells; the any- and
multiple-companion rows use all six cells. Categories fitted as a single
region (inner, outer, massive long-period) are identical under both
constructions. $\Delta$ is the difference in percentage points (pp).}
\end{table*}
\subsection{Intrinsic companion distribution}\label{app:val_intrinsic}

The cell-mean completeness $C_{j,i}$ weights the completeness surface uniformly in $(\log P, \log M)$ within each cell, and the corrected rates inherit this assumption. We recomputed $C_{j,i}$ under two alternatives: the RV-calibrated power law $\mathrm{d}N \propto M^{-0.31}\,P^{+0.26}\,\mathrm{d}\log M\,\mathrm{d}\log P$ of \cite{Cumming2008}, and a distribution rising towards small masses ($\propto M^{-0.7}$) with a break to a flat $\log P$ distribution beyond 10\,d, motivated by the \textit{Kepler} small-planet population \citep{Howard2012}. Table~\ref{tab:intrinsic_sens} shows the results. The high-completeness cells are insensitive (the medium and giant nearby cells and the massive long-period category change by at most $3$ percentage points), whereas the small nearby cell, whose completeness varies most strongly within the cell, increases by up to $24$ ($10$) percentage points for the desert+ridge (savanna) under the bottom-heavy alternative, propagating into the nearby rates. Both populations shift together, and every population comparison is preserved. The calibration simulations of Appendix~\ref{app:val_calibration}, repeated with companions placed according to the bottom-heavy distribution while fitting with the baseline cell means, bound the corresponding bias of the fitted small nearby cell rate at $-7$ to $-14$ percentage points, consistent with the spread in Table~\ref{tab:intrinsic_sens}.

\begin{table*}
\centering
\caption{Sensitivity of the joint-model occurrence fractions to the assumed intrinsic distribution of companions within each cell (posterior medians, uniform-$f$ prior).}
\label{tab:intrinsic_sens}
\begin{tabular}{llcccc}
\hline\hline
\noalign{\smallskip}
Category & Population & Log-uniform & C08 & Rising-small & $\Delta$ [pp] \\
\noalign{\smallskip}\hline\noalign{\smallskip}
Any companion & Desert+ridge & $87.7\%$ & $93.0\%$ & $93.9\%$ & 6.2 \\
Any companion & Savanna & $98.7\%$ & $99.8\%$ & $99.9\%$ & 1.2 \\
Multiple companions & Desert+ridge & $62.0\%$ & $74.4\%$ & $76.9\%$ & 14.9 \\
Multiple companions & Savanna & $93.1\%$ & $98.4\%$ & $99.0\%$ & 5.9 \\
Massive long-period ($M_\mathrm{comp} \geq 1\,\mathrm{M}_\mathrm{Jup}$, $P_\mathrm{comp}>365$\,d) & Desert+ridge & $18.3\%$ & $21.2\%$ & $19.5\%$ & 2.9 \\
Massive long-period ($M_\mathrm{comp} \geq 1\,\mathrm{M}_\mathrm{Jup}$, $P_\mathrm{comp}>365$\,d) & Savanna & $15.4\%$ & $18.0\%$ & $16.4\%$ & 2.6 \\
Nearby ($P_\mathrm{comp}<200$\,d) & Desert+ridge & $35.8\%$ & $50.2\%$ & $58.4\%$ & 22.5 \\
Nearby ($P_\mathrm{comp}<200$\,d) & Savanna & $91.1\%$ & $97.5\%$ & $98.8\%$ & 7.7 \\
Inner nearby ($P_\mathrm{comp}<P_\mathrm{ss}$) & Desert+ridge & $4.5\%$ & $7.2\%$ & $13.3\%$ & 8.7 \\
Inner nearby ($P_\mathrm{comp}<P_\mathrm{ss}$) & Savanna & $47.2\%$ & $63.4\%$ & $83.2\%$ & 35.9 \\
Outer nearby ($P_\mathrm{ss}<P_\mathrm{comp}<200$\,d) & Desert+ridge & $12.6\%$ & $25.2\%$ & $45.3\%$ & 32.7 \\
Outer nearby ($P_\mathrm{ss}<P_\mathrm{comp}<200$\,d) & Savanna & $58.3\%$ & $83.3\%$ & $97.4\%$ & 39.0 \\
Small nearby ($M_\mathrm{comp}<20\,\mathrm{M}_\oplus$) & Desert+ridge & $29.2\%$ & $43.8\%$ & $52.8\%$ & 23.6 \\
Small nearby ($M_\mathrm{comp}<20\,\mathrm{M}_\oplus$) & Savanna & $88.2\%$ & $96.5\%$ & $98.4\%$ & 10.2 \\
Medium nearby ($20\,\mathrm{M}_\oplus \leq M_\mathrm{comp}<80\,\mathrm{M}_\oplus$) & Desert+ridge & $5.1\%$ & $6.5\%$ & $6.7\%$ & 1.7 \\
Medium nearby ($20\,\mathrm{M}_\oplus \leq M_\mathrm{comp}<80\,\mathrm{M}_\oplus$) & Savanna & $13.4\%$ & $16.0\%$ & $15.9\%$ & 2.5 \\
Giant nearby ($M_\mathrm{comp}\geq 80\,\mathrm{M}_\oplus$) & Desert+ridge & $2.1\%$ & $2.3\%$ & $2.4\%$ & 0.3 \\
Giant nearby ($M_\mathrm{comp}\geq 80\,\mathrm{M}_\oplus$) & Savanna & $11.8\%$ & $12.6\%$ & $12.7\%$ & 0.9 \\
\noalign{\smallskip}\hline
\end{tabular}
\tablefoot{$\Delta$ is the maximum spread across the three distributions in percentage points (pp). C08: \cite{Cumming2008}.}
\end{table*}

\subsection{Sensitivity to the assumed mutual inclination distribution}\label{app:val_maps}

We also propagated the principal assumptions on the mutual inclination distribution entering the combined completeness maps into the occurrence rates (Table~\ref{tab:map_sens}). Replacing the mutual-inclination prescription \citep[][evaluated at the known multiplicity including the hypothetical companion]{he2020} with the same relation at the known multiplicity alone, with a fixed Rayleigh scale of $\sigma_i = 5$\,deg, changes the nearby rates by at most $8$ and $3$ percentage points, respectively; even the extreme assumption of isotropic companion orbits, which removes all coplanarity information, preserves every population comparison. 

\begin{table*}
\centering
\caption{Headline occurrence fractions (posterior medians) under variants of the assumed mutual inclination distribution.}
\label{tab:map_sens}
\begin{tabular}{lcccccccc}
\hline\hline
\noalign{\smallskip}
Variant & \multicolumn{2}{c}{Small nearby} & \multicolumn{2}{c}{Nearby} & \multicolumn{2}{c}{Any} & \multicolumn{2}{c}{Multiple}\\
 & D+R & Sav & D+R & Sav & D+R & Sav & D+R & Sav\\
\noalign{\smallskip}\hline\noalign{\smallskip}
Baseline & $29.2\%$ & $88.2\%$ & $35.8\%$ & $91.1\%$ & $87.7\%$ & $98.7\%$ & $62.0\%$ & $93.1\%$ \\
He+2020 with $n = n_\mathrm{known}$ & $36.9\%$ & $91.8\%$ & $43.6\%$ & $94.0\%$ & $89.7\%$ & $99.2\%$ & $66.3\%$ & $95.4\%$ \\
Fixed $\sigma_i = 5$\,deg & $30.2\%$ & $91.4\%$ & $36.8\%$ & $93.6\%$ & $87.9\%$ & $99.2\%$ & $62.4\%$ & $95.4\%$ \\
Isotropic inclinations & $51.4\%$ & $99.7\%$ & $57.7\%$ & $99.8\%$ & $93.6\%$ & $100.0\%$ & $76.1\%$ & $99.8\%$ \\
\noalign{\smallskip}\hline
\end{tabular}
\end{table*}

\FloatBarrier
\section{Occurrence rates by radius class} \label{app:rad}

\begin{figure*}
    \centering
    \includegraphics[width=1.99\columnwidth]{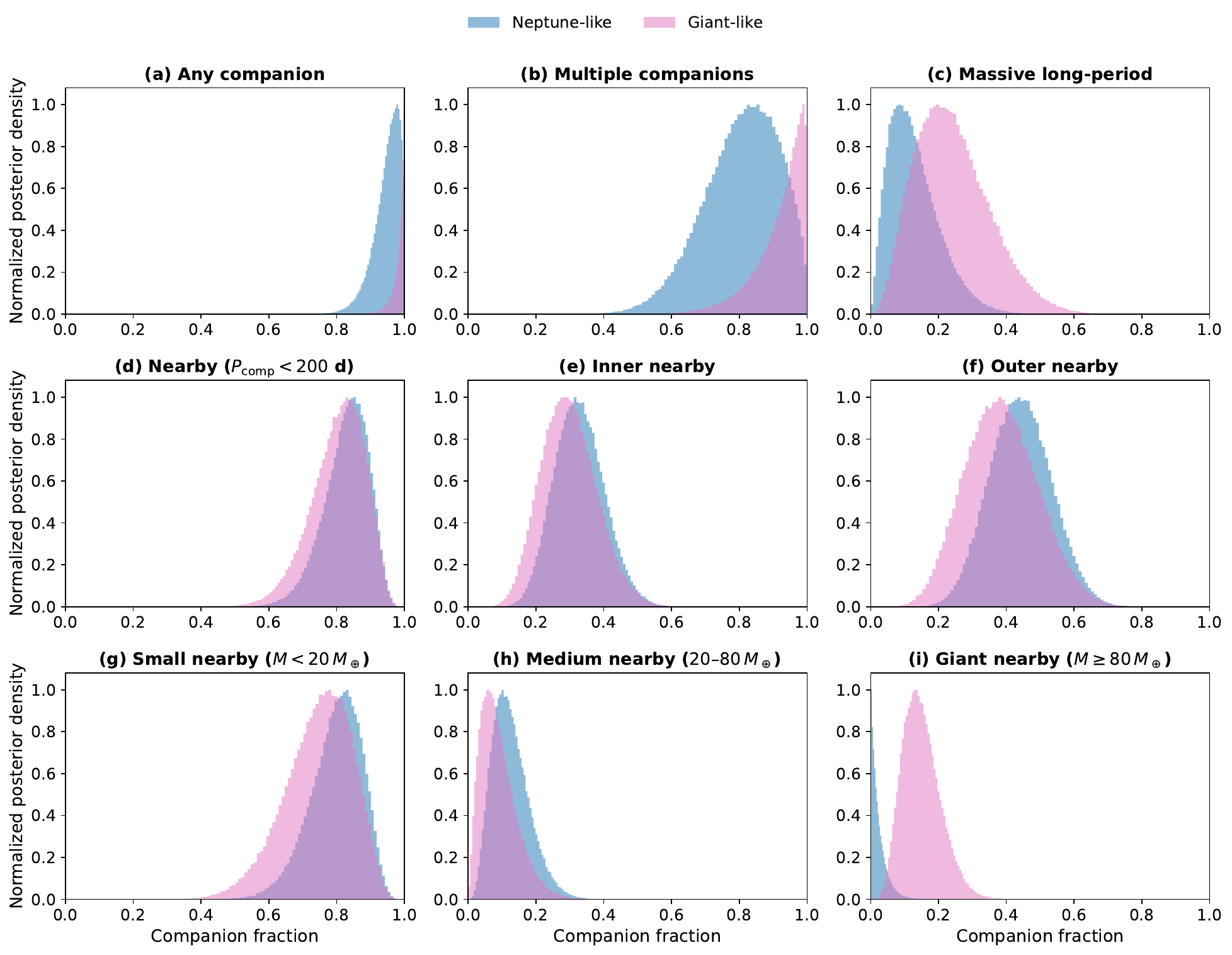} 
    \caption{Posterior distributions of the completeness-corrected occurrence rates of different types of companions for systems with a Neptune-like ($4\,\mathrm{R}_\oplus \leq R_p < 6\,\mathrm{R}_\oplus$) or giant-like ($6\,\mathrm{R}_\oplus \leq R_p \leq 8.5\,\mathrm{R}_\oplus$) sub-Saturn.}
    \label{fig:post_rad}
\end{figure*}

\begin{table*}
\centering
\caption{Completeness-corrected companion fractions by sub-Saturn radius class (boundary $6\,\mathrm{R}_\oplus$).}
\label{tab:rates_radius}
\resizebox{1.99\columnwidth}{!}{
\begin{tabular}{lcccccc}
\hline\hline
\noalign{\smallskip}
 & \multicolumn{3}{c}{Neptune-like ($N=44.5$)} & \multicolumn{3}{c}{Giant-like ($N=41.5$)}\\
Category & All & D+R ($N=21.5$) & Sav ($N=23$) & All & D+R ($N=16$) & Sav ($N=25.5$)\\
\noalign{\smallskip}\hline\noalign{\smallskip}
Any companion & $95.5_{-4.5}^{+2.9}\%$ & $92.3_{-10.1}^{+5.5}\%$ & $97.9_{-2.9}^{+1.5}\%$ & $98.90_{-2.30}^{+0.85}\%$ & $90.6_{-16.3}^{+7.4}\%$ & $99.54_{-1.16}^{+0.37}\%$ \\[2pt]
Multiple companions & $82_{-12}^{+10}\%$ & $73_{-21}^{+17}\%$ & $89.8_{-9.9}^{+6.4}\%$ & $93.9_{-8.9}^{+4.3}\%$ & $68_{-29}^{+22}\%$ & $97.1_{-5.4}^{+2.2}\%$ \\[2pt]
Massive long-period ($M_\mathrm{comp} \geq 1\,\mathrm{M}_\mathrm{Jup}$, $P_\mathrm{comp}>365$\,d) & $11.7_{-6.0}^{+8.6}\%$ & $26_{-13}^{+16}\%$ & $5.9_{-4.4}^{+9.0}\%$ & $23.0_{-9.7}^{+12.2}\%$ & $14_{-10}^{+19}\%$ & $31_{-13}^{+15}\%$ \\[2pt]
Nearby ($P_\mathrm{comp}<200$\,d) & $83.5_{-7.3}^{+5.7}\%$ & $47_{-16}^{+17}\%$ & $92.0_{-5.6}^{+3.7}\%$ & $81.2_{-8.8}^{+6.9}\%$ & $37_{-17}^{+23}\%$ & $88.6_{-7.2}^{+5.0}\%$ \\[2pt]
Inner nearby ($P_\mathrm{comp}<P_\mathrm{ss}$) & $32.5_{-7.1}^{+7.6}\%$ & $7.2_{-4.6}^{+7.9}\%$ & $50.3_{-9.9}^{+9.7}\%$ & $29.4_{-7.8}^{+8.6}\%$ & $6.2_{-4.6}^{+9.3}\%$ & $44_{-11}^{+11}\%$ \\[2pt]
Outer nearby ($P_\mathrm{ss}<P_\mathrm{comp}<200$\,d) & $44.1_{-9.1}^{+9.3}\%$ & $18.7_{-9.3}^{+12.7}\%$ & $61_{-11}^{+11}\%$ & $38_{-11}^{+11}\%$ & $9.8_{-7.3}^{+14.1}\%$ & $54_{-13}^{+13}\%$ \\[2pt]
Small nearby ($M_\mathrm{comp}<20\,\mathrm{M}_\oplus$) & $80.6_{-8.4}^{+6.7}\%$ & $37_{-17}^{+20}\%$ & $89.7_{-7.0}^{+4.7}\%$ & $75.6_{-11.0}^{+8.9}\%$ & $23_{-17}^{+27}\%$ & $82.9_{-10.1}^{+7.4}\%$ \\[2pt]
Medium nearby ($20\,\mathrm{M}_\oplus \leq M_\mathrm{comp}<80\,\mathrm{M}_\oplus$) & $11.9_{-4.7}^{+6.1}\%$ & $7.8_{-5.0}^{+8.4}\%$ & $17.4_{-7.1}^{+9.1}\%$ & $8.1_{-4.2}^{+6.2}\%$ & $7.3_{-5.4}^{+10.9}\%$ & $11.6_{-5.9}^{+8.5}\%$ \\[2pt]
Giant nearby ($M_\mathrm{comp}\geq 80\,\mathrm{M}_\oplus$) & $1.7_{-1.2}^{+2.6}\%$ & $3.4_{-2.6}^{+5.4}\%$ & $3.0_{-2.2}^{+4.7}\%$ & $14.4_{-4.9}^{+6.1}\%$ & $4.8_{-3.6}^{+7.4}\%$ & $21.6_{-7.2}^{+8.6}\%$ \\[2pt]
\noalign{\smallskip}\hline
\end{tabular}
}
\tablefoot{For each class: the full class, its desert+ridge members, and its savanna members. The inner and outer nearby rows use the $P_\mathrm{ss}<200$\,d subsample.}
\end{table*}

\FloatBarrier 
\section{Occurrence rates by density class} \label{app:den}
 
\begin{figure*}
    \centering
    \includegraphics[width=1.99\columnwidth]{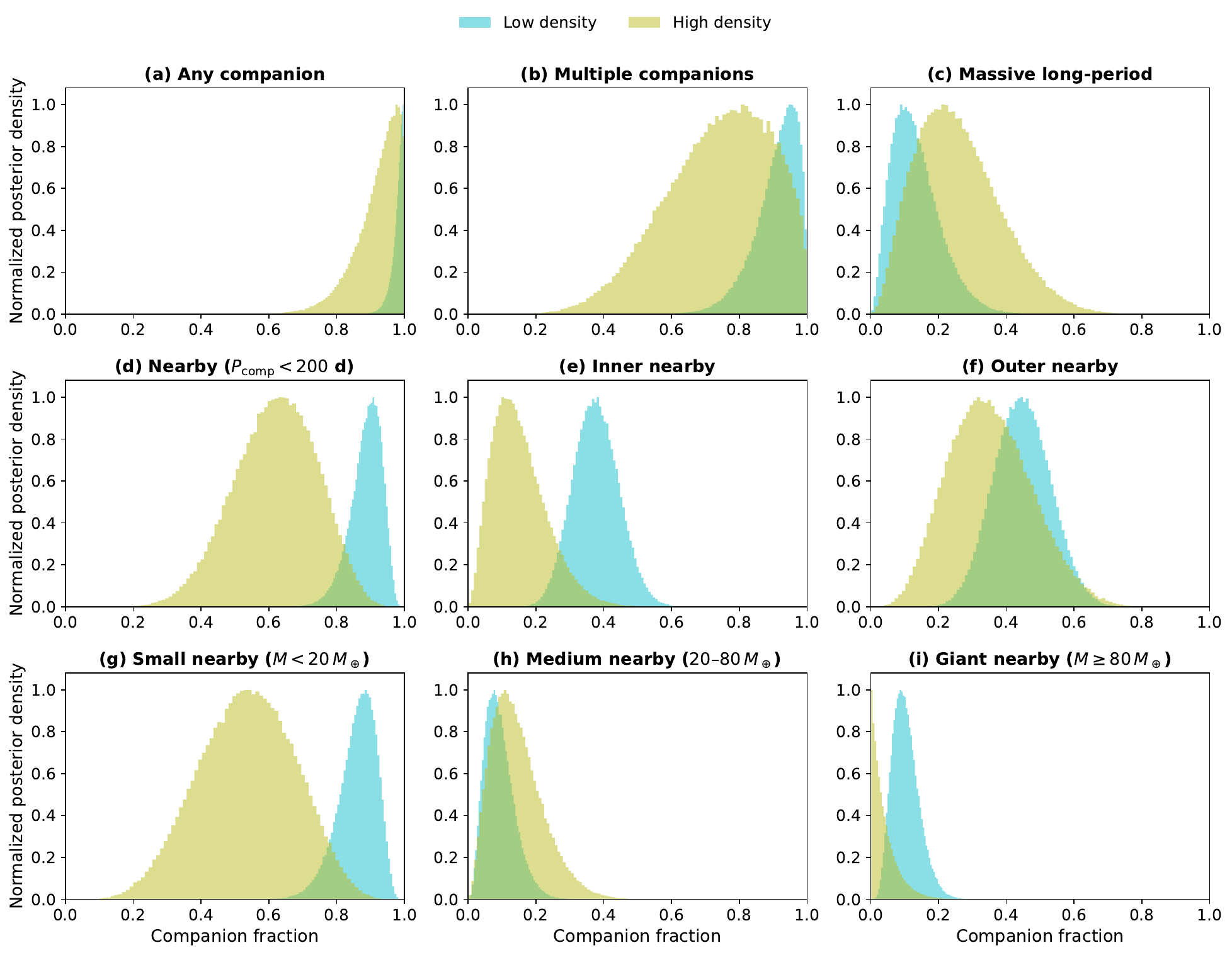} 
    \caption{Same as Fig.~\ref{fig:post_rad}, but for the two density classes: low-density ($\rho< 1.4$\,g\,cm$^{-3}$) and high-density ($\rho \geq 1.4$\,g\,cm$^{-3}$).}
    \label{fig:post_dens}
\end{figure*}

\begin{table*}
\centering
\caption{Same as Table~\ref{tab:rates_radius}, but for the sub-Saturn bulk density classes (boundary $1.4$\,g\,cm$^{-3}$).}
\label{tab:rates_density}
\resizebox{1.99\columnwidth}{!}{
\begin{tabular}{lcccccc}
\hline\hline
\noalign{\smallskip}
 & \multicolumn{3}{c}{Low density ($N=60$)} & \multicolumn{3}{c}{High density ($N=26$)}\\
Category & All & D+R ($N=24.5$) & Sav ($N=35.5$) & All & D+R ($N=13$) & Sav ($N=13$)\\
\noalign{\smallskip}\hline\noalign{\smallskip}
Any companion & $98.5_{-2.0}^{+1.0}\%$ & $90.3_{-14.0}^{+7.3}\%$ & $99.43_{-1.00}^{+0.42}\%$ & $93.4_{-7.8}^{+4.6}\%$ & $93.4_{-10.3}^{+4.9}\%$ & $96.4_{-5.9}^{+2.7}\%$ \\[2pt]
Multiple companions & $92.1_{-7.6}^{+4.9}\%$ & $68_{-26}^{+21}\%$ & $96.5_{-4.6}^{+2.4}\%$ & $75_{-18}^{+15}\%$ & $76_{-22}^{+16}\%$ & $84_{-16}^{+10}\%$ \\[2pt]
Massive long-period ($M_\mathrm{comp} \geq 1\,\mathrm{M}_\mathrm{Jup}$, $P_\mathrm{comp}>365$\,d) & $12.4_{-5.8}^{+8.1}\%$ & $16_{-10}^{+16}\%$ & $14.0_{-7.1}^{+10.0}\%$ & $25_{-11}^{+14}\%$ & $28_{-15}^{+20}\%$ & $27_{-14}^{+19}\%$ \\[2pt]
Nearby ($P_\mathrm{comp}<200$\,d) & $89.3_{-5.1}^{+3.9}\%$ & $36_{-15}^{+19}\%$ & $95.0_{-3.5}^{+2.3}\%$ & $62_{-13}^{+12}\%$ & $51_{-19}^{+20}\%$ & $73_{-14}^{+11}\%$ \\[2pt]
Inner nearby ($P_\mathrm{comp}<P_\mathrm{ss}$) & $37.6_{-6.6}^{+6.9}\%$ & $6.8_{-4.4}^{+7.6}\%$ & $53.2_{-8.3}^{+8.1}\%$ & $14.1_{-6.6}^{+9.1}\%$ & $6.7_{-4.9}^{+10.1}\%$ & $27_{-12}^{+15}\%$ \\[2pt]
Outer nearby ($P_\mathrm{ss}<P_\mathrm{comp}<200$\,d) & $44.6_{-8.4}^{+8.5}\%$ & $10.8_{-6.8}^{+11.3}\%$ & $59.9_{-9.9}^{+9.3}\%$ & $35_{-12}^{+13}\%$ & $22_{-12}^{+16}\%$ & $51_{-18}^{+17}\%$ \\[2pt]
Small nearby ($M_\mathrm{comp}<20\,\mathrm{M}_\oplus$) & $86.8_{-6.2}^{+4.7}\%$ & $26_{-16}^{+22}\%$ & $93.1_{-4.7}^{+3.1}\%$ & $54_{-15}^{+14}\%$ & $40_{-21}^{+23}\%$ & $60_{-17}^{+15}\%$ \\[2pt]
Medium nearby ($20\,\mathrm{M}_\oplus \leq M_\mathrm{comp}<80\,\mathrm{M}_\oplus$) & $8.7_{-3.7}^{+4.9}\%$ & $7.9_{-5.0}^{+8.5}\%$ & $11.2_{-5.0}^{+6.7}\%$ & $13.4_{-6.3}^{+8.7}\%$ & $7.1_{-5.3}^{+10.6}\%$ & $23_{-11}^{+14}\%$ \\[2pt]
Giant nearby ($M_\mathrm{comp}\geq 80\,\mathrm{M}_\oplus$) & $9.9_{-3.4}^{+4.4}\%$ & $3.2_{-2.4}^{+5.0}\%$ & $15.5_{-5.3}^{+6.5}\%$ & $2.9_{-2.1}^{+4.5}\%$ & $5.4_{-4.0}^{+8.3}\%$ & $5.5_{-4.1}^{+8.3}\%$ \\[2pt]
\noalign{\smallskip}\hline
\end{tabular}
}
\end{table*}

\FloatBarrier 
\section{Occurrence rates by eccentricity class} \label{app:ecc}
\begin{figure*}
    \centering
    \includegraphics[width=1.99\columnwidth]{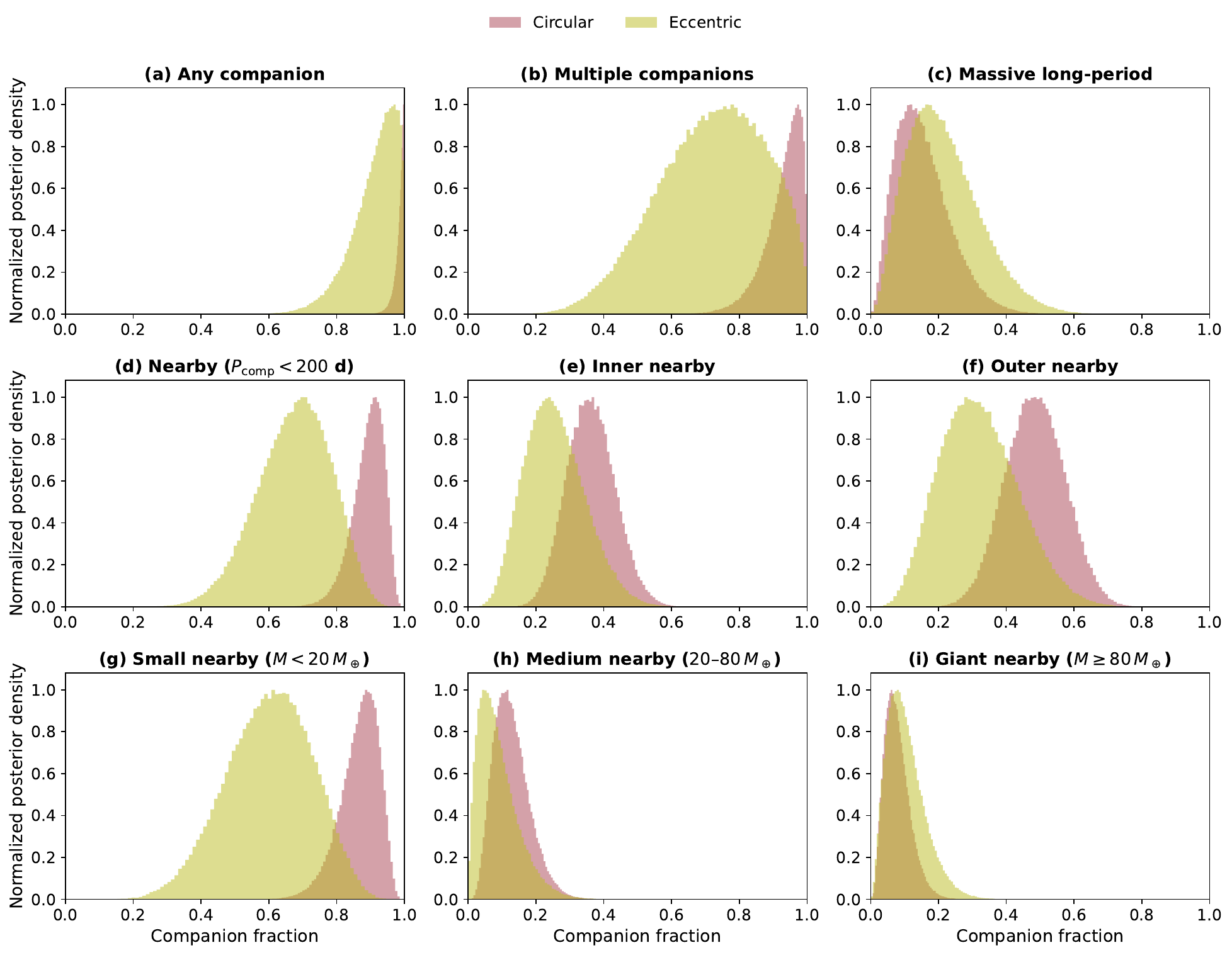} 
    \caption{Same as Fig.~\ref{fig:post_rad}, but for the two eccentricity classes: circular ($e< 0.1$) and eccentric ($e \geq 0.1$).}
    \label{fig:post_ecc}
\end{figure*}
 
\begin{table*}
\centering
\caption{Same as Table~\ref{tab:rates_radius}, but for the sub-Saturn eccentricity classes (boundary $e = 0.1$; the four sub-Saturns without an eccentricity constraint are excluded).}
\label{tab:rates_ecc}
\resizebox{1.99\columnwidth}{!}{
\begin{tabular}{lcccccc}
\hline\hline
\noalign{\smallskip}
 & \multicolumn{3}{c}{Circular ($N=54$)} & \multicolumn{3}{c}{Eccentric ($N=28$)}\\
Category & All & D+R ($N=27.5$) & Sav ($N=26.5$) & All & D+R ($N=9$) & Sav ($N=19$)\\
\noalign{\smallskip}\hline\noalign{\smallskip}
Any companion & $98.94_{-1.64}^{+0.76}\%$ & $92.2_{-11.3}^{+5.8}\%$ & $99.76_{-0.55}^{+0.19}\%$ & $92.4_{-8.0}^{+5.0}\%$ & $91.8_{-13.0}^{+6.2}\%$ & $95.2_{-6.3}^{+3.4}\%$ \\[2pt]
Multiple companions & $94.1_{-6.6}^{+3.8}\%$ & $72_{-23}^{+18}\%$ & $98.3_{-2.9}^{+1.2}\%$ & $73_{-17}^{+15}\%$ & $71_{-25}^{+19}\%$ & $81_{-16}^{+12}\%$ \\[2pt]
Massive long-period ($M_\mathrm{comp} \geq 1\,\mathrm{M}_\mathrm{Jup}$, $P_\mathrm{comp}>365$\,d) & $14.3_{-6.7}^{+9.2}\%$ & $15.4_{-9.6}^{+15.1}\%$ & $17.7_{-8.9}^{+12.2}\%$ & $20.2_{-9.3}^{+12.1}\%$ & $31_{-17}^{+21}\%$ & $18_{-10}^{+14}\%$ \\[2pt]
Nearby ($P_\mathrm{comp}<200$\,d) & $90.0_{-5.2}^{+3.9}\%$ & $45_{-16}^{+18}\%$ & $96.8_{-2.9}^{+1.7}\%$ & $68_{-12}^{+10}\%$ & $44_{-19}^{+22}\%$ & $75.9_{-11.2}^{+9.1}\%$ \\[2pt]
Inner nearby ($P_\mathrm{comp}<P_\mathrm{ss}$) & $36.1_{-6.9}^{+7.3}\%$ & $6.0_{-3.8}^{+6.6}\%$ & $59.2_{-9.5}^{+9.0}\%$ & $25.0_{-8.2}^{+9.6}\%$ & $9.5_{-7.0}^{+13.7}\%$ & $34_{-11}^{+12}\%$ \\[2pt]
Outer nearby ($P_\mathrm{ss}<P_\mathrm{comp}<200$\,d) & $48.5_{-8.8}^{+8.8}\%$ & $16.6_{-8.3}^{+11.5}\%$ & $69.5_{-10.5}^{+9.2}\%$ & $31_{-11}^{+12}\%$ & $13.3_{-9.8}^{+18.3}\%$ & $42_{-14}^{+15}\%$ \\[2pt]
Small nearby ($M_\mathrm{comp}<20\,\mathrm{M}_\oplus$) & $87.5_{-6.4}^{+4.8}\%$ & $37_{-17}^{+20}\%$ & $95.2_{-4.2}^{+2.5}\%$ & $61_{-13}^{+12}\%$ & $25_{-18}^{+28}\%$ & $68_{-14}^{+12}\%$ \\[2pt]
Medium nearby ($20\,\mathrm{M}_\oplus \leq M_\mathrm{comp}<80\,\mathrm{M}_\oplus$) & $12.2_{-4.6}^{+5.9}\%$ & $6.7_{-4.3}^{+7.4}\%$ & $18.9_{-7.3}^{+9.1}\%$ & $7.8_{-4.4}^{+6.9}\%$ & $10.4_{-7.6}^{+14.8}\%$ & $10.3_{-5.8}^{+8.9}\%$ \\[2pt]
Giant nearby ($M_\mathrm{comp}\geq 80\,\mathrm{M}_\oplus$) & $7.3_{-3.1}^{+4.2}\%$ & $2.8_{-2.1}^{+4.5}\%$ & $13.7_{-5.6}^{+7.3}\%$ & $9.5_{-4.5}^{+6.4}\%$ & $7.4_{-5.5}^{+11.1}\%$ & $13.4_{-6.3}^{+8.7}\%$ \\[2pt]
\noalign{\smallskip}\hline
\end{tabular}
}
\end{table*}

\FloatBarrier 
\section{Occurrence rates by stellar mass} \label{app:stellar_mass}
 
\begin{figure*}
    \centering
    \includegraphics[width=1.99\columnwidth]{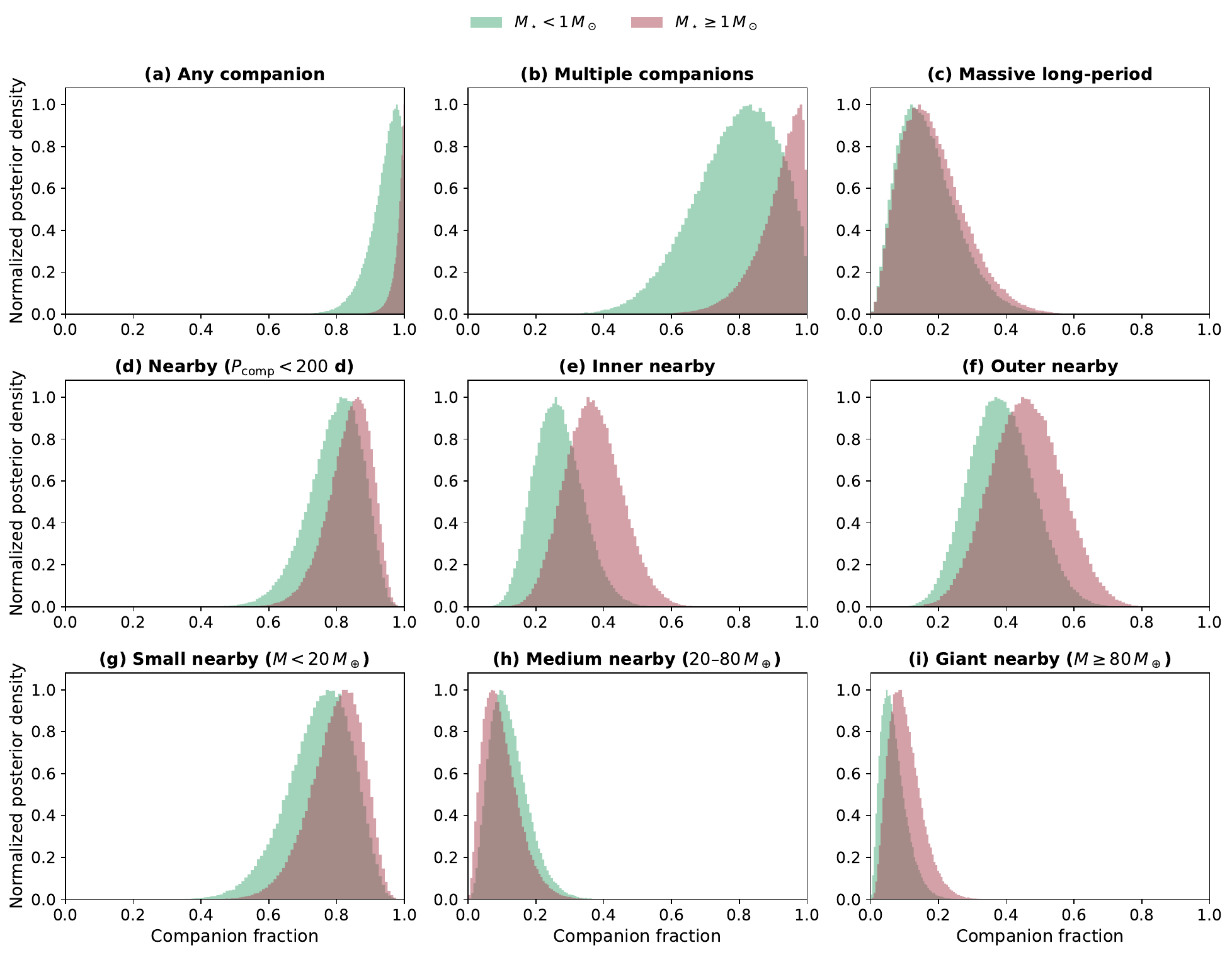} 
    \caption{Same as Fig.~\ref{fig:post_rad}, but for the two host-star mass classes: low-mass ($M_\star<1\,\mathrm{M}_\odot$) and high-mass ($M_\star \geq 1\,\mathrm{M}_\odot$).}
    \label{fig:post_stmass}
\end{figure*}

\begin{table*}
\centering
\caption{Same as Table~\ref{tab:rates_radius}, but for the host-star mass classes (boundary $1.0\,\mathrm{M}_\odot$).}
\label{tab:rates_stmass}
\resizebox{1.99\columnwidth}{!}{
\begin{tabular}{lcccccc}
\hline\hline
\noalign{\smallskip}
 & \multicolumn{3}{c}{$M_\star < 1\,M_\odot$ ($N=46$)} & \multicolumn{3}{c}{$M_\star \geq 1\,M_\odot$ ($N=40$)}\\
Category & All & D+R ($N=27.5$) & Sav ($N=18.5$) & All & D+R ($N=10$) & Sav ($N=30$)\\
\noalign{\smallskip}\hline\noalign{\smallskip}
Any companion & $95.0_{-5.4}^{+3.3}\%$ & $88.3_{-13.5}^{+8.2}\%$ & $98.87_{-2.12}^{+0.85}\%$ & $98.72_{-2.20}^{+0.95}\%$ & $96.0_{-9.5}^{+3.3}\%$ & $99.04_{-1.76}^{+0.72}\%$ \\[2pt]
Multiple companions & $80_{-14}^{+11}\%$ & $63_{-23}^{+22}\%$ & $93.8_{-8.2}^{+4.3}\%$ & $93.1_{-8.3}^{+4.6}\%$ & $83_{-24}^{+13}\%$ & $94.6_{-7.1}^{+3.8}\%$ \\[2pt]
Massive long-period ($M_\mathrm{comp} \geq 1\,\mathrm{M}_\mathrm{Jup}$, $P_\mathrm{comp}>365$\,d) & $15.8_{-7.4}^{+9.9}\%$ & $22_{-11}^{+14}\%$ & $14.0_{-8.8}^{+14.1}\%$ & $17.2_{-8.0}^{+10.7}\%$ & $20_{-15}^{+25}\%$ & $19.8_{-9.1}^{+11.9}\%$ \\[2pt]
Nearby ($P_\mathrm{comp}<200$\,d) & $80.3_{-8.8}^{+7.0}\%$ & $31_{-13}^{+17}\%$ & $94.9_{-5.0}^{+2.9}\%$ & $84.4_{-7.2}^{+5.6}\%$ & $65_{-22}^{+19}\%$ & $86.8_{-6.8}^{+5.1}\%$ \\[2pt]
Inner nearby ($P_\mathrm{comp}<P_\mathrm{ss}$) & $26.3_{-6.7}^{+7.5}\%$ & $6.0_{-3.8}^{+6.6}\%$ & $50_{-11}^{+11}\%$ & $36.7_{-8.1}^{+8.6}\%$ & $9.0_{-6.7}^{+13.1}\%$ & $45.6_{-9.5}^{+9.7}\%$ \\[2pt]
Outer nearby ($P_\mathrm{ss}<P_\mathrm{comp}<200$\,d) & $38.3_{-9.2}^{+9.7}\%$ & $9.1_{-5.8}^{+9.7}\%$ & $65_{-13}^{+11}\%$ & $46_{-10}^{+11}\%$ & $29_{-16}^{+20}\%$ & $52_{-12}^{+12}\%$ \\[2pt]
Small nearby ($M_\mathrm{comp}<20\,\mathrm{M}_\oplus$) & $75.8_{-10.4}^{+8.5}\%$ & $21_{-13}^{+19}\%$ & $92.2_{-7.3}^{+4.3}\%$ & $80.7_{-8.7}^{+6.8}\%$ & $54_{-26}^{+24}\%$ & $82.6_{-8.7}^{+6.6}\%$ \\[2pt]
Medium nearby ($20\,\mathrm{M}_\oplus \leq M_\mathrm{comp}<80\,\mathrm{M}_\oplus$) & $11.4_{-4.7}^{+6.3}\%$ & $6.6_{-4.2}^{+7.2}\%$ & $20.4_{-8.8}^{+11.1}\%$ & $9.0_{-4.3}^{+6.1}\%$ & $10.2_{-7.5}^{+14.6}\%$ & $11.0_{-5.2}^{+7.3}\%$ \\[2pt]
Giant nearby ($M_\mathrm{comp}\geq 80\,\mathrm{M}_\oplus$) & $6.2_{-3.0}^{+4.3}\%$ & $2.8_{-2.1}^{+4.4}\%$ & $13.8_{-6.5}^{+8.9}\%$ & $9.6_{-4.0}^{+5.4}\%$ & $7.2_{-5.3}^{+10.8}\%$ & $12.2_{-5.0}^{+6.7}\%$ \\[2pt]
\noalign{\smallskip}\hline
\end{tabular}
}
\end{table*}

\FloatBarrier 
\section{Occurrence rates by host star metallicity} \label{app:stellar_met}
 
\begin{figure*}
    \centering
    \includegraphics[width=1.99\columnwidth]{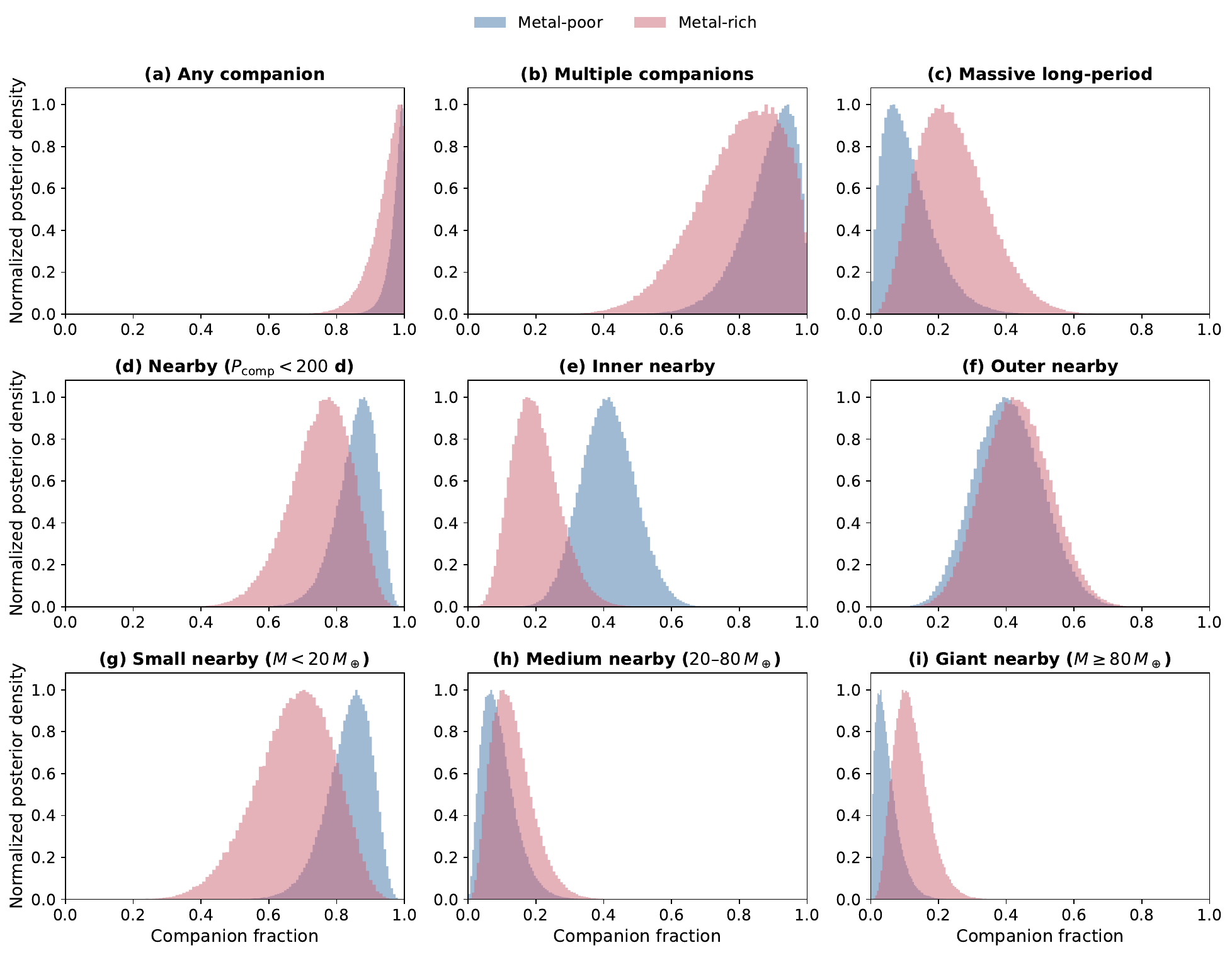} 
    \caption{Same as Fig.~\ref{fig:post_rad}, but for the two host-star metallicity classes: metal-poor ($\mathrm{[Fe/H]} < 0.17$\,dex) and metal-rich ($\mathrm{[Fe/H]} \geq 0.17$\,dex).}
    \label{fig:post_stmet}
\end{figure*}

\begin{table*}
\centering
\caption{Same as Table~\ref{tab:rates_radius}, but for the host-star metallicity classes (boundary at the sample median $\mathrm{[Fe/H]} = 0.17$\,dex).}
\label{tab:rates_met}
\resizebox{1.99\columnwidth}{!}{
\begin{tabular}{lcccccc}
\hline\hline
\noalign{\smallskip}
 & \multicolumn{3}{c}{Metal-poor ($N=43$)} & \multicolumn{3}{c}{Metal-rich ($N=43$)}\\
Category & All & D+R ($N=14$) & Sav ($N=29$) & All & D+R ($N=23.5$) & Sav ($N=19.5$)\\
\noalign{\smallskip}\hline\noalign{\smallskip}
Any companion & $97.9_{-2.7}^{+1.4}\%$ & $90.4_{-14.4}^{+7.2}\%$ & $98.98_{-1.66}^{+0.74}\%$ & $95.5_{-5.5}^{+3.1}\%$ & $92.6_{-11.0}^{+5.5}\%$ & $98.4_{-3.0}^{+1.2}\%$ \\[2pt]
Multiple companions & $90.0_{-9.2}^{+6.1}\%$ & $68_{-26}^{+21}\%$ & $94.3_{-6.7}^{+3.7}\%$ & $82_{-14}^{+11}\%$ & $73_{-23}^{+17}\%$ & $91.6_{-10.4}^{+5.8}\%$ \\[2pt]
Massive long-period ($M_\mathrm{comp} \geq 1\,\mathrm{M}_\mathrm{Jup}$, $P_\mathrm{comp}>365$\,d) & $10.0_{-5.6}^{+8.6}\%$ & $14_{-10}^{+18}\%$ & $12.8_{-7.1}^{+10.8}\%$ & $23.4_{-9.3}^{+11.5}\%$ & $26_{-13}^{+16}\%$ & $25_{-12}^{+16}\%$ \\[2pt]
Nearby ($P_\mathrm{comp}<200$\,d) & $86.5_{-6.5}^{+4.9}\%$ & $36_{-17}^{+22}\%$ & $91.4_{-5.3}^{+3.6}\%$ & $75.8_{-9.8}^{+8.2}\%$ & $47_{-17}^{+18}\%$ & $88.4_{-8.3}^{+5.6}\%$ \\[2pt]
Inner nearby ($P_\mathrm{comp}<P_\mathrm{ss}$) & $41.2_{-7.8}^{+8.0}\%$ & $6.7_{-4.9}^{+10.0}\%$ & $54.2_{-9.2}^{+8.9}\%$ & $19.4_{-6.3}^{+7.5}\%$ & $6.9_{-4.4}^{+7.6}\%$ & $35_{-11}^{+12}\%$ \\[2pt]
Outer nearby ($P_\mathrm{ss}<P_\mathrm{comp}<200$\,d) & $40.5_{-9.7}^{+10.2}\%$ & $9.9_{-7.3}^{+14.3}\%$ & $53_{-12}^{+11}\%$ & $42.9_{-9.7}^{+10.1}\%$ & $18.6_{-9.3}^{+12.6}\%$ & $64_{-13}^{+12}\%$ \\[2pt]
Small nearby ($M_\mathrm{comp}<20\,\mathrm{M}_\oplus$) & $84.3_{-7.4}^{+5.7}\%$ & $22_{-16}^{+26}\%$ & $89.5_{-6.3}^{+4.4}\%$ & $68_{-12}^{+11}\%$ & $39_{-18}^{+20}\%$ & $80.6_{-12.8}^{+9.2}\%$ \\[2pt]
Medium nearby ($20\,\mathrm{M}_\oplus \leq M_\mathrm{comp}<80\,\mathrm{M}_\oplus$) & $8.3_{-4.0}^{+5.6}\%$ & $7.3_{-5.5}^{+11.0}\%$ & $11.2_{-5.3}^{+7.4}\%$ & $12.3_{-5.1}^{+6.7}\%$ & $7.7_{-4.9}^{+8.3}\%$ & $19.9_{-8.5}^{+10.9}\%$ \\[2pt]
Giant nearby ($M_\mathrm{comp}\geq 80\,\mathrm{M}_\oplus$) & $4.2_{-2.4}^{+3.8}\%$ & $5.1_{-3.8}^{+7.9}\%$ & $5.9_{-3.4}^{+5.4}\%$ & $11.4_{-4.3}^{+5.5}\%$ & $3.2_{-2.4}^{+5.2}\%$ & $22.2_{-8.0}^{+9.7}\%$ \\[2pt]
\noalign{\smallskip}\hline
\end{tabular}
}
\end{table*}

\FloatBarrier
\section{RV observations}
 
\begin{table*}
    \centering
    \resizebox{\textwidth}{!}{
    \begin{threeparttable}[b]
    \caption{\label{tab:rv_observations} Number of radial velocity observations per instrument for each sub-Saturn system.}
\begin{tabular}{l*{22}{c}l}
\hline\hline
\noalign{\smallskip}
System & APF & CAFE & CARMENES & CHIRON & CORALIE & ESPRESSO & FEROS & FIES & HARPS & HARPS-N & HIRES & HPF & MaHPS & MAROON-X & NEID & PARAS & PFS & SOPHIE & TRES & iSHELL & NIRPS & Total & References \\
\noalign{\smallskip}
\hline
\noalign{\smallskip}
CoRoT-8 & ... & ... & ... & ... & ... & ... & ... & ... & 19 & ... & ... & ... & ... & ... & ... & ... & ... & 2 & ... & ... & ... & 21 & 1\\
GJ 3470 & ... & ... & ... & ... & ... & ... & ... & ... & 52 & ... & 56 & ... & ... & ... & ... & ... & ... & ... & ... & ... & ... & 108 & 2,3 \\
GJ 436 & ... & ... & 113 & ... & ... & ... & ... & ... & 130 & ... & 356 & ... & ... & ... & ... & ... & ... & ... & ... & ... & ... &599 & 4,5,6\\
HAT-P-11 & ... & ... & ... & ... & ... & ... & ... & ... & ... & ... & 253 & ... & ... & ... & ... & ... & ... & ... & ... & ... & ... &253 & 7 \\
HAT-P-26 & ... & ... & ... & ... & ... & ... & ... & ... & ... & ... & 26 & ... & ... & ... & ... & ... & ... & ... & ... & ... & ... &26 & 8 \\
HATS-38 & ... & ... & ... & ... & ... & ... & 10 & ... & 18 & ... & ... & ... & ... & ... & ... & ... & 4 & ... & ... & ... & ... &32 & 9,10 \\
HD 106315 & 125 & ... & ... & ... & ... & ... & ... & ... & 130 & ... & 352 & ... & ... & ... & ... & ... & 24 & ... & ... & ... & ... &631 & 11,12 \\
HD 219666 & ... & ... & ... & ... & ... & ... & ... & ... & 21 & ... & ... & ... & ... & ... & ... & ... & ... & ... & ... & ... & ... &21 & 13\\
HD 89345 & 9 & ... & ... & ... & ... & ... & ... & 16 & 38 & 12 & 12 & ... & ... & ... & ... & ... & ... & ... & ... & ... & ... &87 & 14,15 \\
HD 149026 & ... & ... & ... & ... & ... & ... & ... & ... & ... & ... & 42 & ... & ... & ... & ... & ... & ... & ... & ... & ... & ... &42 & 8 \\
HD 152843 & ... & ... & ... & ... & ... & ... & ... & ... & ... & 136 & ... & ... & ... & ... & ... & ... & ... & ... & ... & ... & ...  &136 & 16,17 \\
HD 28109 & ... & ... & ... & ... & ... & 44 & ... & ... & 37 & ... & ... & ... & ... & ... & ... & ... & ... & ... & ... & ... & ... &81 & 18\\
HD 95338 & ... & ... & ... & ... & ... & ... & ... & ... & 11 & ... & ... & ... & ... & ... & ... & ... & 81 & ... & ... & ... & ... &92 & 19 \\
K2-19 & ... & ... & ... & ... & ... & ... & ... & 10 & 11 & 9 & 51 & ... & ... & ... & ... & ... & 61 & ... & ... & ... & ... &142 & 21,22,23 \\
K2-24 & ... & ... & ... & ... & ... & ... & ... & ... & 10 & ... & 64 & ... & ... & ... & ... & ... & 16 & ... & ... & ... & ... &90 & 24,25,26 \\
K2-27 & ... & ... & ... & ... & ... & ... & ... & 6 & 6 & 19 & 15 & ... & ... & ... & ... & ... & ... & ... & ... & ... & ... &46 & 27,28 \\
K2-32 & ... & ... & ... & ... & ... & ... & ... & ... & 242 & ... & 31 & ... & ... & ... & ... & ... & 6 & ... & ... & ... & ... &279 & 24,28,30\\
K2-39 & ... & ... & ... & ... & ... & ... & ... & 17 & 7 & ... & 42 & ... & ... & ... & ... & ... & 6 & ... & ... & ... & ... &72 & 28,31\\
K2-55 & ... & ... & ... & ... & ... & ... & ... & ... & ... & ... & 12 & ... & ... & ... & ... & ... & ... & ... & ... & ... & ... &12 & 32\\
K2-79 & ... & ... & ... & ... & ... & ... & ... & ... & ... & 77 & 62 & ... & ... & ... & ... & ... & ... & ... & ... & ... & ... &139 & 20,33,34 \\
K2-280 & ... & ... & ... & ... & ... & ... & ... & 6 & 18 & 14 & 16 & ... & ... & ... & ... & ... & ... & ... & ... & ... & ... &54 & 20,29 \\
K2-121 & ... & ... & ... & ... & ... & ... & ... & ... & ... & ... & 18 & ... & ... & ... & ... & ... & ... & ... & ... & ... & ... &18 & 20 \\
Kepler-4 & ... & ... & ... & ... & ... & ... & ... & ... & ... & ... & 19 & ... & ... & ... & ... & ... & ... & ... & ... & ... & ... &19 & 41\\
Kepler-9 & ... & ... & ... & ... & ... & ... & ... & ... & ... & 30 & ... & ... & ... & ... & ... & ... & ... & ... & ... & ... & ... &30 & 44 \\
Kepler-18 & ... & ... & ... & ... & ... & ... & ... & ... & ... & ... & 25 & ... & ... & ... & ... & ... & ... & ... & ... & ... & ... &25 & 40\\
Kepler-25 & ... & ... & ... & ... & ... & ... & ... & ... & ... & ... & 101 & ... & ... & ... & ... & ... & ... & ... & ... & ... & ... &101 & 40 \\
Kepler-56 & ... & ... & ... & ... & ... & ... & ... & ... & ... & 19 & 24 & ... & ... & ... & ... & ... & ... & ... & ... & ... & ... &43 & 43 \\
Kepler-101 & ... & ... & ... & ... & ... & ... & ... & ... & ... & 40 & ... & ... & ... & ... & ... & ... & ... & ... & ... & ... & ... &40 & 35 \\
Kepler-103 & ... & ... & ... & ... & ... & ... & ... & ... & ... & 60 & ... & ... & ... & ... & ... & ... & ... & ... & ... & ... & ... &60 & 36\\
Kepler-111 & ... & ... & ... & ... & ... & ... & ... & ... & ... & ... & 8 & ... & ... & ... & ... & ... & ... & ... & ... & ... & ... &8 & 37 \\
Kepler-411 & ... & ... & ... & ... & ... & ... & ... & ... & ... & ... & 14 & ... & ... & ... & ... & ... & ... & ... & ... & ... & ... &14 & 40\\
Kepler-539 & ... & 20 & ... & ... & ... & ... & ... & ... & ... & ... & ... & ... & ... & ... & ... & ... & ... & ... & ... & ... & ... &20 & 42\\
Kepler-849 & ... & ... & ... & ... & ... & ... & ... & ... & ... & ... & 16 & ... & ... & ... & ... & ... & ... & ... & ... & ... & ... &16 & 37 \\
Kepler-1656 & ... & ... & ... & ... & ... & ... & ... & ... & ... & ... & 150 & ... & ... & ... & ... & ... & ... & ... & ... & ... & ... &150 & 38,39 \\
KOI-351 & ... & ... & ... & ... & ... & ... & ... & ... & ... & ... & 34 & ... & ... & ... & ... & ... & ... & ... & ... & ... & ... &34 & 40\\
LP 714-47 & ... & ... & 33 & ... & ... & 19 & ... & ... & ... & ... & 14 & ... & ... & ... & ... & ... & 6 & ... & ... & 9 & ... &81 & 45 \\
LTT 9779 & ... & ... & ... & ... & 18 & ... & ... & ... & 32 & ... & ... & ... & ... & ... & ... & ... & ... & ... & ... & ... & ... &50 & 46 \\
NGTS-14 A & ... & ... & ... & ... & ... & ... & ... & ... & 15 & ... & ... & ... & ... & ... & ... & ... & ... & ... & ... & ... & ... &15 & 47\\
TOI-181 & ... & ... & ... & ... & ... & ... & ... & ... & 23 & ... & ... & ... & ... & ... & ... & ... & ... & ... & ... & ... & ... &23 & 61\\
TOI-216 & ... & ... & ... & ... & ... & ... & 25 & ... & 15 & ... & ... & ... & ... & ... & ... & ... & 18 & ... & ... & ... & ... &58 & 65\\
TOI-257 & ... & ... & ... & ... & ... & ... & 6 & ... & 29 & ... & ... & ... & ... & ... & ... & ... & ... & ... & ... & ... & ... &35 & 68 \\
TOI-329 & ... & ... & ... & ... & ... & ... & ... & ... & ... & ... & 43 & ... & ... & ... & ... & ... & ... & ... & ... & ... & ... &43 & 49\\
TOI-333 & ... & ... & ... & ... & ... & ... & 7 & ... & 37 & ... & ... & ... & ... & ... & ... & ... & ... & ... & ... & ... & ... &44 & 69 \\
TOI-421 & ... & ... & ... & ... & ... & ... & ... & ... & 105 & ... & 33 & ... & ... & ... & ... & ... & 9 & ... & ... & ... & ... &147 & 79 \\
TOI-532 & ... & ... & ... & ... & ... & ... & ... & ... & ... & ... & ... & 18 & ... & ... & ... & ... & ... & ... & ... & ... & ... &18 & 82 \\
TOI-672 & ... & ... & ... & ... & ... & ... & ... & ... & 16 & ... & ... & ... & ... & ... & ... & ... & ... & ... & ... & ... & 23 &39 & 87 \\
TOI-674 & ... & ... & ... & ... & ... & ... & ... & ... & 17 & ... & ... & ... & ... & ... & ... & ... & ... & ... & ... & ... & ... &17 & 88 \\
TOI-762 A & ... & ... & ... & ... & ... & 8 & ... & ... & ... & ... & ... & ... & ... & ... & ... & ... & ... & ... & ... & ... & ... &8 & 91 \\
TOI-1136 & 320 & ... & ... & ... & ... & ... & ... & ... & ... & 49 & 155 & ... & ... & ... & ... & ... & ... & ... & ... & ... & ... &524 & 48 \\
TOI-1248 & ... & ... & ... & ... & ... & ... & ... & ... & ... & ... & 19 & ... & ... & ... & ... & ... & ... & ... & ... & ... & ... &19 & 49 \\
TOI-1272 & ... & ... & ... & ... & ... & ... & ... & ... & 94 & ... & 62 & ... & ... & ... & ... & ... & ... & ... & ... & ... & ... &156 & 50,51 \\
TOI-1288 & ... & ... & ... & ... & ... & ... & ... & ... & ... & 57 & 28 & ... & ... & ... & ... & ... & ... & ... & ... & ... & ... &85 & 52 \\
TOI-1338 A & ... & ... & ... & ... & ... & 103 & ... & ... & 58 & ... & ... & ... & ... & ... & ... & ... & ... & ... & ... & ... & ... &161 & 53\\
TOI-1386 & ... & ... & ... & ... & ... & ... & ... & ... & ... & ... & 63 & ... & ... & ... & ... & ... & ... & ... & ... & ... & ... &63 & 54\\
TOI-1439 & ... & ... & ... & ... & ... & ... & ... & ... & ... & ... & 53 & ... & ... & ... & ... & ... & ... & ... & ... & ... & ... &53 & 49\\
TOI-1472 & ... & ... & ... & ... & ... & ... & ... & ... & ... & 52 & 22 & ... & ... & ... & ... & ... & ... & ... & ... & ... & ... &74 & 49,55 \\
TOI-1694 & ... & ... & ... & ... & ... & ... & ... & ... & 89 & ... & 20 & ... & ... & ... & ... & ... & ... & ... & ... & ... & ... &109 & 51,56 \\
TOI-1710 & 178 & ... & ... & ... & ... & ... & ... & ... & 85 & ... & 16 & ... & ... & ... & ... & ... & ... & ... & ... & ... & ... &279 & 49,57,58\\
TOI-1728 & ... & ... & ... & ... & ... & ... & ... & ... & ... & ... & ... & 30 & ... & ... & ... & ... & ... & ... & ... & ... & ... &30 & 59 \\
TOI-1775 & ... & ... & ... & ... & ... & ... & ... & ... & ... & ... & 16 & ... & ... & ... & ... & ... & ... & ... & ... & ... & ... &16 & 49\\
TOI-1803 & ... & ... & ... & ... & ... & ... & ... & ... & ... & 112 & ... & ... & ... & ... & ... & ... & ... & ... & ... & ... & ... &112 & 60 \\
TOI-1823 & 40 & ... & ... & ... & ... & ... & ... & ... & ... & ... & 16 & ... & ... & ... & ... & ... & ... & ... & ... & ... & ... &56 & 49\\
TOI-1836 & ... & ... & ... & ... & ... & ... & ... & ... & ... & ... & 53 & ... & ... & ... & ... & ... & ... & 85 & ... & ... & ... &138 & 49,62 \\
TOI-2000 & ... & ... & ... & 15 & ... & ... & 14 & ... & 41 & ... & ... & ... & ... & ... & ... & ... & ... & ... & ... & ... & ... &70 & 63 \\
TOI-2019 & ... & ... & ... & ... & ... & ... & ... & ... & ... & ... & 43 & ... & ... & ... & ... & ... & ... & ... & ... & ... & ... &43 & 49 \\
TOI-2134 & ... & ... & ... & ... & ... & ... & ... & ... & ... & 113 & ... & ... & ... & ... & ... & ... & ... & 108 & ... & ... & ... &221 & 64\\
TOI-2374 & ... & ... & ... & ... & ... & ... & ... & ... & 21 & ... & ... & ... & ... & ... & ... & ... & 7 & ... & ... & ... & ... &28 & 66 \\
TOI-2498 & ... & ... & ... & ... & ... & ... & ... & ... & 15 & ... & ... & ... & ... & ... & ... & ... & ... & ... & ... & ... & ... &15 & 67\\
TOI-3071 & ... & ... & ... & ... & ... & ... & ... & ... & 14 & ... & ... & ... & ... & ... & ... & ... & ... & ... & ... & ... & ... &14 & 66\\
TOI-3568 & ... & ... & ... & ... & ... & ... & ... & ... & ... & ... & ... & ... & ... & 34 & ... & ... & ... & ... & ... & ... & ... &34 & 70 \\
TOI-3629 & ... & ... & ... & ... & ... & ... & ... & ... & ... & ... & 9 & 23 & ... & ... & 5 & ... & ... & ... & ... & ... & ... & 37 &71,72 \\
TOI-3785 & ... & ... & ... & ... & ... & ... & ... & ... & ... & ... & ... & 29 & ... & ... & 10 & ... & ... & ... & ... & ... & ... &39 & 73 \\
TOI-3862 & ... & ... & ... & ... & ... & ... & ... & ... & 28 & ... & ... & ... & ... & ... & ... & ... & ... & ... & ... & ... & ... &28 & 74\\
TOI-3884 & ... & ... & ... & ... & ... & 2 & ... & ... & ... & ... & ... & 17 & ... & ... & ... & ... & ... & ... & ... & ... & ... &19 & 75,76 \\
TOI-3984 A & ... & ... & ... & ... & ... & ... & ... & ... & ... & ... & ... & 35 & ... & ... & 6 & ... & ... & ... & ... & ... & ... &41 & 77 \\
TOI-4010 & ... & ... & ... & ... & ... & ... & ... & ... & ... & 112 & ... & ... & ... & ... & ... & ... & ... & ... & ... & ... & ... &112 & 78 \\
TOI-5005 & ... & ... & ... & ... & ... & ... & ... & ... & 38 & ... & ... & ... & ... & ... & ... & ... & ... & ... & ... & ... & ... &38 & 80\\
TOI-5108 & ... & ... & ... & ... & ... & ... & ... & ... & ... & ... & ... & ... & 120 & ... & ... & ... & ... & 28 & ... & ... & ... &148 & 81 \\
TOI-5678 & ... & ... & ... & ... & 14 & ... & ... & ... & 23 & ... & ... & ... & ... & ... & ... & ... & ... & ... & ... & ... & ... &37 & 83 \\
TOI-5795 & ... & ... & ... & ... & ... & ... & ... & ... & 19 & ... & ... & ... & ... & ... & ... & ... & ... & ... & 12 & ... & ... &31 & 84 \\
TOI-6038 A & ... & ... & ... & ... & ... & ... & ... & ... & ... & ... & ... & ... & ... & ... & ... & 29 & ... & ... & ... & ... & ... &29 & 85 \\
TOI-6651 & ... & ... & ... & ... & ... & ... & ... & ... & ... & ... & ... & ... & ... & ... & ... & 27 & ... & ... & ... & ... & ... &27 & 86 \\
TOI-7510 & ... & ... & ... & ... & 30 & 7 & 27 & ... & 16 & ... & ... & ... & ... & ... & ... & ... & ... & ... & ... & ... & ... &80 & 89,90 \\
WASP-148 & ... & ... & ... & ... & ... & ... & ... & ... & ... & ... & ... & ... & ... & ... & ... & ... & ... & 116 & ... & ... & ... &116 & 92 \\
WASP-156 & ... & ... & ... & ... & 13 & ... & ... & ... & ... & ... & 20 & ... & ... & ... & ... & ... & ... & 44 & ... & ... & ... &77 & 49,93 \\
WASP-166 & ... & ... & ... & ... & 39 & ... & ... & ... & 27 & ... & ... & ... & ... & ... & ... & ... & ... & ... & ... & ... & ... &66 & 94 \\
\noalign{\smallskip}
\hline
\noalign{\smallskip}
Total & 672 & 20 & 146 & 15 & 114 & 183 & 89 & 55 & 1607 & 911 & 2484 & 152 & 120 & 34 & 21 & 56 & 238 & 383 & 12 & 9 & 23 & 7344 \\
\noalign{\smallskip}
\hline
\end{tabular}
        \begin{tablenotes}
       \item References: 1 \cite{borde2010}; 2 \cite{Bonfils2012}; 3 \cite{Kosiarek2019}; 4 \cite{Lanotte2014}; 5 \cite{Butler2017}; 6 \cite{trifonov2018carmenes}; 7 \cite{Yee2018}; 8 \cite{Knutson2014}; 9 \cite{jordan2020}; 10 \cite{retamal2024}; 11 \cite{Barros2017}; 12 \cite{Kosiarek2021}; 13 \cite{Esposito2019}; 14 \cite{vaneylen2018}; 15 \cite{yu2018}; 16 \cite{Eisner2021}; 17 \cite{Nicholson2014}; 18 \cite{Barone2025}; 19 \cite{diaz2020}; 20 \cite{Howard2025}; 21 \cite{Dai2016}; 22 \cite{Nespral2017}; 23 \cite{Petigura2020}; 24 \cite{Dai2016}; 25 \cite{Petigura2016}; 26 \cite{Petigura2018a}; 27 \cite{vaneylen2016}; 28 \cite{Petigura2017}; 29 \cite{Nowak2020}; 30 \cite{lillo2020}; 31 \cite{vaneylen2016b}; 32 \cite{Dressing2018}; 33 \cite{Nava2022}; 34 \cite{Bonomo2023}; 35 \cite{Bonomo2014}; 36 \cite{Dubber2019}; 37 \cite{Dalba2024}; 38 \cite{Brady2018}; 39 \cite{Angelo2022}; 40 \cite{Weiss2024}; 41 \cite{Borucki2010}; 42 \cite{Mancini2016}; 43 \cite{Otor2016}; 44 \cite{Borsato2019}; 45 \cite{Dreizler2020}; 46 \cite{Jenkins2020b}; 47 \cite{2Smith2021}; 48 \cite{Beard2024}; 49 \cite{Polanski2024}; 50 \cite{MacDougall2022}; 51 \cite{Mancini2026}; 52 \cite{Knudstrup2023}; 53 \cite{Standing2023}; 54 \cite{Hill2024}; 55 \cite{Carleo2026}; 56 \cite{zandt2023}; 57 \cite{konig2022}; 58 \cite{orell2024}; 59 \cite{Kanodia2020}; 60 \cite{Zingales2025}; 61 \cite{Mistry2023}; 62 \cite{Heidari2025}; 63 \cite{Sha2023}; 64 \cite{Rescigno2024}; 65 \cite{Dawson2021}; 66 \cite{Hacker2024}; 67 \cite{Frame2023}; 68 \cite{Addison2021}; 69 \cite{Alves2026}; 70 \cite{Martioli2024}; 71 \cite{canas2022}; 72 \cite{Hartman2023}; 73 \cite{Powers2023}; 74 \cite{Carleo2026b}; 75 \cite{Almenara2022}; 76 \cite{libby2023}; 77 \cite{canas2023}; 78 \cite{Kunimoto2023}; 79 \cite{Carleo2020}; 80 \cite{castro2024b}; 81 \cite{Thomas2025}; 82 \cite{Kanodia2021}; 83 \cite{ulmer2023}; 84 \cite{Manni2025}; 85 \cite{Baliwal2025}; 86 \cite{Baliwal2024}; 87 \cite{Osborn2026}; 88 \cite{Murgas2021}; 89 \cite{Brahm2025}; 90 \cite{Almenara2025}; 91 \cite{Hartman2024}; 92 \cite{hebrard2020}; 93 \cite{Demangeon2018}; 94 \cite{Hellier2019}.
       \end{tablenotes}
     \end{threeparttable}}
\end{table*}

\FloatBarrier
\end{appendix}
\end{document}